\documentclass[11pt]{article}

\usepackage{amssymb}
\usepackage{amsmath}
\usepackage{amscd}

\begin{document}

\newtheorem{theo}{\itshape \bfseries Theorem}[section]
\newtheorem{prop}[theo]{\itshape \bfseries Proposition}
\newtheorem{lem}[theo]{\itshape \bfseries Lemma}
\newtheorem{cor}[theo]{\itshape \bfseries Corollary}
\newtheorem{defin}[theo]{\itshape \bfseries Definition}
\newtheorem{remark}[theo]{\itshape\bfseries Remark}
\newtheorem{proof}{\sc Proof:}
\newtheorem{example}[theo]{\itshape \bfseries Example}

\renewcommand{\theequation}{\thesection.\arabic{equation}}
\renewcommand{\thesubsubsection}
{\thesubsection.\arabic{subsubsection}}

\newcommand{\dl}{\partial}
\newcommand{\la}{\lambda}
\newcommand{\al}{\alpha}
\newcommand{\T}{T^*}
\newcommand{\ph}{\varphi}
\newcommand{\R}{{\mathbb R}}
\newcommand{\C}{{\mathbb C}}
\newcommand{\Z}{{\mathbb Z}}
\newcommand{\g}{{\mathfrak g}}
\newcommand{\gs}{{\mathfrak g^*}}
\newcommand{\h}{{\mathfrak h}}
\newcommand{\ow}{\omega}
\newcommand{\gh}{\hat g}
\newcommand{\liou}{\vartheta}
\newcommand{\ti}{\tilde}
\newcommand{\tH}{\ti H}
\newcommand{\tx}{\ti x}
\newcommand{\tp}{\ti p}
\newcommand{\tr}{\ti r}
\newcommand{\ty}{\ti y}
\newcommand{\Nh}{{\hat N}}
\newcommand{\Uh}{W}
\newcommand{\Vh}{{\hat V}}
\newcommand{\fih}{{\hat \phi}}
\newcommand{\fio}{\phi}
\newcommand{\Fgr}{\mathfrak F}
\newcommand{\fas}{F}
\newcommand{\Eic}{{E}}
\newcommand{\quer}{L}
\newcommand{\UH}{\vec W}
\newcommand{\UZH}{\vec U}
\newcommand{\YtU}{Y^{\vec v} \times \Upsilon}
\newcommand{\HpH}{\hat \psi_H}
\newcommand{\Fol}{A}
\newcommand{\rfo}{r}
\newcommand{\Vo}{\vec{V}_0}
\newcommand{\Uo}{\vec{\Wod}_0}
\newcommand{\UZo}{\vec U_0}
\newcommand{\Hmaph}{\hat \psi_H}
\newcommand{\Hmap}{\psi_H}
\newcommand{\lion}{\vartheta}
\newcommand{\liox}{\vartheta^X}
\newcommand{\baro}{\bar \rho}
\newcommand{\rok}{\bar \rho^\sharp}
\newcommand{\rovv}{\bar \rho^{\vec v}}
\newcommand{\roo}{\bar \rho^0}
\newcommand{\gar}{I}
\newcommand{\mau}{\theta}
\newcommand{\formHO}{\Omega}
\newcommand{\formHOH}{\hat \Omega}
\newcommand{\ios}{\tilde i_O}

\newcommand{\Ahs}{\hat A^{0}}
\newcommand{\Fhl}{{\hat F}}
\newcommand{\Fhs}{{\hat F^{0}}}
\newcommand{\fhl}{F}
\newcommand{\fhs}{{F^{0}}}
\newcommand{\Povy}{\upsilon}
\newcommand{\OOH}{\ow_H^O}
\newcommand{\HOH}{\hat \ow_H^O}

\newcommand{\beq}{\begin{equation}}
\newcommand{\eeq}{\end{equation}}
\newcommand{\bes}{\begin{equation*}}
\newcommand{\ees}{\end{equation*}}
\newcommand{\bea}{\begin{eqnarray}}
\newcommand{\eea}{\end{eqnarray}}

\newcommand{\qed}{\hfill $\square$}
\newcommand{\bems}{\hfill $\lozenge$}

\newcommand{\ZQ}{Z^Q}
\newcommand{\piz}{\pi_Z}
\newcommand{\etaz}{\eta_Z}
\newcommand{\pr}{p}
\newcommand{\Wek}{\mathfrak X}
\newcommand{\Frm}{\Omega}
\newcommand{\Casi}{\mathfrak Z}
\newcommand{\Poi}{Pois}
\newcommand{\Hae}{\mathfrak{Ham}}
\newcommand{\Sekt}{\Gamma}
\newcommand{\Nuli}{\mathfrak C}

\newcommand{\Pno}{\mathfrak {Pr}}
\newcommand{\li}{li}
\newcommand{\gag}{Gau}
\newcommand{\mom}{\la}
\newcommand{\gaa}{\mathfrak{gau}}
\newcommand{\Uver}{U^\circ}
\newcommand{\TWV}{T\Uh|_{\Vh}/T\Vh}

\newcommand{\TWO}{(T \Vh)^0}

\newcommand{\po}{\varpi}
\newcommand{\Inc}{{\mathfrak A}}
\newcommand{\Kru}{\Phi}

\newcommand{\Lie}{{\rm L}}

\newcommand{\XO}{(\underline{TX})^0}
\newcommand{\eha}{\frac{1}{2}}
\newcommand{\Wod}{W}

\newcommand{\co}{Q}
\newcommand{\ch}{\hat Q}
\newcommand{\cn}{N}

\newcommand{\Co}{\mathfrak C}
\newcommand{\Ch}{\mathfrak {\hat C}}
\newcommand{\fra}{\mathfrak}
\newcommand{\hp}{\hspace{0.06cm}}
\newcommand{\ha}{\mathfrak{c}}
\newcommand{\hs}{\mathfrak{c}^0}

\newcommand{\mc}{\varrho}
\newcommand{\Fk}{F}
\newcommand{\kal}{\eta}
\newcommand{\frag}{tr}
\newcommand{\RS}{{\sf R}}
\newcommand{\BS}{{\sf B}}
\newcommand{\PS}{{\sf P}}

\newcommand{\UX}{U'}

% Bedeutung geaendert:

\newcommand{\TSO}{{\fb}}
\newcommand{\rev}{r_S}
\newcommand{\reb}{\bar r_S}

\newcommand{\cb}{\ti \Uh}
\newcommand{\cf}{\ti \fas}
\newcommand{\hig}{\ti U}
\newcommand{\cp}{\ti \rho}
\newcommand{\Ha}{\bar H}
\newcommand{\TU}{\jb}
\newcommand{\TVO}{\fb}
\newcommand{\TXo}{\jx}
\newcommand{\TxU}{\T Z|_X}
\newcommand{\Ahl}{A}
\newcommand{\Pa}{R}
\newcommand{\pa}{r}
\newcommand{\Pha}{\mathfrak {Ph}}

% Name und Bedeutung geaendert:

\newcommand{\TZS}{\eb}
\newcommand{\qq}{q}
\newcommand{\cm}{\hat q}

% neu kopiert:

\newcommand{\cre}{\mu}
\newcommand{\joH}{\kaq}
\newcommand{\jh}{\ti{jh}}
\newcommand{\Ga}{\ti \KK}
\newcommand{\Lg}{G}
\newcommand{\Lc}{C}
\newcommand{\ka}{\ti{\fra{g}}}
\newcommand{\Na}{N_{Aut(\g)}(\ha)}
\newcommand{\N}{N}
\newcommand{\Pred}{\Pa_X^{\ha}}
\newcommand{\G}{G}

% neu kopiert und Bedeutung geaendert:

\newcommand{\Fred}{\ti \fb}

% neu kopiert, Name und Bedeutung geaendert:

\newcommand{\nq}{\ti{\fra n}}

% ganz neu im artikel:

\newcommand{\eb}{L^*}
\newcommand{\fb}{L}
\newcommand{\jb}{E}
\newcommand{\db}{E^*}
\newcommand{\fx}{L_X}
\newcommand{\jx}{E_X}
\newcommand{\dxb}{E_X^*}
\newcommand{\dfb}{L_X^*}
\newcommand{\ZW}{Z}
\newcommand{\gl}{{\g_{\tt L}}}
\newcommand{\gsl}{{\g_{\tt L}^*}}
\newcommand{\HB}{H_0}
\newcommand{\HBH}{H_{\hat 0}}
\newcommand{\HP}{K_0}
\newcommand{\HX}{H'_0}
\newcommand{\HXH}{H'_{\hat 0}}
\newcommand{\HSW}{H'_1}
\newcommand{\HEM}{H'_2}
\newcommand{\KK}{K}
\newcommand{\KEM}{K'_2}
\newcommand{\ZX}{Z'}
\newcommand{\econ}{\al}
\newcommand{\emet}{\gamma}
\newcommand{\bmet}{\gamma}
\newcommand{\pmet}{\kappa}
\newcommand{\ymet}{\kappa}
\newcommand{\poka}{\mu}
\newcommand{\pomo}{\psi}
\newcommand{\scalg}{\chi}
\newcommand{\napro}{\bp_{\gsl}}
\newcommand{\canco}{\hat \rho^*}
\newcommand{\lanco}{\hat \la^*}
\newcommand{\cx}{A_X}
\newcommand{\jq}{\ti E_X}
\newcommand{\fq}{\ti L_X}
\newcommand{\djq}{\ti E_X^*}
\newcommand{\dfq}{\ti L_X^*}
\newcommand{\Pq}{\ti \Pa_X}
\newcommand{\Nq}{\ti{N}}
\newcommand{\Iq}{\mathcal C}
\newcommand{\Hq}{\ti H}
\newcommand{\Zhig}{\ti Z}
\newcommand{\Ho}{H_{\Zhig}}
\newcommand{\HL}{H'_{20}}
\newcommand{\Ihq}{\hat{\mathcal C}}
\newcommand{\yp}{p}
\newcommand{\acon}{\al}
\newcommand{\ank}{-\po^\sharp}
\newcommand{\bP}{P^\pi}
\newcommand{\bp}{p^\pi}
\newcommand{\bPa}{\Pa_X^\pi}
\newcommand{\bfx}{\fx^\pi}
\newcommand{\vef}{\mathcal X} % statt \Xi
\newcommand{\vefal}{\mathcal Y}
\newcommand{\vfrm}{\beta}
\newcommand{\vct}{X}
\newcommand{\cov}{q}
\newcommand{\Za}{\mathfrak Z}
\newcommand{\Noa}{\mathfrak N}
\newcommand{\No}{\fra N}

\newcommand{\Noh}{\hat \No}
\newcommand{\OHa}{J}
\newcommand{\kaq}{\ti \kal}
\newcommand{\KA}{\ti{\fra P}}
\newcommand{\PA}{\fra P}
\newcommand{\Ahut}{\hat \Ahl}
\newcommand{\inn}{{\rm i}}
\newcommand{\Tan}{T}
\newcommand{\jet}{J}
\newcommand{\epro}{\tau}
\newcommand{\Cox}{\mathfrak C_X}
\newcommand{\chela}{l}
\newcommand{\SX}{S'}
\newcommand{\baSE}{\bar\Sekt}
\newcommand{\oli}{\overline}
\newcommand{\Nau}{N}
\newcommand{\Zen}{Z}
\newcommand{\Ze}{\Nau_{\g/\ha}}
\newcommand{\admo}{k}
\newcommand{\Stab}{K}
\newcommand{\stab}{\fra k}
\newcommand{\Te}{a}
\newcommand{\tila}{\ti\la}
\newcommand{\hor}{h^\acon}
\newcommand{\lio}{\mathcal I}
\newcommand{\mY}{\mathcal Y}
\newcommand{\adj}{hor_H}
\newcommand{\ada}{\oli{ad}}
\newcommand{\wep}{\pi}

\title{Gauge fields and Sternberg-Weinstein approximation of Poisson manifolds}
\vspace{1cm}

\author{\vspace{1.0cm}\\
Oliver Maspfuhl
\vspace{0.8cm}\\
EPF Lausanne - DMA\\
CH-1015 Lausanne\\
Switzerland\\
e-mail: oliver.maspfuhl@epfl.ch\\
\vspace{0.7cm}}

\maketitle

\vspace{2cm}

\abstract{The motion of a classical particle in a gravitational and a
Yang-Mills field was described by S. Sternberg and A. Weinstein by a
particular Hamiltonian system on a Poisson manifold known under the
name of Sternberg-Weinstein phase space. This system leads to
the generalization of the Lorentz equation of motion first discovered by Wong.

The aim of this work is to show that inversely, a Hamiltonian $H$ on a general Poisson manifold, with the property that its differential vanishes on a Lagrangian submanifold $X$ of a symplectic leaf and is generic in any other direction, naturally defines a metric on $X$, as well as a principal connection form on a canonical principal fiber bundle on $X$. These fields, which are credited to model a gravitational and a Yang-Mills field on $X$, respectively, define a linearized Hamiltonian system of Wong type on a canonical linearized Poisson manifold at $X$ locally isomorphic to a Sternberg-Weinstein phase space. In addition, $H$ is shown to define scalar fields which first appeared in a theory of Einstein and Mayer. In the presence  of a coisotropic constraint, the reduced system can be regarded as the phase space of particles in gravitational, Yang-Mills and Higgs fields. We further show that all our constructions are locally related to usual gauge and Kaluza-Klein theory via symplectic realization.

%We show further that H defines canonical fields at the second order, which play the rôle of Higgs fields if the system is coisotropically constrained. In that case, we find the gauge fields produced by dimensional reduction in generalized Kaluza-Klein theory, which we interpret in terms of symplectic realization. Finally, we deduce the nonlinear generalization of the equations of motion we started with.}

\pagebreak

\tableofcontents

\section*{\itshape \bfseries  Introduction}
\addcontentsline{toc}{section}{\itshape Introduction}

The purpose of the present work is to argue that classical Yang-Mills theory of elementary particles emerges naturally from Hamiltonian dynamics on Poisson manifolds. This is mainly suggested by the fact that a special Poisson manifold, which goes under the name of {\it Sternberg-Weinstein space} and is obtained as the quotient of
the cotangent bundle of a principal fiber bundle $P\to B$ by the lifted action of the structure group, is known to be 
the universal phase space for classical particles moving in a gravitational and Yang-Mills field, modelled by a metric on $B$ and a principal connection on $P$, respectively (\cite{Sternberg,Wein7}, see also \cite{RatiuPerl}). The corresponding equations of motion where first discovered by Wong (\cite{Wong}, see also \cite{Mont}). 

On the other hand, it is shown here that to any generic Hamiltonian system on a Poisson manifold, there is a natural approximation with respect to a Lagrangian submanifold on which the differential of the Hamiltonian vanishes, which is locally equivalent to a Wong system on a Sternberg-Weinstein space. This relation between Poisson geometry and gauge theory is already apparent from the observation that Poisson manifolds are, in an important special case common in physics, locally isomorphic to a product of a cotangent bundle and the dual space of a finite dimensional semi-simple Lie algebra.

We will proceed as follows. In section 1, we collect the needed results on the geometry of a Poisson manifold $Z$, reduction, linear approximation (cf \cite{Vais}) and the construction of Sternberg and Weinstein. Section 2 is dedicated to the study of the most important object making the link between Poisson geometry and gauge theory, i.e., the Lie algebroid associated to every Poisson manifold, and in particular, its transitive restriction $\jb$ to a symplectic leaf $S\subset Z$. The last allows to define the notion of an $\jb$-connection form, which is intimately related to gauge fields. Similar results have earlier been obtained by Y. Vorobjev \cite{Voro}. In section 3, we construct, for any Hamiltonian system defined by a generic Hamiltonian $H$ on $Z$ and any Lagrangian submanifold $X\subset S$ such that $dH|_X=0$, a Poisson manifold $Z'$, and a canonical Hamiltonian system defined by a Hamiltonian $\HSW$ on $Z'$, which can be considered as natural linear approximation of $Z$ and $H$, respectively, and in addition, is precisely determined by a metric and a Yang-Mills field on $X$, as for the system describing the Wong equations. 

Furthermore, we show that another canonical system on $Z'$ can be seen as a natural quadratic approximation and is related to Einstein-Mayer theory. This system contains additional scalar fields, which are credited to introduce Higgs fields into the Poisson geometric model. However, in order to obtain interesting models, we have to constrain the construction to a coisotropic 
submanifold $\co\subset Z$. This is done in section 4, leading to results well-known from dimensional reduction (\cite{Coquereaux1,Coquereaux}). Finally, we show in section 5 that our constructions can be locally related to usual gauge theoretic objects via the choice of a locally minimal symplectic realization $\rho:\Uh\to U\subset Z$ of $Z$. We recognize in this construction the spirit of
Kaluza-Klein theory (\cite{Kaluza,Klein,Kerner}).

Clearly, our construction raises the question whether it is possible to construct classical gauge theories involving non-linear Poisson structures. The Yang-Mills equations for such a theory should be similar to those of the Poisson sigma model of N. Ikeda (\cite{Ikeda}) and Schaller-Strobl (\cite{SchallerStrobl}). We make some comments related to this question for Hamiltonians defined by a metric in the last chapter. Since the Lie algebras of usual gauge theories lead to linearizable Poisson structures, classical gauge theory turns out to be a generic setting in the Poisson geometric framework. Notice also that the results of this work should partly fit into the more general technical framework developed in \cite{Kara2} for studying linearized dynamics near invariant isotropic submanifolds. Since our constructions are mostly elementary, we tried to avoid obvious explanations, and rather to give a natural and hopefully useful presentation of the results.

\subsection*{\itshape \bfseries  General Notation}

For any manifold $M$, we call $\tau_M:TM\to M$ and $\pi_M:\T M\to M$ the tangent and cotangent projection, respectively. The  exterior algebra and contravariant exterior algebra of $M$ are denoted by $\Frm(M):=\oplus_{k=0}^{\dim M} \Frm^k(M)$ and $\Wek(M):=\oplus_{k=0}^{\dim M} \Wek^k(M)$, respectively, with
$\Frm^k(M):=\Sekt(\wedge^k \T M)$ and $\Wek^k(M):=\Sekt(\wedge^k TM),0\leq k\leq\dim M$. The exterior derivation on $\Frm(M)$ and the Schouten-Nijenhuis
bracket on $\Wek(M)$ are denoted by $d$ and $[\cdot,\cdot]$. If $f:M\to \ti M$ is a map, then the usual push-forward and pull-back induced by the tangent and cotangent maps $Tf$ and $\T f$ (if defined) are denoted by  $f_*:\Wek(M)\to \Wek(\ti M)$ and $f^*:\Frm(\ti M)\to \Frm(M)$. The contraction and Lie derivative with respect to a vector field $\vef$ are denoted by $\inn_{\vef}$ and $\Lie_{\vef}$, respectively.
 
For any fiber bundle $p:E\to M$ over $M$, 
$\Sekt(E)$ denotes the global sections of $E$, and $VE$ the vertical tangent bundle of $E$. If $\ti p:\ti E\to \ti M$ is another fiber bundle and $M=\ti M$, then the product bundle of $E$ and $\ti E$ over $M$ is denoted by $E\times_M \ti E$, and we write $(pr_1,pr_2):E\times_M \ti E\to E\times \ti E$ for the component projections. If $E$ is a vector bundle, the dual bundle is denoted by $E^*$, and the dual of a linear bundle map $\phi:E\to \ti E$ over $f:M\to \ti M$ by $\phi^*:\ti E^*|_{f(M)}\to E^*$. This is consistent with the pull-back since $\phi^*\bar s= \oli{\phi^*s}=\oli{\phi^*\circ s\circ f}$ if $\bar s\in C^\infty(\ti E)$ denotes the fiber-linear function defined by a section $s\in \Sekt(\ti E^*)$. If $M=\ti M$, we denote $E\oplus_M \ti E$ the Whitney sum, and $E\otimes_M \ti E$ the fiberwise tensor product of $E$ and $\ti E$.  
For $q\in \Sekt(E\otimes_M \ti E)$ and $\sigma\in \Sekt(E^*\otimes_M \ti E^*)$, we denote $q^\sharp:\ti E^*\to E$ and   
$\sigma^\flat:E\to \ti E^*$ the corresponding bundle morphisms. 
For a principal fiber bundle $P\to M$, the associated fiber bundle to $P$ with standard fiber $F$ will be denoted by $F(P)$.

\section{Fundamental results on Poisson manifolds}

\begin{defin}  \label{defPoismani}
\begin{rm}
A $C^\infty$-manifold $Z$ is called a {\it Poisson manifold} if $C^\infty(Z)$ is
endowed with the structure of a {\it Poisson algebra}, that is, a Lie algebra structure
\beq
\{\cdot,\cdot\}: C^\infty(Z)\times C^\infty(Z)\to C^\infty(Z)
\label{dasPoisbracket}
\end{equation}
satisfying the {\it Leibnitz identity}, i.e., such that for all $h\in C^\infty(Z)$, $\{\cdot,h\}$ is a derivation of the associative algebra $C^\infty(Z)$. The map (\ref{dasPoisbracket}) is called the {\it Poisson bracket}.
\end{rm}
\end{defin}

\subsection{General structure theory and reduction}

\begin{theo}\label{musicmor}(\cite{Poistheo})
A Poisson manifold determines, and is given by, a pair
$(Z,w)$ consisting of a $C^\infty$-manifold
$Z$ and a bivector-field $w\in\Wek^2(Z)$ inducing a
 bundle morphism $w^\sharp:\T Z\to TZ$, 
%by $\langle\al,w^\sharp\circ \beta\rangle = w(\al,\beta)$ for $\al,\beta \in \Frm^1(Z)$, 
if one defines 
%the Poisson bracket 
$\{\cdot,\cdot\}$ 
for all $f,g\in C^\infty(Z)$ by
\bes
\{f,g\}=w(df,dg)=X_g\cdot f=-X_f\cdot g,\qquad\text{where} \quad X_ f =w^\sharp\circ df,
\end{equation*}
and if one of the following conditions, both equivalent to the Jacobi identity, holds:  
\bes
\lbrack w,w\rbrack = 0\qquad \Longleftrightarrow \qquad X_{\{f,g\}}=-[X_f,X_g]\qquad \forall f,g \in C^{\infty}(Z).
\end{equation*}
The image of  $w^\sharp$ is a completely integrable general distribution on $Z$, that is, through each point $x\in Z$, there exists a maximal integral manifold $S_x$ such that $T_x(S_x)= w^\sharp(\T_xZ)$, and $\mathcal S=\cup_{x\in Z}S_x$ is a generalized foliation of $Z$. Furthermore, the Poisson structure induces symplectic structures on the leaves of $\mathcal S$. 
\end{theo}

\begin{defin}
\begin{rm}
A field $w\in\Wek^2(Z)$ with $[w,w]=0$ is called {\it Poisson structure}, and $-w^\sharp$  is called the {\it anchor map}. If $w=-[w,\mathcal T]$ for some $\mathcal T\in \Wek^1(Z)$, it is called {\it exact}. $X_f$ is called the {\it Hamiltonian vector field} generated by the {\it Hamiltonian} $f$. The triple $(Z,w,f)$ is called {\it a Hamiltonian system} on $Z$. The subspace of Hamiltonian vector fields is denoted by $\Hae (Z)$. The leaves of $\mathcal S$ are called the {\it symplectic leaves} of $Z$, and
$\mathcal S$ and $T\mathcal S$ are called the {\it symplectic foliation} and the {\it symplectic
distribution} of $Z$, respectively. A Poisson manifold $Z$ is called {\it regular} ({\it symplectic}) if
its symplectic distribution has constant (maximal) rank. If $Z$ is symplectic, then $\ow\in \Frm^2(Z)$ defined by 
\bes
\forall \vef\in \Wek^1(Z)\qquad \inn_{\vef} \ow =\ow^\flat\circ\vef,\qquad \text{where}\quad \ow^\flat=(w^\sharp)^{-1}:T Z\to \T Z, 
\end{equation*}
is a symplectic form on $Z$, i.e., $\ker \ow^\flat=0=d\ow$, and $\inn_{X_f}\ow=df$ for all $f\in C^\infty(Z)$.
\end{rm}
\end{defin}

\begin{defin}
\begin{rm}
A map $\phi:(Z_1,w_1)\to (Z_2,w_2)$ between two Poisson manifolds is called
{\it Poisson} map (or morphism) if $\phi^*$ is an morphism of Poisson algebras,
i.e., if for all $f,g\in C^\infty(Z_2)$,
\beq
\{\phi^*f,\phi^*g\}_1=\phi^*\{f,g\}_2.
\label{CondfuerPoismor}
\end{equation}
If $\phi$ is a diffeomorphism, then $\phi^{-1}$ is obviously a
Poisson morphism, too, and $\phi$ is called {\it Poisson isomorphism} or {\it equivalence}. Two Hamiltonian systems $(Z_i,w_i,f_i)$, $i=1,2$ are called {\it $\phi$-related} or {\it equivalent} if in addition, $\phi_*X_{f_1}=X_{f_2}$. A vector field $\vef$ on a Poisson manifold $(Z,w)$ will be called {\it Poisson} vector field iff $\Lie_{\vef} w=0$. 
\end{rm}
\end{defin}

\begin{prop}
The condition (\ref{CondfuerPoismor}) is equivalent to all of the conditions 
\beq
\phi_*w_1=w_2\quad\Leftrightarrow\quad \phi_*X_{\phi^*f}=X_f \quad\Leftrightarrow\quad (w_2^\sharp)_{\phi(z)}=T_z\phi\circ (w_1^\sharp)_z\circ (T_z\phi)^{*} \label{nochPoiscond}
\end{equation}
for all $f\in C^\infty(Z_2)$, $z\in Z_1$. Furthermore, the local group of diffeomorphisms defined by a Poisson vector field acts by Poisson automorphisms. 
\end{prop}

\begin{prop}
Let $\phi:(Z_1,w_1)\to (Z_2,w_2)$ and $\psi:(Z_2,w_2)\to (Z_3,w_3)$ be maps
of Poisson manifolds. If $\phi$ and $\psi$ are Poisson, so is $\psi\circ \phi$. If $\phi$ is a surjective Poisson mapping, and if $\psi\circ \phi$ is Poisson, then $\psi$ is Poisson, too. Furthermore, the Poisson vector fields form a subalgebra of $\Wek^1(Z)$, containing $\Hae(Z)$ as an ideal.
\end{prop}

\begin{defin}
\begin{rm}
Let $(Z,w)$ be a Poisson manifold. An embedded submanifold $i_{\ti Z}:\ti Z \to Z$ is called a {\it Poisson submanifold} if
there exists a Poisson structure $\ti w$ on $\ti Z$ such that
$i_{\ti Z}$ is a Poisson mapping. If it exists, $\ti w$ is unique.
\end{rm}
\end{defin}

\begin{prop} \label{wannconid}
Let $(Z_1,w_1)$ be a Poisson manifold, and
let $\phi:(Z_1,w_1)\to Z_2$ be a surjective mapping. There exists a Poisson
structure $w_2$ on $Z_2$ for which $\phi$ is a Poisson map iff for all
$f,g\in C^\infty(Z_2)$, $\{\phi^*f,\phi^*g\}$ is constant on the fibers of
$\phi$.
In addition, $w_2$ is uniquely defined. We say that it is {\rm coinduced} by
$\phi$.
\end{prop}

\begin{defin} \label{symprealdefin}
\begin{rm}
Let $(Z,w)$ be a Poisson manifold.
A surjective submersive
Poisson map $\rho:W\to Z$ from a symplectic manifold $(W,\ow)$, endowed with its canonical Poisson structure denoted by $\ow^{-1}$, to $Z$ is
called a {\it symplectic realization} of $(Z,w)$.
\end{rm}
\end{defin}

\begin{theo} {\rm (\cite{WeinReal,KaraReal})} \label{globsymprealexistence}
Every Poisson manifold has a symplectic realization.
\end{theo}

\begin{prop}
Let $(W,\ow)$ be a symplectic manifold, and
let the (associative) subalgebra $\Fgr$ of $C^\infty(W)$
consist of the functions constant along the leaves of a foliation $\mathcal \fas$ on $W$ such that the quotient $W/\mathcal \fas$ is a manifold. Then $\Fgr$ is a Poisson subalgebra and thus, the natural projection $\rho:W\to W/\mathcal \fas$ coinduces a unique Poisson
structure, iff
the symplectically orthogonal subbundle $T\mathcal \fas^\bot$ is involutive. In this case, this subbundle is integrable and defines a foliation $\mathcal \fas^\bot$ and a subalgebra $\Fgr^\bot$. If it exists as a manifold, there is a unique coinduced Poisson structure on quotient $W/\mathcal\fas^\bot$.
Consequently, $W$ is simultaneously a symplectic realization of  $W/\mathcal \fas$ and $W/\mathcal \fas^\bot$.
\end{prop}

\begin{defin}
\begin{rm}
Two Poisson manifolds $(Z,w)$ and $(Z_\bot,w_\bot)$
will be said to be {\it dual} to each other, or to form a {\it dual pair},
if they are realized by the same symplectic manifold $(W,\ow)$, and the
tangent spaces to the fibers of two realization mappings are symplectic
orthogonal complements at each point in $W$. Then, the
corresponding subalgebras $\Fgr$ and $\Fgr^\bot$ are called {\it polar} to each other.
\end{rm}
\end{defin}

\begin{lem}\label{sympleavesbij}
Let $Z_\bot \stackrel{\rho_\bot}{\leftarrow} W \stackrel{\rho}{\rightarrow} Z$
be a dual
pair. For each point $z\in Z$,
each connected component of $\rho^{-1}(z)$ maps under $\rho_\bot$ to a
symplectic leaf in $Z_\bot$, and vice versa. This gives a locally bijective
correspondence between the symplectic leaves of $Z$ and $Z_\bot$. If $\rho$
and $\rho_\bot$ have connected fibers, then this correspondence is bijective.
\end{lem}

\begin{theo} \label{transverseareantiiso}
Let $Z_\bot \stackrel{\rho_\bot}{\leftarrow} W \stackrel{\rho}{\rightarrow} Z$
be a dual pair. For each $w\in W$, the transverse Poisson structures on $Z$
and $Z_\bot$ at $\rho(w)$ and $\rho_\bot(w)$ are anti-isomorphic. If
$\dim Z \geq \dim Z_\bot$, then $Z$ is locally anti-isomorphic to the
product of $Z_\bot$ with a symplectic manifold.
\end{theo}

\begin{defin}
\begin{rm}
Let $\rho:W \rightarrow Z$ be a symplectic realization. The elements of the center $\Za(C^\infty(Z)))$ are called {\it Casimir functions}, and we set $\Casi:=\rho^*\Za(C^\infty(Z))$.
The subalgebra of $C^\infty(W)$ consisting of functions whose Hamiltonian vector fields is projectable to a Hamiltonian vector field on $Z$ will be called {\it projectable Hamiltonians} and denoted by $\Pha(\mathcal \fas)$. 
\end{rm}
\end{defin}

\begin{lem}\label{Aufspaltungslemma}
Let $Z_\bot \stackrel{\rho_\bot}{\leftarrow} W \stackrel{\rho}{\rightarrow} Z$
be a dual pair, and denote $\mathcal\fas$, $\mathcal\fas^\bot$ and $\Fgr$, $\Fgr^\bot$
the corresponding orthogonal foliations and dual function groups. We have
$\rho^*:C^\infty(Z)\to \Fgr
 =\Za_{C^\infty(W)}(\Fgr^\bot)$ and $\rho_\bot^*:C^\infty(Z_\bot) \to\Fgr^\bot =\Za_{C^\infty(W)}(\Fgr)$, and
\beq
\Za(C^\infty(Z))\cong\Casi=\Za(\Fgr)=\Fgr \cap \Fgr^\bot=\Za(\Fgr^\bot)\cong \Za(C^\infty(Z_\bot)),
\label{socasisein}
\end{equation}
denoting by $\Za$ and $\Za_{C^\infty(W)}$ the center and  the centralizer in $C^\infty(W)$, respectively.
Furthermore, let $\KK\in \Pha(\mathcal \fas)$ be such that 
$\rho_{*} X_\KK=X_{H}$ for some $H\in C^\infty(Z)$. Then, there are decompositions
\beq
\KK=\rho^* H+ \rho_\bot^* H_\bot \qquad\quad
X_\KK=X_{\rho^*H}+X_{\rho_\bot^*H_\bot}\label{Gdecomposition}
\end{equation}
for some $H_\bot\in C^\infty(Z_\bot)$.
In addition, the decomposition of $X_\KK$ is unique, while
the decomposition of $\KK$ is unique up an element of $\Casi$. This yields exact sequences
\begin{eqnarray}
&0\longrightarrow \Fgr^\bot\stackrel{\rho_\bot^*}{\longrightarrow}
\Pha(\mathcal \fas) \longrightarrow
\Fgr/\Casi\cong \Hae(Z)\longrightarrow 0&\nonumber\\
&0\longrightarrow \Fgr\stackrel{\rho^*}{\longrightarrow}
 \Pha(\mathcal \fas)\longrightarrow
\Fgr^\bot/\Casi\cong \Hae(Z_\bot)\longrightarrow 0&,\nonumber
\end{eqnarray}
and isomorphisms
$\Hae(Z)\cong \Pha/\Fgr^\bot$ and
$\Hae(Z_\bot)\cong \Pha/\Fgr$, with $\Pha=\Fgr+\Fgr^\bot$.
\end{lem}

\noindent
{\sc Proof}: The first part follows from the fact $T\mathcal \fas^\bot=(T\mathcal \fas)^\bot$. For the second, we define $\hat H_\bot\in C^\infty(W)$ by
$\hat H_\bot=\KK-\rho^*H$.
By assumption, $\rho_*X_{\hat H_\bot}=0$, and thus, $\Za_{C^\infty(W)}(\Fgr)=\Fgr^\bot\ni \hat H_\bot=\rho_\bot^*H_\bot$ for some $H_\bot\in C^\infty(Z_\bot)$.
The symmetry between $\mathcal\fas$ and $\mathcal\fas^\bot$ implies $\rho_{\bot*}X_\KK=X_{H_\bot}$.
Furthermore, $X_{H}$ and $X_{H_\bot}$
define $H$ and $H_\bot$ up to an element of
$\Za(C^\infty(Z))$ and $\Za(C^\infty(Z_\bot))$, respectively. Thus, (\ref{socasisein}) implies that the
decomposition of $\KK$ in (\ref{Gdecomposition}) is unique up to
an element of $\Casi$. \hfill$\square$

\begin{defin}
\begin{rm}\cite{Mackenzie}
A {\it Lie algebroid} over a manifold $X$ is a (real) vector bundle $E$ over $X$ together with a map $\rho:E\to TX$ and a (real) Lie algebra structure $[\cdot,\cdot]_E$ on $\Sekt(E)$ such that the induced map $\Sekt(\rho):\Sekt(E)\to \Wek^1(X)$ is a Lie algebra homomorphism, and such that
the following Leibnitz identity holds for $f\in C^\infty(X)$, $\mathcal V_1, \mathcal V_2\in \Sekt(E)$:
\bes
[\mathcal V_1,f\mathcal V_2]_E = (\rho(\mathcal V_1)\cdot f)\mathcal V_2 + f[\mathcal V_1,\mathcal V_2]_E.
\end{equation*}
If $\rho$ is surjective, the Lie algebroid is called {\it transitive}.
\end{rm}
\end{defin}

\begin{prop}\label{dualliealgebroid}
The dual bundle $E^*$ of a Lie algebroid $E\to X$ carries a natural Poisson structure completely determined by $[\cdot,\cdot]_E$ and $\rho$.
\end{prop}

\begin{prop} \label{bracketononeforms}
Let $(Z,w)$ be a Poisson manifold. Then, there exists a unique
$\R$-bilinear, skew-symmetric operation
$\{\cdot,\cdot\}:\Frm^1(Z)\times \Frm^1(Z)\to \Frm^1(Z)$ such that
\begin{eqnarray}
\{df,dg\}&=&d\{f,g\} \qquad \qquad \hspace{1.8cm}\forall f,g\in
C^{\infty}(Z)\nonumber\\
\{\al,f\beta\}&=& f\{\al,\beta\} - (w^\sharp\circ \al)(f)\beta \qquad
\forall f\in C^\infty(Z), \al,\beta\in \Frm^1(Z).
\label{dercon}
\end{eqnarray}
This operation is given by the general formulas
\begin{eqnarray}
\{\al,\beta\}&=&\Lie_{w^\sharp\circ\beta}\al - \Lie_{w^\sharp\circ \al}\beta- dw(\al,\beta) \nonumber\\
&=& \inn_{w^\sharp\circ\beta}d\al -  \inn_{w^\sharp\circ\al}d\beta +
dw(\al,\beta).
\nonumber
\end{eqnarray}
Furthermore, it provides $\Frm^1(Z)$ with a Lie algebra structure such
that $w^\sharp \circ$ is a Lie algebra anti-homomorphism, i.e.
\begin{equation}
w^\sharp\circ \{\al,\beta\}=-[w^\sharp\circ\al,w^\sharp\circ\beta].
\label{formbrakhomo}
\end{equation}
\end{prop}

\begin{cor} (\cite{WeinNC},
%\cite{fuenf})
\label{tangentPstruc}
The triple $(\T Z,-w^\sharp,\{\cdot,\cdot\})$ is a Lie algebroid over $Z$. Consequently, the tangent bundle $TZ$ carries a natural Poisson structure completely determined by the relations (\ref{dercon}).
\end{cor}

\begin{defin}
\begin{rm}
This structure is called the {\it tangent} Poisson structure on $TZ$, while $(\T Z,-w^\sharp,\{\cdot,\cdot\})$ is called the {\it tangent Lie algebroid} (see \cite{Alvarez}).
\end{rm}
\end{defin}

\paragraph{Poisson reduction.}

Poisson structures can also be coinduced by a reduction procedure.

\begin{prop} \label{soprjectierts}
Let $Q$ be a submanifold of the Poisson manifold $(Z,w)$ such that
${\sf N}(Q):=w^\sharp((TQ)^0)\cap TQ$ is a distribution of constant dimension
along $Q$. Then ${\sf N}(Q)$ is differentiable and involutive. Furthermore,
if  ${\sf N}(Q)$ is transversal to the symplectic leaves of $\mathcal S(Z)$,
then ${\sf N}(Q)$ is a sub-distribution of
$T\mathcal S(Z)\cap TQ$, which is transversally symplectic along each leaf
of the latter.
\end{prop}

\begin{prop}\label{reductionexists}
Let $Q$ is a submanifold such that ${\sf N}(Q)$ satisfies the hypothesises of
proposition \ref{soprjectierts}.
If the integral foliation $\Iq(\co)$ of ${\sf N}(Q)$ is given by the
fibers of a submersion $\phi:Q\to \ti Z$, then $T\mathcal S(Z)\cap TQ$ projects to
a distribution whose integral manifolds form a
symplectic foliation $\phi(\mathcal S)$ of $\ti Z$ which is the
symplectic foliation
of a well-defined
Poisson structure $\ti w$ on $\ti Z$.
\end{prop}

\begin{defin}
\begin{rm}
The distribution $\sf N(Q)$ and its integral foliation $\Iq(Q)$ are called the
{\it sub-characteristic distribution} and the {\it sub-characteristic foliation} of $Q$, respectively. The Poisson structure $\ti w$
is called the {\it (leafwise) reduction} of $w$ via $Q$.
\end{rm}
\end{defin}

\begin{defin}
\begin{rm}
A submanifold $Q$ of a Poisson manifold $(Z,w)$ is called {\it coisotropic}
if $w|_{(TQ)^0}\equiv 0$, or, equivalently,
$w^\sharp((TQ)^0)\subset TQ$. The functions in the subspaces ${\Nuli_Q}=\{f\in C^\infty(Z)|f|_Q=0\}$ and $\Noa_{C^\infty(Z)}(\Nuli_Q)$ are called {\it constraints} and {\it admissible functions}, respectively. Here, $\Noa$ denotes the idealizer subalgebra. 
\end{rm}
\end{defin}

\begin{prop} \label{lokgeschlossen}
Let $Q$ be a locally closed submanifold of a Poisson manifold $(Z,w)$. Then, $Q$ is coisotropic
iff one of the following equivalent conditions is satisfied:
(i) $\forall f\in \Nuli_Q$ the Hamiltonian vector field $X_f|_Q$ it tangent
to $Q$;
(ii) $\{\Nuli_Q,\Nuli_Q\}\subset\Nuli_Q$, that is, $\Nuli_Q$ is a
subalgebra. In this case, the first assumption of proposition \ref{soprjectierts} is satisfied.
\end{prop}

\begin{prop}\label{admissible}
Suppose that $Q$ is locally closed and coisotropic. We have $C^\infty(Z)/\Nuli_Q\cong C^\infty(Q)$, the projection being given by $i_Q^*$.  If the reduced manifold $\ti Z=Q/\Iq(Q)$ exists, we have a canonical identification 
\bes
C^\infty(\ti Z)\cong\Noa_{C^\infty(Z)}(\Nuli_Q)/\Nuli_Q.
\end{equation*}
The functions in $\Noa_{C^\infty(Z)}(\Nuli_Q)$ are precisely those which are constant along the leaves of $\Iq(Q)$, and whose Hamiltonian vector fields are tangent to $\co$. In particular, $Q$ is a Poisson submanifold manifold iff  $\Nuli_Q$ is an ideal.
\end{prop}

\paragraph{Local structure.}

In local coordinates $(x^i,i=1,\dots,m=\dim Z)$, the Poisson structure is determined
by the component functions $w^{ij}=\{x^i,x^j\}$ of $w$, which are called {\it structure functions}.
The Jacobi identity $[w,w]=0$ reads
\begin{equation}
w^{lj}\frac{\dl w^{ik}}{\dl x^l}+w^{li}\frac{\dl w^{kj}}{\dl x^l}
+w^{lk}\frac{\dl w^{jl}}{\dl x^l}=0.
\label{Jacobiforcoord}
\end{equation}

\begin{theo} {\sc (Splitting theorem \cite{Wein1})} \label{splittingtheo}
Let $(Z,w)$ be a Poisson manifold, $z\in Z$ and the rank of $w_z$ be $2h$.
Then $z$ has an open neighborhood $U$ in $Z$ such that $(U,w|_U)$ is Poisson
equivalent by a mapping $sp$ to a product $V\times N$, where $V$ is a
$2h$-dimensional symplectic manifold, and $N$ is a Poisson manifold of rank 0
at $sp(z)$. Moreover, the factors $V$ and $N$ are unique up to local
equivalence.
\end{theo}

There is a distinguished candidate for the symplectic factor $V$
given by the intersection of $U$ with the symplectic leaf $S_x\subset Z$
passing through $x$, and we will always identify $V$ with this natural
representative. One can show that the
intersection of $U$ (restricted if necessary) with every submanifold of $Z$
which is transverse to the symplectic leaf $S_x$ at $x$ has an induced Poisson structure which realizes the Poisson structure of $N$, called the
{\it transverse structure} to the symplectic leave $S_x$ at $x$, but their is in general no natural representative.

\begin{cor} \label{Darbcord}
On a Poisson manifold $(Z,w)$, any point $x\in Z$ has a coordinate
neighborhood with {\rm Darboux} coordinates $(x^\mu,p_\mu,r_a)$,
$\mu=1,\dots,m=\frac{1}{2}rk(w_x)$
and $a=1,\dots,n=\dim Z-2m$, centered at $x$, such that
\begin{equation*}
w=\frac{\dl}{\dl x^\mu}\wedge \frac{\dl}{\dl p_\mu} + \eha w_{ab}(r_a))\frac{\dl}{\dl r_a}\wedge \frac{\dl}{\dl r_b} \qquad \mbox{and}
\qquad w_{ab}(r_a)=w^c_{ab}(r_a)r_c,
\end{equation*}
i.e., satisfying the following fundamental Poisson bracket relations
\begin{eqnarray}
\{x^\mu,x^\nu\}=0 & \{p_\mu,p_\nu\}=0 & \{x^\mu,p_\nu\}=\delta^\mu_\nu
\nonumber\\
\{x^\mu,r_a\}=0   & \{p_\mu,r_a\}=0   & \{r_a,r_b\}=w^c_{ab}(r_a)r_c. \nonumber
\end{eqnarray}
\end{cor}

\begin{theo} \label{locsymprealexistence}
Any point $z$ of a Poisson manifold $(Z,w)$ has an open neighborhood $U$
such that $(U,w|_U)$ is realizable by a symplectic manifold of (locally minimal) dimension
$d_{min}=2(\dim Z - (1/2) {\rm rk}(w_z))$.
\end{theo}

\begin{theo} \label{essentialunique}
Let $\rho_1:W_1\to Z$ and $\rho_2:W_2\to Z$ be
symplectic realizations of a Poisson manifold $Z$. Then, for all $z\in Z$ and
for all $w_k\in W_k$ such that $\rho_k(w_k)=z$ (k=1,2), there are
neighborhoods $U\ni z, \hat U_k\ni w_k$, symplectic realizations $\rho'_k:U'_k\to (U,w|_U)$, symplectic embeddings $i_k:U_k'\to U_k$ and a symplectomorphism $i:U'_1\to U'_2$ such that $\rho'_k=\rho_k\circ i_k$ and $\rho'_1=\rho'_2\circ i$. In particular, if the $W_k$ have both dimension $d_{min}$, then $U_1$ and $U_2$ are symplectomorphic.  
\end{theo}

Let us consider a fixed Poisson manifold $(Z,\po)$, and for a fixed point $z_0\in Z$, let $\rho:(\Uh,\ow)\to (U,\po)\subset (Z,\po)$ (omitting the restriction) be a symplectic realization of dimension $d_{min}$ of an open neighborhood $U\ni z_0$. The fibers of $\rho$ define the foliation $\mathcal \fas$ which we will suppose to have {\it connected leaves}. Equally, we assume that the dual foliation $\mathcal \fas^\bot$ has connected leaves, and is such that the image of the quotient map $\la:\Uh\to \Uh/\mathcal \fas^\bot=:\Upsilon$ is a manifold with coinduced Poisson structure $\Povy$.  Then, we get a dual pair
\beq
(\Upsilon,\Povy)\stackrel{\la}{\longleftarrow} (\Uh,\ow) \stackrel{\rho}{\longrightarrow} (U,\po).
\label{minimalreal}
\end{equation}
Let $S$ denote the symplectic  leaf through $z_0$. We set
\bes
V:=S\cap U, \qquad\Vh:=\rho^{-1}(V) \qquad  \text{and}\qquad \fas:=\rho^{-1}(z_0),
\end{equation*}
assuming again that these manifolds are {\it connected}. We shall say that our minimal local realization is {\it split} if in addition, $U$ is such that the splitting theorem applies. 

\begin{lem} \label{Faserbuendel}
The manifold $\hat V$ is a coisotropic submanifold of $\Uh$, and the
characteristic foliation is given by the fibers of the restricted projection
\beq
\rho|_{\hat V}:\hat V\to V.
\label{fibration3}
\end{equation}
In particular, the fibers are isotropic submanifolds of $\Uh$, and $\fas$ is a Lagrangian submanifold of $\hat N=\rho^{-1}(N)$ for any transverse submanifold $N$ to $V$ representing the transverse structure at $z_0$. 
%If the fibers are all diffeomorphic, 
Furthermore, $U$ can be chosen such that (\ref{fibration3}) is a fiber bundle over $V$ with standard fiber $\fas$.
%If the transverse structure at $V$ has arbitrary small neighborhoods of $z_0$ which are invariant under all automorphisms, then (\ref{fibration3}) is a fiber bundle over $V$ with standard fiber $\fas$. 
\end{lem}

\noindent
{\sc Proof}: Since the assertions are local, let us suppose that our realization is split by a splitting map $sp:U\to V\times N$,  where $N$ is an arbitrary transverse submanifold through $z_0$ to $V$ representing the transverse structure. Then, in \cite{Wein1}, p.542f., it is proven that $\Vh$ is coisotropic with characteristic foliation given by the fibers of $\rho$, which are thus isotropic. It is also shown that the restriction $\rho|_{\Nh}:\Nh\to N$ is a symplectic realization of $N$ of minimal dimension at $z_0$, and repeating the argument for the special case $V=\{z_0\}$, we see that $\fas$ is Lagrangian in $\Nh$. 
Furthermore, since the $sp$ is Poisson, we know that
$sp^{-1}\circ (Id_V,\rho|_{\hat N}): V\times
\hat N\to U$ is a realization of minimal dimension. After restricting $U$ and $W$, theorem
\ref{essentialunique} implies that there is a symplectomorphism
$\widehat {sp}:W\to V\times \Nh$ such that $\rho=sp^{-1}\circ (Id_V,\rho|_{\hat N})\circ \widehat {sp}$. 
%which makes the following diagram commutative:
%\bea
%\Uh & \stackrel{\widehat {sp}}{\longrightarrow} & V\times \hat N %\nonumber\\
%{\scriptstyle \rho}\downarrow &    &\quad\downarrow %{\scriptstyle Id_V \times
%\rho|_{\hat N}} \nonumber\\
%\label{liftedsplitting}\\
%U & \stackrel{sp}{\longrightarrow} & V \times N \nonumber
%\end{eqnarray}
The restriction of $\widehat{sp}$ to $\Vh$ yields a diffeomorphism
$\widehat{sp}|_{\Vh}: \Vh\to V\times \fas$,
that is, a trivialization of the bundle $\Vh$ with standard fiber $\fas$.
%In the general case, we can cover $U$ by open subsets $\{U_i,i\in I\}$ such that the restriction of $\rho$ to each subset $U_i$ with $V_i=U_i\cap V\neq \emptyset$ is split. Thus, we can apply the lemma to each restriction $\rho|_{\Vh_i}:\Vh_i\to V_i$, where $\Vh_i=\rho^{-1}(V_i)$. 
%In particular, we have $\Vh_i\cong V_i\times \fas_i$, where $\fas_i=\rho^{-1}(z_i)$ for some point $z_i\in V_i$, and if the fibers of $\rho|_{\Vh}$ are all diffeomorphic, this yields local trivializations of $\Vh_i\cong V_i\times \fas$ for all $i\in I$. 
\hfill$\square$

\begin{lem} \label{Upsilonistnull}
For all $w_1,w_2\in \Vh$, we have
$\la(w_1)=\la(w_2)=:l_0\in \Upsilon$.
Furthermore, ${\rm rk}_{l_0}\Povy=0$,
that is, $l_0$ is a symplectic leaf, and $\Upsilon$ is it's own transverse structure at $l_0$.
\end{lem}

\noindent
{\sc Proof:} Let $w\in \fas$, and set $l_0:=\la(w)$. Because of theorem \ref{transverseareantiiso}, the transverse Poisson structure of $\Upsilon$ at $l_0$ is anti-isomorphic to any transverse manifold $N$ to $V$ representing the transverse structure of $Z$ at $z_0$. But since $\dim \Uh=d_{min}$, we have
$\dim \Upsilon=\dim \Uh-\dim \mathcal \fas^\bot=\dim \mathcal \fas=\dim W-\dim Z=2(\dim Z-(1/2){\rm rk}_{z_0}Z)-\dim Z=
\dim N$, and thus, $\Upsilon$ coincides with its transverse structure and is of rank zero at $l_0$. Thus, $l_0$ is a symplectic leaf, and lemma \ref{sympleavesbij} implies
$\la^{-1}(l_0)=\rho^{-1}(V)=\Vh$.
\hfill $\square$

\subsection{Linearized Poisson structures and moment maps}

For canonical coordinates centered at $z_0\in Z$ adapted to a splitting $sp$, the functions $w_{ab}$ are the structure functions of the transverse structure at $z_0$ which is of rank zero at $n_0=pr_2\circ sp(z_0)$.  We can Taylor expand these functions at $n_0$, obtaining
\begin{equation}
w_{ab}=w^c_{ab}(r_a)r_c=c^c_{ab}r_c+O(r^2) \qquad \text{where} \qquad c^c_{ab}=w^c_{ab}(0).\label{structureconstdefin}
\end{equation}
From the coordinate expression (\ref{Jacobiforcoord}) of the Jacobi identity, it is easy to show that the truncated functions
$w'_{ab}=c^c_{ab}r_c$
also define a Poisson structure, and that the Jacobi identify means that the
$c^c_{ab}$ are the structure constants of a $\dim N$-dimensional Lie
algebra $\g$. Conversely, if $\g$ is any finite-dimensional Lie algebra, then
the
Lie bracket generates a Poisson structure on the dual vector space $\g^*$.

\begin{lem} \label{Kirillow}
The Poisson bracket on the dual $\g^*$ of a given finite dimensional
Lie algebra $(\g,[\cdot,\cdot])$ is given for all $f,g\in
C^\infty(\g^*)$ by
\bes
\{f,g\}(l)=\langle l,[d_l f,d_l g]\rangle \qquad 
\text{where}\quad d_l f,d_l g \in (T_l\g^*)^*\cong \g \quad\forall l\in \g^*.
\end{equation*}
\end{lem}

\begin{defin}
\begin{rm}
The canonical Poisson structures on the duals of a finite dimensional
Lie algebras are called {\it linear} or {\it Lie-Poisson structures}. The Poisson bracket defined in lemma \ref{Kirillow} is called the {\it Kirillov-Kostant-Souriau bracket} on $\g^*$. The Poisson structure
defined by the $w'_{ab}$ is called the {\it linearized Poisson structure}
or {\it linear approximation}
of $w_{ab}$ at $z_0$. The associated Lie algebra is called the
{\it transverse Lie algebra} at $z_0$.
\end{rm}
\end{defin}

\begin{lem} \label{thelinearapprox}
The transverse Lie algebra is intrinsically identified with the annihilator subspace  
$(T_{z_0}S_{z_0})^0\cong (T_{z_0}Z/T_{z_0}S)^*$ in $\T_{z_0} Z$
to the symplectic leaf $S_{z_0}$ through $z_0$, endowed with a bracket well-defined for $\al,\beta \in (T_{z_0}S_{z_0})^0$ by setting
\begin{equation*}
[\al,\beta]=d\{f,g\}_{z_0},
\end{equation*}
where $f,g\in C^\infty(Z)$ are any functions such that $df_{z_0}=\al$ and   $dg_{z_0}=\beta$.
Consequently, the linearized Poisson structure is
naturally defined on the normal space $NS_{z_0}=T_{z_0}Z/T_{z_0}S$ to the leaf.
\end{lem}

\begin{defin}
\begin{rm}
The transverse structure is called {\it linearizable
at $z_0$} there is a neighborhood of $z_0$ in a representative of $N$ which
is Poisson equivalent to the linearized Poisson structure at $z_0$.
This is equivalent to the existence of canonical coordinates $(r_a)$ such that $w_{ab}(r_a)=c^c_{ab}r_a$
for the structure constants $c^c_{ab}$ of some finite-dimensional
Lie algebra $\g$.
%It is {\it regular} at $z_0$ if, in addition, $\g$ is abelian, i.e.
%$w_{ab}(r_a)=c^c_{ab}=0$ in a neighborhood of $z_0$.
\end{rm}
\end{defin}

\begin{theo}\label{linearization1}(cf \cite{Conn1,Conn2,WeinNC})
Let $\g$ be a semi-simple Lie algebra (of compact type or isomorphic to a semidirect product $\g^R\rtimes \R$, where $\g^R$ is semi-simple of compact type). Then all analytic or formal ($C^\infty-$) transverse structures with a given
transverse Lie algebra $\g$ are
linearizable by an analytic or formal ($C^\infty$-) change of coordinates, respectively. If $\g$ is of non-compact type and has
real rank of at least 2, this fails to be true in general.
\end{theo}

Let $(Z,\po)$ be a Poisson manifold,
and let $\g$ denote the transversal Lie algebra at a point $z_0\in Z$. Let $(\Upsilon,\Povy)\stackrel{\la}{\longleftarrow} (\Uh,\ow) \stackrel{\rho}{\longrightarrow} (U,\po)$ be the dual pair associated to a minimal local symplectic realization at $z_0$ of an open neighborhood $U\ni z_0$ as above . Lemma \ref{Upsilonistnull} and theorem \ref{transverseareantiiso} imply that the linear approximation to $\Upsilon$ at $l_0$ is given by
\begin{equation}
T_{l_0}\Upsilon\cong \g_{\tt L}^*\label{wegenantiiso}
\end{equation}
if $\g_{\tt L}$ denotes the Lie algebra with opposite Lie bracket. If $\Upsilon$ is linearizable at $l_0$ (or, equivalently, $N$ is linearizable at $z_0$), we obtain a (local) dual pair
\beq
\g_{\tt L}^*\stackrel{\la}{\longleftarrow}\Uh\stackrel{\rho}{\longrightarrow}U.
\label{tranverslineardp}
\end{equation}

\begin{defin}
\begin{rm}
A Poisson map $\phi:(Z,w)\to (\bar Z,\bar w)$ is {\it complete} if, for each
$h\in C^\infty(\bar Z)$, $X_{\phi^*h}$ is a complete vector field whenever $X_h$ is. In particular, let $\gl$ be a finite dimensional Lie algebra, and $w^{\gl}$ the Lie-Poisson structure on $\gsl$. A complete Poisson map $\la:(\ZW,w)\to (\gsl,w^{\gl})$ is called a {\it moment map}.
\end{rm}
\end{defin}

\begin{prop} (\cite{Palais}) \label{momentintegriert}
Let $\la:(\ZW,w)\to (\gsl,w^{\gl})$ be a moment map, and identify $\gl$ with
$(\gsl)^*\subset C^\infty(\gsl)$.
The maps
$\la^*:\gl\to C^\infty(\ZW)$ and $(-w^\sharp\circ
d):C^\infty(\ZW)\to \Hae(\ZW)$
are homomorphisms of Lie algebras. Their composition
%\bes
$(-w^\sharp\circ d)\circ \la^*:\gl\rightarrow \Hae(\ZW)$
%\end{equation*}
induces an infinitesimal action of $\gl$ on $\ZW$ by Hamiltonian vector fields.
Furthermore, this action integrates to a right action of the connected,
simply connected Lie group $\Lg_1$ with Lie algebra $\gl$ on $\ZW$ by Poisson
automorphisms. 
\end{prop}

\begin{defin}
\begin{rm} 
Let $\Lg_Z=\cap_{z\in Z}\Lg_z$, where $\Lg_z$ denotes the stabilizer subgroup of $z$. For every normal subgroup $\bar \Lg_1\subset\Lg_1$ with $\bar\Lg_1\subset \Lg_Z$, there is an induced action of the quotient group $\Lg=\bar\Lg_1\backslash\Lg_1$ on $Z$. We call $\la$ the {\it moment map of the right $\Lg$-action}, and say that the right $\Lg$-action is {\it generated by $\la$}.  
\end{rm}
\end{defin}

\begin{prop} \label{sympleavescoadact}(\cite{Vais}, p.32f, \cite{Wein1}, p.534)
The identity map $id:\gsl\to \gsl$ or the inclusion $i:O\to \gsl$
of a coadjoint orbit are complete Poisson maps. They
generate precisely the right
coadjoint action of any $G$ with $\gl=Lie(G)$ on $\gsl$, and thus, the symplectic leaves of the Lie-Poisson structure are the orbits of
the coadjoint action. More precisely, denoting the coadjoint representation by $Ad^*$, we have
\bes
X_{-D}(l)=(\ank\circ d)(D)=d/dt|_0 Ad^*(\exp(tD))(l)\quad \forall D\in\gl\subset C^\infty(\gsl), l\in \gsl.
\end{equation*}
Furthermore, the transverse Lie algebra at $l\in \g^*$ is given by
$\g_l:=N_{\g}(l)=Lie(N_G(l))$, the stabilizer subalgebra of $l$ under the
coadjoint representation.
\end{prop}

\begin{cor}
The moment map $\la:\ZW\to \gsl$ of a right $\Lg$-action is $\Lg$-equivariant with respect to $Ad^*$, that is, $\la(zg)=Ad^*(z)(\la(z))$ for all $z\in Z, g\in \Lg$. 
\end{cor}

\subsection{The Sternberg-Weinstein phase space}

\begin{prop} \label{linreal}
Let $(W,\ow)$ be a symplectic manifold and $\la: W\to \gsl$ a surjective
submersive moment map generating the right
action of a Lie group $G$ on $W$. Suppose in addition that the
quotient $W/G$ is a manifold. Then, the natural projection $\rho:W\to W/G$
coinduces a Poisson structure $W/G$, and the $G$-orbits are
symplectically orthogonal to the fibers of $\la$. Then,  
$\gsl\stackrel{\la}{\leftarrow}W\stackrel{\rho}{\rightarrow}
W/G$ forms a dual pair.
\end{prop}

\begin{cor}
Let $p:P\to B$ be a principal fiber bundle over the manifold $B$ with connected structure group $G$ and $Lie(G)=\gl$.
The canonical lift of the (right) $G$-action on $P$ to
$(\T P,\ow)$, where $\ow$ is the canonical symplectic form, 
is generated by a moment map given by
\beq
\mom:T^*P \to \g_{\tt L}^* \qquad
\alpha \mapsto (\li_y)^* (\alpha|_{V_yP}) \qquad y=\pi_P(\alpha),
\label{Speicher:vert}
\end{equation}
with $li: P\times \g_{\tt L}\to VP$, $(y,D) \mapsto (D_P)(y)$, $D_P$ being the fundamental vector field of $D \in \g_{\tt L}$.
Assuming that the quotient $\T P/G$ is a manifold, 
there is a coinduced Poisson structure $\po$ on it. If $\rho$ denotes the projection map, we obtain a dual pair
\beq
(\g_{\tt L}^*,w^{\g_{\tt L}})\stackrel{\la}{\longleftarrow}(\T P,\ow)\stackrel{\rho}{\longrightarrow}(\T P/G,\po).
\label{Sternweindef}
\end{equation}
\end{cor}

\begin{defin}
\begin{rm}
We call the quotient space $(\T P/G,\po)$ the {\it Sternberg-Weinstein phase space}. Since $G$ was supposed connected, lemma \ref{sympleavesbij} implies that the leaves of the symplectic foliation of $\T P/G$ are in bijection with the symplectic leaves of $\g_{\tt L}^*$, which by proposition \ref{sympleavescoadact} are precisely the coadjoint orbits.
A coadjoint orbit $O \subset \g_{\tt L}^*$ endowed with its symplectic structure and moment map 
is called a {\it generalized charge}, in analogy with the special case $G=U(1)$ related to electrodynamics.
\end{rm}
\end{defin}

\begin{remark}\label{SternbergWeinstein}
\begin{rm}
Sternberg (\cite{Sternberg}) and Weinstein (\cite{Wein7}) directly constructed the phase space corresponding to a specific choice of a coadjoint orbit, performing a Marsden-Weinstein reduction. Notice that the charge 0 corresponds to a particular symplectic leaf
of minimal dimension and
which is naturally identified with the base phase space $\T B$.
\hfill $\lozenge$
\end{rm}
\end{remark}

Any metric $\bmet$ on $B$ naturally defines a
Hamiltonian $\HB$ on $\T B$, quadratic on the fibers and describing geodesic motion in a gravitational field modelled by $\bmet$ on $B$. Given in addition a Yang-Mills field modelled by a principal connection 1-form $\Ahut$ on $P$, we dispose of the associated projections
\begin{equation}
\poka_{\Ahut}:\T P\to \T B \qquad \text{and} \qquad \check \poka_{\Ahut}:T^*P/G\to T^*B
\label{scheckap}
\end{equation}
since $\Ahut$ is $G$-equivariant. Then, $H=\check \poka_{\Ahut}^*\HB$ defines a Hamiltonian system on $\T P/G$. This construction is called {\it minimal coupling}. Let us consider the bundle
\begin{equation*}
\gsl(\bP)=(\bP\times \gsl)/G \qquad \mbox{where} \qquad
\bP:=T^*B\times_B P\stackrel{\bp=pr_1}{\longrightarrow}\T B,
\end{equation*}
with the induced and coadjoint right $G$-action.
Then, the map $\hat{\psi}_{\Ahut}$ defined by
\bes
\hat{\pomo}_{\Ahut}=(\hat \poka_{\Ahut},\la)=(\poka_{\Ahut}\times_B \pi_P,\la): \hspace{0.45cm}\T P\longrightarrow \bP \times \gsl 
\end{equation*}
is a $G$-equivariant diffeomorphism. Thus, it induces
a diffeomorphism of the quotient spaces
$\pomo_{\Ahut}:T^*P/G\longrightarrow \gsl(\bP)$.
Denoting $\napro:\gsl(\bP)\to \T B$ the natural projection, we can summarize the situation in a commutative diagram:
\bea
T^*B \stackrel{\poka_{\Ahut}}{\longleftarrow}
& \T P \stackrel{\hat{\pomo}_{\Ahut}}{\longrightarrow}
\bP\times \gsl
& \stackrel{\bp \circ pr_1}{\longrightarrow} T^*B\nonumber\\
\qquad \nwarrow \check \poka_{\Ahut}
& \downarrow \rho \qquad \quad \downarrow &
\nearrow \napro \nonumber\\
%\label{diag1}\\
&T^*P/G \stackrel{\pomo_{\Ahut}}{\longrightarrow} \gsl(\bP)&\nonumber
\end{eqnarray}
If we define a Poisson structure and a Hamiltonian on $\gsl(\bP)$ by setting
\bes
\po_{\Ahut}=(\pomo_{\Ahut})_*\po \qquad \text{and}\qquad
\HBH=(\napro)^* \HB,
\end{equation*}
then the Hamiltonian systems $(T^*P/G,\po,H)$ and 
$(\gsl(\bP),\po_{\Ahut},\HBH)$
are $\pomo_{\Ahut}$-related. In addition, $\gsl(\bP)$ is naturally fibred over $\T B$, the phase space corresponding to the charge 0, while the gauge field influences the dynamics only via the Poisson structure. This is the natural phase space for writing down the equations of motion of particles in a gravitational and Yang-Mills fields, by calculating $\po_{\Ahut}$ explicitly. 

\begin{defin}
\begin{rm}
The equivalent systems $(T^*P/G,\po,H)$ and 
$(\gsl(\bP),\po_{\Ahut},\HBH)$ are called the {\it Wong system} and the {\it left gauged Wong system}, respectively.
\end{rm}
\end{defin} 

It is well-known that {\it Kaluza-Klein theory} provides a description of particle dynamics in gauge fields on the realization space $\T P$ (\cite{Kaluza},\cite{Klein},\cite{Kerner}).

\begin{defin} \label{KKmetric1}
\begin{rm}
Let $p:P\to B$ be a principal fiber bundle with structure
group $G$. Let $\gl=Lie(G)$, and let $\iota$ be a scalar product on $\gl$.
To any metric $\bmet$ on $B$, and to any principal connection form
$\Ahut$ on P, we associate a {\it
Kaluza-Klein metric} (or {\it bundle metric}) $\pmet$ by setting
\beq
\pmet=p^*\bmet+\iota \circ {\Ahut}
\label{KKmetric2}
\end{equation}
where 
$(\iota \circ {\Ahut})_s(U,V)=\iota(\Ahut_s(U),\Ahut_s(V))$ for all $s\in P;U,V\in T_sP$.
\end{rm}
\end{defin}

\begin{prop}\label{KKmetricprop}
Let $p:P\to B$, $\bmet$, $\Ahut$, $\iota$ and $\pmet$ be as in definition
(\ref{KKmetric1}), and let $\HB$ and $\HP$ be the Hamiltonians defined by
$\bmet$ and $\pmet$ on $\T B$ and $\T P$, respectively. Let $\rho:(\T P,\ow)\to (\T P/G,\po)$ be the Poisson projection as in (\ref{Sternweindef}), define $\check \poka_{\Ahut}:T^*P/G\to T^*B$ as in (\ref{scheckap}) and set $H:=\check \poka_{\Ahut}^* \HB$.
Then, the Hamiltonian systems $(\T P,\ow, \HP)$ and $(\T P/G,\po,H)$ are $\rho_*$-related. Furthermore, the projections by $p$ of the geodesics of the
Kaluza-Klein metric (\ref{KKmetric2}) coincide with the projections
by $\pi_B \circ \check \poka_{\Ahut}:\T P/G\to B$ of the solutions of the
Hamiltonian equations defined by $H$
on $(\T P/G,\po)$, if $\pi_B:\T B\to B$ is the cotangent projection.
\end{prop}

\section{The Lie algebroid over a symplectic leaf}

Let $(Z,\po)$ be a Poisson manifold. The similarity in the defining relations for the tangent Lie algebroid of corollary \ref{tangentPstruc} and the transverse Lie algebra at a symplectic leaf $S\subset Z$ of lemma \ref{thelinearapprox} suggests that these are related. We see here that there is a well-defined restriction of the Lie algebroid to every leaf $S$, which in addition is {\it transitive}. This allows to construct the geometric notion of an $\jb$-connection form, which is the first step towards gauge theory. Throughout this section, let $S\subset Z$ be a fixed symplectic leaf. We assume $S$ to be a submanifold.

\subsection{$\jb$-connection forms and curvature}

\begin{defin}
\begin{rm}
The mutually dual vector bundles over $S$ defined by
\bes
\jb=\T Z|_S\longrightarrow S \qquad \text{and} \qquad\db=TZ|_S\longrightarrow S.
\end{equation*}
are called the {\it Lie algebroid} and the {\it dual Lie algebroid} associated to  $S$, respectively. Furthermore, the mutually dual vector bundles
\bes
\fb=(TS)^0\longrightarrow S \qquad \text{and} \qquad \eb=TZ|_S/TS\longrightarrow S,
\end{equation*}
where $(TS)^0$ denotes the annihilator subbundle in $\T Z|_S$,
are called the {\it Lie algebra bundle} and the {\it dual Lie algebra bundle} to $Z$ at $S$, respectively.
\end{rm}
\end{defin}

\begin{lem}
\label{klammerexiatiert}
There is a well-defined bracket on the space of sections of $\jb$
\beq
\{\cdot,\cdot\}: \Sekt(\jb)\times \Sekt(\jb)\to \Sekt(\jb)
\quad \qquad\{\al|_S,\beta|_S\}=\{\al,\beta\}|_S
\label{LiealgebraaufTU}
\end{equation}
for all $\al,\beta \in \Frm^1(Z)$, where the bracket on the right is the
standard bracket defined in proposition \ref{bracketononeforms}, and
the restrictions means the restriction of the maps $Z\to \T Z$ to
$S\to \jb$. The restriction of this bracket to $\Sekt(\fb)$ is given by the fiberwise Lie bracket defined in lemma \ref{thelinearapprox}. Furthermore, $\Sekt(\fb)\subset \Sekt(\jb)$ is an ideal. 
\end{lem}

\noindent
{\sc Proof}: In the lemma, we defined the bracket of two sections of
$\jb$ be restriction to $S$ of the standard bracket of to arbitrary
extensions of the sections to $Z$. Thus, we have to show that this definition doesn't depend on the extensions. Set $\al^\sharp= \po^\sharp\circ \al$ etc., and let $\vef\in \Wek^1(Z)$. We have
\bea
\{\al,\beta\}(\vef) &=&\inn_\vef(-\inn_{\al^\sharp}d\beta +
\inn_{\beta^\sharp}d\al +
d(\po(\al,\beta)))\nonumber\\
&=& \inn_{\al^\sharp}(\inn_\vef d\beta) - \inn_{\beta^\sharp}(\inn_\vef d\al) +
\vef \cdot \po(\al,\beta)\nonumber\\
&=&(\Lie_\vef\beta)(\al^\sharp)- d(\beta(\vef))(\al^\sharp)  -(\Lie_\vef\al)(\beta^\sharp)+ d(\al(\vef))(\beta^\sharp)+ \nonumber\\
&& +\po(\Lie_\vef\al,\beta)+\po(\al,\Lie_\vef\beta)+ (\Lie_\vef\po)(\al,\beta)
\nonumber\\
&=&d(\al(\vef))(\beta^\sharp)-d(\beta(\vef))(\al^\sharp)+ (\Lie_\vef\po)(\al,\beta)
\nonumber\\
&=&\beta^\sharp \cdot \al(\vef)- \al^\sharp\cdot \beta(\vef)+ (\Lie_\vef\po)(\al,\beta)
\nonumber
\end{eqnarray}
because of the definition of $\po^\sharp$ in theorem \ref{musicmor}. Since the image of
$\po^\sharp$ is tangent to $S$, this shows that the restriction of
$\{\al,\beta\}$ to $S$ depends only on the values of $\al$ and
$\beta$ on $S$. Thus, the bracket is
well-defined.

Furthermore, for $\al\in \Sekt(\jb),\beta\in \Sekt(\fb)$, it follows from the definition and $\beta^\sharp=0$ that (omitting the extensions) $\{\al,\beta\}=\Lie_{\beta^\sharp}\al-\Lie_{\al^\sharp}\beta
-d(\al(\beta^\sharp))\in \Sekt(\fb)$. Thus, $\Sekt(\fb)$ is an ideal. In particular, it is closed under the bracket, and by
lemma \ref{thelinearapprox}, the restriction
of the bracket (\ref{LiealgebraaufTU}) to $\Sekt(\fb)$ is given by point-wise Lie algebra brackets on the fibers of $\fb$, inducing Lie-Poisson structures on the fibers of $\eb$.  
\qed

\begin{cor}
The bundle $\jb$ with the bracket (\ref{LiealgebraaufTU}) on $\Sekt(\jb)$ and the anchor map $\ank$ is a transitive Lie algebroid. Furthermore, $\fb$ is a natural Lie subalgebroid of $\jb$, i.e. a subbundle and a Lie algebroid with the restricted bracket and anchor map.
\end{cor}

Let $i_S:S\to Z$ be the natural inclusion, and $i_S^*:\jb\to \T S$ the natural projection. We denote by $\ow_S$ the symplectic form of $S$. Since $i_S$ is Poisson, (\ref{nochPoiscond}) implies that
\beq
(\ank)_{v}
%(-\po^\sharp)_v
=T_vi_{S}\circ (-\ow_S^{\flat-1})_v\circ (T_vi_S)^*
%=(\ow_S^\sharp)_v\circ (i_S^*)_v
\qquad \forall v\in S,
\label{connectproject}
\end{equation}
which is a Lie algebra isomorphism by (\ref{formbrakhomo}).
Writing $\ank$ for $\ank|_S$, the maps (\ref{connectproject}) induce an exact sequence of vector bundles over $S$
\beq
0\longrightarrow
\fb \hookrightarrow \jb  \stackrel{\ank}
{\longrightarrow} T S \longrightarrow 0.
\label{sequenceofPoisconnect}
\end{equation}

\begin{defin}\label{invformdefin1}
\begin{rm}
A 1-form $\theta:S\to \T S\otimes_S \jb$ on
$S$ with values in $\jb$ is called an {\it
$\jb$-connection form} iff the associated bundle morphism over $S$
\beq
\theta^\flat: TS\to \jb \qquad \mbox{satisfies} \qquad \ank \circ \theta^\flat= Id_{TS},
\label{bedsplittPoiscon}
\end{equation}
that is, iff it is
a splitting of the sequence (\ref{sequenceofPoisconnect}). The subset
of $\Sekt(\T S\otimes_V \jb)$ consisting of the
$\jb$-connection forms is denoted by $\Inc$.
\end{rm}
\end{defin}

The $\jb$-connection forms indeed locally define connections in a fiber bundle. There is a natural global object corresponding to the curvature of these connections.

\begin{prop}\label{curvadefin}
Let $\theta$ be an $\jb$-connection form.
There is a well-defined 2-form on $S$ with values in $\fb$ satisfying
\beq
\Kru^\theta(\vef_1,\vef_2)= \{\theta(\vef_1),\theta(\vef_2)\}-\theta([\vef_1,\vef_2])
\qquad \qquad\forall \vef_1,\vef_2
\in \Wek^1(S).\label{Kruemung}
\end{equation}
\end{prop}

\noindent
{\sc Proof:} Let $f\in C^\infty(S)$. Then,
\bea
&&\Kru^\theta(\vef_1,f\vef_2)=
\{\theta(\vef_1),\theta(f\vef_2)\}-\theta([\vef_1,f\vef_2])\nonumber\\
&&\hspace{0.3cm}=\{\theta(\vef_1),f\theta(\vef_2)\}-\theta(f[\vef_1,\vef_2]-(\vef_1\cdot
f)\vef_2) \nonumber\\
&&\hspace{0.3cm}=f\{\theta(\vef_1),\theta(\vef_2)\}-(\po^\sharp\circ
\theta(\vef_1)\cdot f)\theta(\vef_2)
-f\theta([\vef_1,\vef_2])-(\vef_1\cdot f)\theta(\vef_2) \nonumber\\
&&\hspace{0.3cm}=f\{\theta(\vef_1),\theta(\vef_2)\}-f\theta([\vef_1,\vef_2])+
(\vef_1\cdot f)\theta(\vef_2)-(\vef_1\cdot f)\theta(\vef_2)\nonumber\\
&&\hspace{0.3cm}=f(\{\theta(\vef_1),\theta(\vef_2)\}-\theta([\vef_1,\vef_2]))=f\Kru^\theta(\vef_1,\vef_2)
\nonumber
\end{eqnarray}
where we used (\ref{dercon}) and the fact that
$\theta$ is an $\jb$-connection form. Because of
skew-symmetry, this shows that $\Kru^\theta$ is well-defined by
(\ref{Kruemung}) as a 2-form. Furthermore, (\ref{connectproject}) implies
$i_S^*\theta=-\ow_S^\flat\circ i_{S*}^{-1}\circ (\ank)\circ \theta=-\ow_S^\flat$, and thus
\bea
(i_S^*\circ \Kru^\theta)(\vef_1,\vef_2)&=&i_S^*
(\{\theta(\vef_1),\theta(\vef_2)\}-\theta([\vef_1,\vef_2]))\nonumber\\
&=&\{i_S^*\theta(\vef_1),i_S^*\theta(\vef_2)\}_S
-i_S^* \theta([\vef_1,\vef_2])\nonumber\\
&=&\{\ow_S^\flat(\vef_1), \ow_S^\flat(\vef_2)\}_S+\ow_S^\flat([\vef_1,\vef_2])
\nonumber\\
&=&-\ow_S^\flat([\vef_1,\vef_2])+\ow_S^\flat([\vef_1,\vef_2])=0\nonumber,
\end{eqnarray}
where we used the fact that $i_S$ is a Poisson morphism and the naturality
of the bracket on 1-forms, and denoted by $\{\cdot,\cdot\}_S$ the
bracket on 1-forms of the symplectic manifold $(S,\ow_S)$.
\qed

\begin{defin}
\begin{rm}
The 2-form $\Kru^\theta$ is called the {\it curvature} of $\theta$.
\end{rm}
\end{defin}

\subsection{Splitting transformations}

Let $\phi:Z\to Z$ be a Poisson automorphism such that $\phi(S)=S$. Let us call such morphisms {\it $S$-Poisson morphisms}. Its cotangent map induces a bundle isomorphism $\T \phi|_S:\jb\to \jb$, which can be used to define an action of $\phi$ on forms with values in $\jb$, analogous to active gauge transformations for usual principal connections. Since the local trivializations of Poisson manifolds are the splitting maps, we call such transformations {\it splitting transformations}.

\begin{defin}\label{gaugeactiononforms}
\begin{rm}
Let $\phi$ be an $S$-Poisson morphism. The action $\phi^*:\Sekt(\wedge^k\T S\otimes_S \jb)\to \Sekt(\wedge^k\T S\otimes_S \jb)$ is defined by 
\bes
(\phi^*\al)^\flat=(\T \phi|_S)^{-1}\circ \al^\flat \circ
\wedge^kT(\phi|_S)\qquad \forall \al\in
\Sekt(\wedge^k\T S\otimes_S \jb),
\end{equation*}
%where $T^{\otimes k}(\phi|_S)$ denotes the tangent map of $\phi|_S$ applied to all arguments (with $T^{\otimes 0}\phi:= \phi$), and 
where $\al^\flat$ denotes
the bundle morphism associated to the form $\al$. In particular, it satisfies $(\phi\circ \psi)^* \al= \psi^*\circ \phi^*$
for any two $S$-morphisms $\phi$ and $\psi$.
\end{rm}
\end{defin}

\begin{lem} \label{brakcommphi}
The group of $S$-Poisson morphisms acts on the space $\Inc$ by the action of definition \ref{gaugeactiononforms}. Furthermore, the action preserves the bracket defined in lemma \ref{LiealgebraaufTU}, i.e.
$\phi^*\{\al,\beta\}=\{\phi^*\al,\phi^*\beta\}$
for $\al,\beta \in \Sekt(\jb)$.
\end{lem}

\noindent
{\sc Proof:} Let $\theta$ be an $\jb$-connection form, and $\phi$
an $S$-Poisson morphism.
The only fact to verify is that $\phi^*\theta$ satisfies
condition (\ref{bedsplittPoiscon}).
Since $\phi$ is Poisson, (\ref{nochPoiscond}) implies
\bea
%(\ank)_u\circ (\phi^*)_v &=& (\phi_*^{-1})_v\circ (\ank)_v \qquad\qquad\qquad ,\nonumber\\
(\ank \circ (\phi^*\theta)^\flat)_u
&=& \ank \circ (\T_u \phi)^{-1}\circ \theta^\flat \circ
T_u(\phi|_S) \qquad \quad v=\phi(u)\nonumber\\
&=& (T_u\phi)^{-1}\circ (\ank)_v
\circ
(\theta^\flat)_v  \circ T_u(\phi|_{S})=Id_{T_uS}\nonumber
%&=& (\phi_*^{-1})_v\circ (\phi|_{S*})_u =\nonumber
\end{eqnarray}
and thus, $\phi^*\theta$ is an $\jb$-connection form. For $k=0$, let $\al=\al'|_S$, $\beta=\beta'|_S$ for
$\al',\beta'\in \Frm^1(Z)$, then
\bea
\{\phi^*\al,\phi^*\beta\}
&=&\{\phi^*(\al'|_S),\phi^*(\beta'|_S)\}
=\{(\phi^*\al')|_S,(\phi^*\beta')|_S\}\nonumber\\
&=&\{\phi^*\al',\phi^*\beta'\}|_S=(\phi^*\{\al',\beta'\})|_S
=\phi^*(\{\al',\beta'\}|_S)\nonumber\\
&=&\phi^*\{(\al'|_S),(\beta'|_S)\}=\phi^*\{\al,\beta\}\nonumber,
\end{eqnarray}
where we used the definition of the bracket on $\Sekt(\jb)$ and the
fact that $\phi$ is Poisson. ($\phi^*$ denotes either the map
of definition \ref{gaugeactiononforms} or the usual pull-back,
by a consistent abuse of notation.) \qed

\begin{prop}
Let $\theta$ be an $\jb$-connection form, and $\Kru^{\theta}$ the
associated curvature form.
For an $S$-Poisson morphism $\phi$, we have
$\Kru^{\phi^*\theta}=\phi^* \Kru^{\theta}$. 
\end{prop}

\noindent
{\sc Proof:} We compute
\bea
\Kru^{\phi^*\theta}(\vef_1,\vef_2)&=&\{(\phi^*\theta)(\vef_1),(\phi^*\theta)(\vef_2)\}-
(\phi^*\theta)([\vef_1,\vef_2])\nonumber\\
&=&\{\T\phi^{-1}\circ \theta^\flat \circ T\phi\circ \vef_1,
\T\phi^{-1}\circ \theta^\flat \circ T\phi\circ \vef_2\}-\nonumber\\
&&\hspace{4.5cm}-\T\phi^{-1}\circ \theta^\flat \circ T\phi\circ [\vef_1,\vef_2]\nonumber\\
&=&\{\T\phi^{-1}\circ \theta^\flat
\circ (T\phi \circ \vef_1\circ \phi^{-1})\circ \phi,\nonumber\\
&&
\hspace{3cm}\T\phi^{-1}\circ \theta^\flat\circ (T\phi \circ \vef_2\circ \phi^{-1})\circ
\phi\}- \nonumber\\
&&\hspace{3cm}
-\T\phi^{-1}\circ \theta^\flat \circ (T\phi \circ [\vef_1,\vef_2]\circ \phi^{-1})\circ
\phi \nonumber\\
&=& \{\phi^*(\theta(\phi_*\vef_1)),
\phi^*(\theta(\phi_*\vef_2))\}-
\phi^*(\theta(\phi_*[\vef_1,\vef_2]))\nonumber\\
&=& \phi^*(\{\theta(\phi_*\vef_1),
\theta(\phi_*\vef_2)\}-
\theta([\phi_*\vef_1,\phi_*\vef_2]))\nonumber\\
&=& \phi^*(\Kru^{\theta}(\phi_*\vef_1,\phi_*\vef_2))\nonumber\\
(\phi^*\Kru^{\theta})(\vef_1,\vef_2)&=& \T\phi^{-1}\circ (\Kru^{\theta})^\flat\circ (\wedge^2T\phi\circ
(\vef_1,\vef_2)\circ \phi^{-1}) \circ \phi\nonumber,
\end{eqnarray}
where we used lemma \ref{brakcommphi} and the definition of the
push-forward. \qed

\paragraph{Local splitting transformations.}

As in gauge theory, it is possible to write down splitting transformations as transitions between splittings. 

\begin{lem}\label{splitt-form}
Every splitting map $sp:U\to V\times N$ determines an $\jb$-connection form $\theta_{sp}:V\to \T V\otimes_V \jb_{|V}$ on $V$. Furthermore, $\Kru^{\theta_{sp}}=0$.
\end{lem}

\noindent
{\sc Proof:}
Let $sp:U\to V\times N$ be a splitting map, and $i_V=i_S|_V$, $\ow_V=\ow_S|_V$.  Denote by $\mathcal N$ the foliation given by the fibers of $pr_1\circ sp$. By restriction to the annihilator subbundle $(T{\mathcal N})^0|_V$, $(Ti_V)^*$ becomes an isomorphism, so we can define 
\beq
I_{sp}=(Ti_V)^*|_{(T{\mathcal N})^0|_V}^{-1}:
\T V \rightarrow (T{\mathcal N})^0|_V\qquad \text{and}\qquad
\theta_{sp}^\flat=-I_{sp}\circ\ow_V^\flat.
\label{Idefini}
\end{equation}
The corresponding $\jb$-connection form $\theta_{sp}\in \Sekt(\T V\otimes_V \T U|_V$ trivially satisfies (\ref{bedsplittPoiscon}) because of (\ref{connectproject}). Furthermore, we have
\bea
\Kru^{\theta_{sp}}(\vef_1,\vef_2)
&=&\{\theta_{sp}(\vef_1),\theta_{sp}(\vef_2)\}-
\theta_{sp}([\vef_1,\vef_2])\nonumber\\
&=&\{I_{sp}\circ\ow_V^\flat(\vef_1),I_{sp}\circ\ow_V^\flat(\vef_2)\}+
I_{sp}\circ\ow_V^\flat([\vef_1,\vef_2])\nonumber\\
&=&I_{sp}\circ\left( \{\ow_V^\flat(\vef_1),\ow_V^\flat(\vef_2)\}_V+
\ow_V^\flat([\vef_1,\vef_2])\right),\nonumber
\end{eqnarray}
where $\{\cdot,\cdot\}_V$ is the Poisson bracket on $V$, since  $(pr_1\circ sp)^*:C^\infty(V)\to C^\infty(U)$ is a Poisson morphism onto the function which are constant along the leaves of $\mathcal N$. Now, property (\ref{formbrakhomo}) for a symplectic Poisson tensor implies that $\Kru^{\theta_{sp}}=0$. \hfill $\square$

\begin{cor}
Let $sp:U\to V\times N$ be a fixed splitting map and $\theta_{sp}$ the
associated $\jb$-connection form. There is a bijection $\Sekt(\T V\otimes_V \fb_{|V})\to \Inc$ given by $\al\mapsto \theta_\al=\al+ \theta_{sp}$. Thus, $\Inc$ is an affine space modelled over the vector
space $\Sekt(\T V\otimes_V \fb_{|V})$.
\end{cor}

\begin{defin}\label{trivialalpha}
\begin{rm}
Let $\theta$ be a connection form,
and let $\al$ be the corresponding 1-form $\al$
such that $\theta=\al+\theta_{sp}$. We can associate to $\al$ a
$\g$-valued 1-form $\bar \al$ on $V$, given by
\bes
\bar \al^\flat =
pr_2 \circ \T sp|_V
\circ \theta^\flat\circ T(sp^{-1}|_V):\hspace{0.2cm}TV\to V\times
\g,
\end{equation*}
where $\g=\T_{z_0}N$ is a Lie algebra because of lemma \ref{thelinearapprox}.
\end{rm}
\end{defin}

Let $sp:U\to V\times N$ and
$\ti{sp}:U\to V\times N$ be two splitting maps. The composition
$\phi=\ti{sp}^{-1}\circ sp:U\mapsto U$
is a Poisson morphism, and writing $\ti{sp}=sp\circ \phi^{-1}$, we can regard the action of $\phi$ as a
transition between splitting maps. The map induced the splitting transformation
$\phi$ in $V\times N$ is denoted by
$\bar\phi= sp \circ \phi \circ sp^{-1}$. It induces the bundle automorphism
\beq
\T \bar \phi|_V= 
\begin{pmatrix} {\sf P} & {\sf 0} \\ {\sf B} & {\sf R} \end{pmatrix} :\hspace{0.2cm}
\T V\oplus \g \to \T V\oplus \g.
\label{Tfialsmatrix}
\end{equation}
where the matrix elements are given by sections
\beq
{\sf P}: V\to T V\otimes_V \T V \qquad\qquad {\sf B}: V\to TV\otimes 
\g \qquad\qquad {\sf R}: V\to Aut(\g),
\label{diebuendelmorphismen}
\end{equation}
$Aut(\g)$ denoting the automorphism group of $\g$. The maps ${\sf P}$ and ${\sf P^{*-1}}$ are given by
%(if we identify $\T V$ with $\T V \oplus 0$)
\bea
{\sf P}\quad &=& pr_1 \circ \T \bar \phi|_{\T V}= (Ti_V)^* \circ \T \bar
\phi|_{\T V}= \T(\bar \phi|_V) \circ (Ti_V)^*|_{\T V}\nonumber\\
&=& \T(\bar \phi|_V)=\T(sp|_V)\circ \T (\phi|_V)\circ
\T (sp|_V)^{-1}\nonumber\\
{\sf P^{*-1}}&=& T(\bar\phi|_V)\hspace{0.2cm}
= T(sp|_V)\circ T(\phi|_V)\circ T(sp|_V)^{-1} :\hspace{0.2cm}TV \to TV.
\label{dualesPdef}
\end{eqnarray}
Finally, the inverse bundle morphism of (\ref{Tfialsmatrix}) is
easily seen to be given by
\bes
\T \bar \phi^{-1}=
\begin{pmatrix}
{\sf P}^{-1} & {\sf 0} \\ -{\sf R}^{-1}\circ {\sf B}\circ {\sf
P}^{-1}  & {\sf R}^{-1}
\end{pmatrix}.
\end{equation*}

\begin{prop} \label{hierhabenwirdastrfovehalten}
Let $\theta$ be an $\jb$-connection form on $V$. Let $sp:U\to
V\times N$ be a splitting map and $\bar\al$ the associated $\g$-valued
1-form to $\theta$ on $V$ as in definition \ref{trivialalpha}.
Furthermore, let $\phi$ be a splitting transformation. Then, the
$\g$-valued 1-form associated to $\phi^*\theta$ via $sp$ is given by
\bes
\bar{\ti\al}^\flat={\sf R}^{-1}\circ (\bar \al^\flat -
{\sf B}\circ {\sf P}^{-1}\circ \ow_V^\flat)\circ {\sf P^{*-1}}:
\hspace{0.2cm} TV\to
V\times \g,
\end{equation*}
where the bundle morphisms ${\sf R}$, ${\sf B}$ and ${\sf P^*}$ are
defined by (\ref{Tfialsmatrix}) and (\ref{dualesPdef}).
\end{prop}

\noindent
{\sc Proof:}
By definition,
\bea
\bar{\ti \al}^\flat &=& pr_2 \circ \T sp
\circ (\phi^*\theta)^\flat\circ T(sp^{-1}|_V)\nonumber\\
&=& pr_2 \circ \T sp
\circ (\T \phi)^{-1}\circ \theta^\flat \circ T(\phi|_V)\circ T(sp^{-1}|_V)
\nonumber\\
&=&pr_2 \circ \T \bar \phi^{-1}\circ \T sp
\circ \theta^\flat \circ T(sp^{-1}|_V)\circ T(\bar \phi|_V)\nonumber\\
&=& ({\sf R}^{-1}\circ pr_2 \circ \T sp
\circ \theta^\flat \circ T(sp^{-1}|_V) -\nonumber\\
&&\hspace{2cm}-
{\sf R}^{-1}\circ {\sf B}\circ {\sf P}^{-1}\circ pr_1 \circ \T sp
\circ \theta^\flat \circ T(sp^{-1}|_V)) \circ {\sf P^{*-1}}\nonumber\\
&=& {\sf R}^{-1}\circ  (\bar \al^\flat -
{\sf B}\circ {\sf P}^{-1}\circ \T(sp|_V)\circ (Ti_V)^*
\circ \theta^\flat\circ T(sp^{-1}|_V) ) \circ {\sf P^{*-1}} \nonumber\\
&=& {\sf R}^{-1}\circ  (\bar \al^\flat  -
{\sf B}\circ {\sf P}^{-1}\circ \T(sp|_V)\circ \ow_V^\flat\circ
T(sp^{-1}|_V)) \circ {\sf P^{*-1}}\nonumber\\
&=& {\sf R}^{-1}\circ  (\bar \al^\flat  -
{\sf B}\circ {\sf P}^{-1}\circ \ow_V^\flat) \circ {\sf P^{*-1}}\nonumber
\end{eqnarray}
where we used
%(\ref{trivialtheta}),
the fact that $\theta$ was an
$\jb$-connection form, and that $(sp|_V)$ is a symplectomorphism of
$(V,\ow_V)$.
\hfill $\square$

\begin{cor}
Let $\theta$ be a $\jb$-connection form on $V$ and
let $sp$ and $\ti{sp}$ two splitting maps. Let the $\g$-valued 1-forms
associated to $\theta$ by $sp$ and $\ti{sp}$ be denoted by $\bar \al_{sp}$
and ${\bar\al_{\ti{sp}}}$, respectively. Then,
$\bar\al_{\ti{sp}}^\flat={\sf R}^{-1}\circ (\bar\al_{sp}^\flat -
{\sf B}\circ {\sf P}^{-1}\circ \ow_V^\flat)\circ {\sf P^{*-1}}$,
where the bundle morphisms ${\sf R}$, ${\sf B}$ and ${\sf P^*}$ are
defined by (\ref{Tfialsmatrix}) and (\ref{dualesPdef}) for
$\phi^{-1}= sp^{-1}\circ\ti{sp}$.
\end{cor}

\subsection{Principal connections in the Lie frame bundle}

The resemblance of the local splitting transformations with gauge transformations suggests that there should be associated principal connection forms taking values in $Lie(Aut(\g))$.  
In deed, lemma \ref{thelinearapprox} states
that $\fb$ is a Lie algebra bundle with all fibers belonging to the isomorphism class of the fixed element $\g$.

\begin{defin}
\begin{rm}
The bundle $\pa:\Pa\to S$ whose fiber at $x\in S$ consists of the Lie algebra isomorphisms $f:\g\to\fb_x$
is called the {\it Lie frame bundle} at $S$.
That is, $\Pa$ is the bundle of frames ({\it rep\`eres}) in $\fb$ which respect the Lie algebra structure.
\end{rm}
\end{defin}

\begin{lem}\label{TVOisass}
$\Pa$ is a principal bundle over $S$ with structure group $Aut(\g)$, where the group action is given by $(f,a)\to f\circ a$ for $(f,a)\in\Pa\times Aut(\g)$. With respect to the diagonal (right) action of $Aut(\g)$ on $\Pa\times \g$, we have a canonical isomorphism
\beq
\fb\cong (\Pa\times \g)/Aut(\g)=\g(\Pa). \label{TViso}
\end{equation}
\end{lem}

\noindent
{\sc Proof:} This is a standard result on vector bundles. See for example \cite{Kobayashi1}. \qed

\begin{prop}
Let $\theta$ be an  $\jb$-connection form. It defines a covariant
derivative in the vector bundle $\fb\cong\g(\Pa)$ given by
\bes
\nabla^\theta: \Sekt(\fb)\longrightarrow \Sekt(\T S\otimes_S
\fb)\qquad\quad
\eta \longmapsto \{\theta,\eta\}.
\end{equation*}
Here, the brackets denote the bracket of section of $\jb$ defined in lemma \ref{klammerexiatiert}.
\end{prop}

\noindent
{\sc Proof:} Recall from lemma \ref{klammerexiatiert} that $\Sekt(\fb)\subset \Sekt(\jb)$
is an ideal for the bracket, such that the definition makes sense. Now, one can easily check that this defines a covariant derivative corresponding to a linear connection on $\fb$. In particular,
\bes
 \{\theta,f\eta\}=f\{\theta,\eta\}- (\po^\sharp\circ\theta^\flat)(f)\eta=
f\{\theta,\eta\}+ df\wedge \eta,
\end{equation*}
where we used property (\ref{dercon}), and the fact that $\theta$ is
an $\jb$-connection form.  \qed

\begin{cor}\label{thetainner}
Let $\theta$ be an $\jb$-connection form. Then, it defines a
principal connection form $\Ahl^\theta$ in the principal bundle $\Pa$,
which takes values in the inner derivations of $\g$.
\end{cor}

\noindent
{\sc Proof:}
A covariant derivative in an associated bundle naturally induces a
principal connection in the corresponding frame bundle. Since
$\nabla^\theta$ acts by the Lie bracket of forms, it is a derivation
of the Lie algebra structure in the fibers, and thus, parallel
transport will act by automorphisms. Consequently, the  connection in
the principal frame bundle reduces to the Lie frame bundle $\Pa$.
Furthermore, the action of $\nabla^\theta$ is by inner derivations of the fibers of $\g(P)$. Thus, the principal connection form takes values in the inner derivations of $\g$. \qed

\begin{prop}\label{Finad}
The curvature form of the covariant derivative $\nabla^\theta$ is
given by
\bea
\Fk_\nabla^\theta(\vef_1,\vef_2)&=&\nabla^\theta_{\vef_1}\nabla^\theta_{\vef_2}-
\nabla^\theta_{\vef_2}\nabla^\theta_{\vef_1}-\nabla^\theta_{[\vef_1,\vef_2]}\nonumber\\
&=&(ad\circ
\Kru^\theta)(\vef_1,\vef_2) \quad\qquad\qquad\qquad\forall \vef_1,\vef_2\in \Wek^1(S).\nonumber
\end{eqnarray}
Here, $ad$ denotes the adjoint action of the Lie algebra structure of
the fibers of $\fb$.  In particular, the curvature form takes its values in the inner derivations on the fibers of $\fb$, and the
corresponding curvature 2-form
$\Fk^\theta=d\Ahl^\theta+\eha[\Ahl^\theta,\Ahl^\theta]$ on
$\Pa$ values in the inner derivations of $\g$.
\end{prop}

\noindent
{\sc Proof:} We have, using the Jacobi identity, for all $\al\in\Sekt(\fb)$,
\bea
\Fk_\nabla^\theta(\vef_1,\vef_2)(\al)
&=&\nabla^\theta_{\vef_1}\nabla^\theta_{\vef_2}(\al)-
\nabla^\theta_{\vef_2}\nabla^\theta_{\vef_1}(\al)
-\nabla^\theta_{[\vef_1,\vef_2]}(\al)\nonumber\\
%&=&\{\theta(\vef_1),\{\theta(\vef_2),\al\}\} -\{\theta(\vef_2),\{\theta(\vef_1),\al\}\}-
%\{\theta([\vef_1,\vef_2],\al)\}\nonumber\\
&=&\{\{\theta(\vef_1),\theta(\vef_2)\},\al\} -
\{\theta([\vef_1,\vef_2],\al)\}\nonumber\\
(ad\circ
\Kru^\theta)(\vef_1,\vef_2)(\al)&=&\{(\{\theta(\vef_1),\theta(\vef_2)\} -\theta([\vef_1,\vef_2])),\al\}.\nonumber
\end{eqnarray}
Thus, the corresponding two form $\Fk^\theta$ on $\Pa$ takes values in the inner derivation of $\g$, and $\Fk_\nabla^\theta$ takes values in
the inner derivations of the fibers. \qed

\section{Sternberg-Weinstein approximation}

Up to now, we constructed several gauge theoretic objects related to $\jb$, but completely ignored the Poisson structure of proposition \ref{dualliealgebroid} on the dual bundle $\db$. We now turn to this structure in a context where it turns out to generically provide a natural linear approximation to any Hamiltonian system on $Z$ by a Wong system on a Sternberg-Weinstein phase space, involving a metric and a Yang-Mills field. The construction involves the choice of an embedded {\it Lagrangian submanifold} $i_X:X\to S$ of a fixed  leaf $S\subset Z$.

\subsection{The Sternberg-Weinstein Poisson structure}
\label{sound}

Let $(TX)^0$ and $\XO$ denote the subbundles of $\TxU$ and $\T S|_X$ given by the annihilator spaces of the fibers of $TX$ as a subbundle of $TZ|_X$ and $TS|_X$, respectively.  Since $X$ is Lagrangian, the restriction $\ank_X=\ank|_{(TX)^0}$ yields a map
\beq
\ank_X=Ti_{S}\circ (-\ow_S^\flat)^{-1}\circ (Ti_S)^*|_{(TX)^0}:(TX)^0 \to
\XO\to TX.
\label{eingeschraemkterpo}
\end{equation}
If we define the {\it (dual) Lie algebroid} and {\it (dual) Lie bundle at $X$} by
\bea
\jx&=&(TX)^0 \subset \T Z|_X=\jb|_X \qquad \qquad
\dxb = TZ|_X/ TX =\db|_X/TX\nonumber\\
\fx&=&(TS)^0|_X=\fb|_X \qquad\qquad \qquad \quad \hspace{0.2cm}
\dfb = TZ|_X/TS|_X= \eb|_X, \nonumber
\end{eqnarray}
respectively, $\jx$, $\dxb$ and $\fx$, $\dfb$ being dual bundles, we obtain an exact sequence
\beq
0\longrightarrow \fx \longrightarrow \jx \longrightarrow TX \longrightarrow 0.
\label{seq67}
\end{equation}  

\begin{lem} \label{bracketrestrictedtoX}
There is a well-defined bracket on the space of sections of $\jx$
\beq
\{\cdot,\cdot\}: \Sekt(\jx)\times \Sekt(\jx)\to \Sekt(\jx)
\quad \qquad\{\al|_X,\beta|_X\}=\{\al,\beta\}|_X \label{LAaufJX}
\end{equation}
for all $\al,\beta \in \Frm^1(Z)$ such that $\al|_X$ and $\beta|_X$
take values in $\jx$, where the bracket on the right is the
standard bracket defined in proposition \ref{bracketononeforms}, and
the restrictions mean the restriction of the maps $Z\to \T Z$ to $X$.
The restriction of this bracket to $\Sekt(\fx)$ is given by the point-wise Lie bracket in the fibers of $\fx$
defined in lemma \ref{thelinearapprox}. Furthermore, $\Sekt(\fx)\subset \Sekt(\jx)$ is an ideal.
\end{lem}

\noindent
{\sc Proof:}
Let $\al,\beta\in \Frm^1(Z)$ such that $\al|_X$ and $\beta|_X$ take values in $\jx$. In the proof of lemma \ref{klammerexiatiert}, we saw that for $\vef\in \Wek^1(Z)$,
\bea
\{\al,\beta\}(\vef)&=&\inn_{\al^\sharp}(\inn_\vef d\beta) -
\inn_{\beta^\sharp}(\inn_\vef d\al) +
\vef \cdot \po(\al,\beta)\label{ofthebracket}\\
&=&\beta^\sharp \cdot \al(\vef)- \al^\sharp\cdot \beta(\vef)+ (\Lie_\vef\po)(\al,\beta),\nonumber
\end{eqnarray}
where $\al^\sharp=\po^\sharp\circ \al$ etc. Now, (\ref{eingeschraemkterpo})
shows that if $\al|_X$ and $\beta|_X$ take their
values in $\jx$, then $\al^\sharp|_X$ and $\beta^\sharp|_X$ take their
values in $TX$, and thus, the restriction to $X$ of the bracket
(\ref{ofthebracket}) depends only on the values of $\al$ and $\beta$
on $X$. Furthermore, assume that $\vef|_X$ is tangent to $X$.
Then, because of $\po(\al,\beta)=\al
(\beta^\sharp)$, the last
term in the first line (\ref{ofthebracket}) vanishes on $X$ due to
(\ref{eingeschraemkterpo}). On the other hand, the first term
\bes
\inn_{\al^\sharp}(\inn_\vef d\beta)= d\beta(\vef,\al^\sharp)
=\al^\sharp\cdot
\beta(\vef)-\vef\cdot\beta(\al^\sharp)-\beta([\vef,\al^\sharp]),
\end{equation*}
and this vanishes on $X$ because the Lie bracket of vector fields that are
tangent to $X$ is tangent to $X$. Finally, the second term of
(\ref{ofthebracket}) vanishes
for similar reasons. This shows that the bracket of
$\jx$-valued forms is $\jx$-valued. The remaining assertions follow from lemma \ref{klammerexiatiert}. \qed

\begin{cor}
The bundle $\jx$ with the bracket (\ref{LAaufJX}) on $\Sekt(\jx)$ and the anchor map $\ank_X$ is a transitive Lie algebroid, and $\fx$ is a natural Lie subalgebroid.
\end{cor}

\begin{defin}
\begin{rm}
A 1-form $\acon:X\to \T X\otimes_X \jx$ on
$X$ with values in $\jx$ is called an {\it
$\jx$-connection form} if the associated bundle morphism over $X$
\beq
\acon^\flat: TX\to \jx \qquad \mbox{satisfies} \qquad
\ank_X \circ \acon^\flat= Id_{TX},
\label{nochconnbed}
\end{equation}
i.e., if it is
a splitting of (\ref{seq67}).
The set of $\jx$-connection forms is denoted by $\Inc_X$.
\end{rm}
\end{defin}

\begin{lem} \label{invariant-restricted}
For any $\theta\in \Inc$, $\acon=i_X^*\theta\in \Inc_X$, and $\Inc_X=i_X^*\Inc$. 
\end{lem}

\noindent
{\sc Proof:} Take $\vct_1,\vct_2\in TX$. We have
\bea
 (\acon(\vct_1))(\vct_2) &=& (i_S^*(\theta(\vct_1))(\vct_2)
= ((Ti_S)^*\circ \theta^\flat) (\vct_1)(\vct_2)\nonumber\\
&=& (\ow_S^\flat \circ \ow_S^{\flat-1} \circ (Ti_S)^*\circ \theta^\flat)
(\vct_1)(\vct_2)= -\ow_S^\flat(\vct_1)(\vct_2)=0\nonumber
\end{eqnarray}
because of (\ref{eingeschraemkterpo}) and since $X$ is Lagrangian, and thus, $\acon$ takes values in
$\jx$. Property (\ref{nochconnbed}) follows from
$\ank\circ \theta^\flat=Id_{TS}$ and
$(i_X^*\theta)^\flat=\theta^\flat\circ i_{X*}$.
On the other hand, any $\jx$-connection form can be extended to an $\jb$-connection form. \qed

\begin{cor}
Let $\acon$ and $\theta$ be $\jx$- and $\jb$-connection forms  such that $\acon=i_X^*\theta$. There is a well-defined 2-form on $X$ with values in $\fx$ satisfying
\beq
\Kru^\acon(\vef_1,\vef_2)=i_X^*\Kru^\theta= \{\acon(\vef_1),\acon(\vef_2)\}-\acon([\vef_1,\vef_2])
 \qquad\forall \vef_1,\vef_2
\in \Wek^1(X).
\label{restrictedKruemmung}
\end{equation}
\end{cor}

\begin{defin}
\begin{rm}
The 2-form $\Kru^\acon$ is called the {\it  curvature} of $\acon$.
%($\theta$), for every $\theta$ such that
%$\acon=i_X^*\theta$.
\end{rm}
\end{defin}

\begin{cor} \label{gaugeactrestricted}
Let $\phi$ be an $S$-Poisson isomorphism preserving $X$. Its cotangent map yields a bundle morphism $\T \phi|_X:\jx\to \jx$, inducing an action of $\phi^*$ on $\Inc_X$ as in definition \ref{gaugeactiononforms}  by setting 
$\phi^* (i_X^*\theta)= i_X^*(\phi^*\theta)$
for all $\theta\in \Inc$. Furthermore,
the action preserves the bracket defined in lemma \ref{bracketrestrictedtoX}.
If $\acon \in \Inc_X$, then
$\Kru^{\phi^*\acon}=\phi^* \Kru^{\acon}$.
\end{cor}

\begin{cor}
An $\jx$-connection form induces a principal connection
in the restricted principal bundle $\Pa_X:=\Pa|_X$ and the associated bundle $\g(\Pa_X)=\g(\Pa)|_X$.
\end{cor}

Now we can turn to the Poisson structure on $\dxb$ and show that it is a natural linear approximation to $Z$ near $X$. Let us give a detailed proof of proposition \ref{dualliealgebroid} in this special case since it will be useful in the sequel. We denote the projection induced by $\tau_Z:T Z\to Z$ by 
\bes
\epro:\dxb\to X.
\end{equation*}

\begin{prop}\label{podot}
There is a natural extension of the bracket defined in lemma \ref{bracketrestrictedtoX} on the sections of $\jx$, seen as vertically linear functions on the dual bundle $\dxb$, to an exact Poisson bracket $\{\cdot,\cdot\}'$ on $C^\infty(\dxb)$.
\end{prop}

\noindent
{\sc Proof:} In order to define a Poisson structure on $\dxb$, it suffices to specify the Poisson bracket on a characterizing family of functions which is large enough to ensure that the differentials of its elements span the cotangent space to $\dxb$ at each point and thus, their Poisson brackets determine the Poisson tensor.

Clearly, we can naturally define the Lie bracket $\{\cdot,\cdot\}'$ to be the trivial Lie bracket on the subspace $\epro^*C^\infty(X) \subset C^\infty(\dxb)$ of vertically constant functions. On the other hand, $\Sekt(\jx)$ can be identified with the subspace $\baSE(\jx)\subset C^\infty(\dxb)$ whose elements are linear on the fibers, and the Lie algebroid bracket on $\Sekt(\jx)$ induces a Lie bracket $\{\cdot,\cdot\}'$ on this subspace. Furthermore, one easily verifies that
\bes \T_t(\dxb)=d(\baSE(\jx)\oplus \epro^*C^\infty(X))_t\qquad  \qquad  \forall t\in \dxb.
\end{equation*}
Thus, the bivector on $\dxb$ defining $\{\cdot,\cdot\}'$ on the whole of $C^\infty(\dxb)$ will be determined by specifying the bracket of an element of $\baSE(\jx)$ with an element of $\epro^*C^\infty(X)$.

From the Leibnitz identities for the Lie algebroid brackets (for clarity denoted by $[\cdot,\cdot]_{\jx}$ below) and the bracket $\{\cdot,\cdot\}'$, it follows that for $\mathcal V_1, \mathcal V_2\in \Sekt(\jb)$, denoting the corresponding functions as $\bar{\mathcal V}_1, \bar{\mathcal V}_2\in \baSE(\jx)$, and $f\in C^\infty(X)$, we obtain
\bea
\{\bar{\mathcal V}_1,(\epro^*f)\bar{\mathcal V}_2\}'=\oli{[\mathcal V_1,f\mathcal V_2]}_{\jx}&=&\oli{(((\ank_X\circ \mathcal V_1)\cdot f)\mathcal V_2)} +\oli{(f[\mathcal V_1,\mathcal V_2]_{\jx})}\nonumber\\
\{\bar{\mathcal V}_1,\epro^*f\}'\bar{\mathcal V}_2 +(\epro^*f)\{\bar{\mathcal V}_1,\bar{\mathcal V}_2\}'&=&\epro^*((\ank_X\circ \mathcal V_1)\cdot f)\bar{\mathcal V}_2 +(\epro^*f)\{\bar{\mathcal V}_1,\bar{\mathcal V}_2\}'.\nonumber
\end{eqnarray}
This implies that we must set
\bea
\{\bar{\mathcal V}_1,\epro^*f\}'=-X_{\bar{\mathcal V}}\cdot\epro^*f&=&\epro^*((\ank_X\circ \mathcal V_1)\cdot f)
\label{derdefin}\\
&=&((\ank_X)^*\circ df)\cdot \bar{\mathcal V}_1=X_{\epro^*f}\cdot \bar{\mathcal V}_1\nonumber,
\end{eqnarray}
where the section $(\ank_X)^*\circ df:X\to \T X\to \dxb$ of $\dxb$ is seen as a constant vertical vector field on $\dxb$, which is thus the Hamiltonian vector field $X_{\epro^*f}$ of $\epro^*f$. In fact, (\ref{derdefin}) defines a map $\al:\baSE(\jx)\to \fra{der}(\epro^*C^\infty(X))$, and the bracket $\{\cdot, \cdot\}'$ on $\epro^*C^\infty(X)\oplus \baSE(\jx)\subset C^\infty(\dxb)$ is the semidirect sum $\epro^*C^\infty(X)\rtimes_\al \baSE(\jx)$. This completely determines an extension to a natural bracket $\{\cdot,\cdot\}'$ on $C^\infty(\dxb)$.

It remains to verify that this bracket satisfies the Jacobi identity. This is most easily done in local coordinates. In deed, let us choose Darboux coordinates $(x^\mu,p_\mu, r_a)$ on $Z$ centered at $x_0\in X$. They induce coordinates $(x^\mu,\dot x^\mu,\dot p_\mu,\dot r_a)$ on $TX|_X$ with respect the vector fields $(\dl/\dl x^\mu,\dl/\dl p_\mu,\dl/\dl r_a)$ whose values at $x$ span $T_xZ$ for each $x\in X$, and coordinates $(x^\mu,[\dot p_\mu],[ \dot r_a])$ on $\dxb=TZ|_X/TX$ with respect to the sections $([\dl/\dl p_\mu],[\dl/\dl r_a])$ of $\dxb$ whose values at $x$ span $(\dxb)_x$ for all $x\in X$.

On the other hand, the values at $x$ of the sections $(dp_\mu|_X, dr_a|_X)$ of $\T Z_X$ span $\jx=(TX)^0$ at each $x\in X$. By definition, $x^\mu\in \epro^*C^\infty(X)$, while $[\dot p_\mu]=\bar{d p_\mu}|_X,[\dot r_a]=\bar{d r_a}|_X\in \baSE(\jx)$. Thus, we have
\bea
\{x^\mu,x^\nu\}'&=&0\nonumber\\
\{[\dot p_\mu],[\dot p_\nu]\}'&=&\oli{[dp_\mu|_X,dp_\nu|_X]}_{\jx}=\oli{(d\{p_\mu,p_\nu\}|_X)}=0\nonumber\\
\{x^\mu,[\dot p_\nu]\}'&=&((\po^\sharp_X\circ dp_\nu|_X)\cdot x^\mu=\dl x^\mu/\dl x^\nu =\delta^\mu_\nu\nonumber\\
\{x^\mu,[\dot r_b]\}'&=&((\po^\sharp_X\circ dr_b|_X)\cdot x^\mu=0\nonumber\\
\{[\dot p_\mu],[\dot r_b]\}'&=&\oli{[dp_\mu|_X,dr_b|_X]}_{\jx}=\oli{(d\{p_\mu,r_b\}|_X)}=0\nonumber\\
\{[\dot r_a],[\dot r_b]\}'&=&\oli{[dr_a|_X,dr_b|_X]}_{\jx}=\oli{(d\{r_a,r_b\}|_X)}=c^c_{ab}\bar{dr_c}|_X =c^c_{ab}[\dot r_c]\nonumber
\end{eqnarray}
where we used the definitions of corollary \ref{Darbcord} and (\ref{structureconstdefin}). Hence, with respect to the coordinates $(x^\mu,[\dot p_\mu],[ \dot r_a])$, the bivector $\po'$ defined by the bracket $\{\cdot,\cdot\}'$ on $C^\infty(\dxb)$ reads
\beq
\po'=\frac{\dl}{\dl x^\mu}\wedge \frac{\dl}{\dl [\dot p_\mu]} + \eha c^c_{ab}[\dot r_c]\frac{\dl}{\dl [\dot r_a]}\wedge \frac{\dl}{\dl [\dot r_b]}. \label{SWw}
\end{equation}
Since the $c^c_{ab}$ are structure constants of the Lie algebra $\g$, it follows that $\po'$ satisfies the Jacobi identity $[\po',\po']=0$ as required.  If
%let $\sigma'$ denote the contravariant exterior differential defined by the Poisson bracket on $\dxb$, and 
$\lio\in \Wek(\dxb)$ is the linear vertical vector field on $\dxb$ corresponding to the identity bundle morphism $Id_{\dxb}\in \Sekt(\dxb\otimes_X\jx)$, then $\lio=[\dot p_\mu]\dl/\dl [\dot p_\mu]+[\dot r_a] \dl/\dl [\dot r_a]$, and one easily verifies that $\po'=-[\po',\lio]$, that is, $\po'$ is even exact. In addition, we note that $(x^\mu,[\dot p_\mu],[ \dot r_a])$ are local Darboux coordinates for $\po'$ on $\dxb$ at $x$. \qed

\begin{cor}\label{HamVFlin}
Let $\Sekt(S^\bullet\jx)$ be the symmetric algebra of the sections of $\jx$. There is a canonical identification (denoted again by a bar) of this algebra with the subalgebra $\baSE(S^\bullet\jx)\subset C^\infty(\dxb)$ consisting of those functions which are polynomial on the fibers. On the other hand, if $\Cox=\{f\in C^\infty(Z)|f|_X=0\}$, there is a natural map
\bes
\Cox\to \baSE(\jx)=\baSE(S^1\jx)\subset C^\infty(\dxb)\qquad\quad f\mapsto \bar{df}|_X.
\end{equation*}
For any $f\in \Cox$, the adjoined action of $\bar{df}|_X$ on $\baSE(S^\bullet\jx)$ is given by
\beq
\{\bar P,\bar{df}|_X\}'=X_{\bar{df}|_X}\cdot \bar P=\oli{{(\Lie_{X_f}\hat P)}|_X} \qquad \quad\forall P\in \Sekt(S^\bullet\jx),
\label{adacfunc}
\end{equation}
where $\hat P \in \Sekt(S^\bullet \T Z)$ is an arbitrary extension of $P$ on a neighborhood of $X$.
\end{cor}

\noindent
{\sc Proof:} The identifications mentioned in the corollary are obviously canonical. Since $\{\cdot,X_{\bar{df}|_X}\}'$ and $\Lie_{X_f}$ are both derivatives of the respective associative algebra structures, it suffices to verify (\ref{adacfunc}) for elements of $\Sekt(S^0\jx)=C^\infty(X)$ and $\Sekt(S^1\jx)=\Sekt(\jx)$. For $P\in C^\infty(X)$, we have $\bar P=\epro^*P$, and thus
\bes
\{\bar P,\bar{df}|_X\}'=-\oli{((\ank_X\circ df|_X)\cdot P)}
=\oli{(X_f|_X \cdot P)} =\oli{(\Lie_{X_f} \hat P)|_X}
\end{equation*}
for any extension $\hat P$ of $P$ since $X_f|_X$ is tangent to $X$ for $f\in \Cox$. For $P\in \Sekt(\jx)$ and any extension $\hat P$, we have further
\bea
\{\bar P,\bar{df}|_X\}'&=&\oli{[P,df|_X]}_{\jx}=\oli{(\inn_{\po^\sharp\circ df}d \hat P -\inn_{\po^\sharp\circ \hat P}d(df)+d(\po(\hat P,df)))|_X}\nonumber\\
&=& \oli{(\inn_{X_f}\circ d \hat P+d\circ \inn_{X_f}\hat P)|_X}=\oli{(\Lie_{X_f}\hat P)|_X}.\nonumber
\end{eqnarray}
This proves the corollary. \qed

\begin{remark}
\begin{rm}
Recall that for every Lie algebroid $(E,\rho,[\cdot,\cdot]_E)$, $\Sekt(\wedge^\bullet E)$ is provided with a Gerstenhaber algebra structure involving the Schouten-Nijenhuis extension of $[\cdot,\cdot]_E$ (cf \cite{WeinNC}). \bems
\end{rm}
\end{remark}  

\begin{cor}\label{horistproj}
There is a canonical injection 
$\Sekt(\wedge^n\T X \otimes_X\jx)\to 
%\bar\Sekt(\wedge^n\T X  \otimes_X\jx)\subset 
\Frm^n(\dxb)$, defined by the tensor product of the pull-back by $\epro$ with the identification of corollary \ref{HamVFlin}, which we denote again by a bar. A projection $\hor:\Wek^1(\dxb)\to \Wek^1(\dxb)$ with $\epro_*\circ \hor=\epro_*$ is defined by 
\bes
\hor=\ada(\acon) \circ \epro_*, \qquad \text{where}\quad \ada(\acon)(\vef)=-X_{\oli{\acon(\vef)}} \qquad\forall \vef\in \Wek(X).
\end{equation*}  
For $\acon\in \Inc_X$, the following structure equation and Bianchi identity 
\bes
d\bar  \acon\circ \hor=\bar \Kru^\acon+\eha\{\bar\acon,\bar\acon\}'\qquad \text{and}\qquad d\bar\Kru^\acon\circ \hor=\eha\{\bar\acon,\bar\Kru^\acon\}',
\end{equation*}
where $\{\bar\acon,\bar\acon\}'(\mathcal Y_1,\mathcal Y_2)=\{\bar\acon(\mathcal Y_1),\bar\acon(\mathcal Y_2)\}'-\{\bar\acon(\mathcal Y_2),\bar\acon(\mathcal Y_1)\}'$ etc., hold good.     
\end{cor}

\noindent
{\sc Proof:} Because of (\ref{derdefin}), we have in deed $\epro_*\circ\hor(\mathcal Y)=\ank\circ \acon^\flat\circ\epro_ *(Y)=\epro_*(\mathcal Y)$. Let $\mathcal Y_i\in \Wek^1(\dxb)$, $\vef_i=\epro_*\mathcal Y_i, i=0,1,2$.  Then,
\bes
\hor(\mathcal Y_i)\cdot\bar\acon(\hor(\mathcal Y_j))=-X_{\oli{\acon^\flat\circ\epro_*(\mathcal Y_i)}}\cdot\oli{\acon(\epro_*\circ \hor(\mathcal Y_j ))} =\{\oli{\acon(\vef_i)},\oli{\acon(\vef_j)}\}'.
\end{equation*}
Thus, we have
\bea
&&\bar \Kru^\acon(\mathcal Y_1,\mathcal Y_2)
=\oli{\{\acon(\vef_1),\acon(\vef_2)\}-\acon([\vef_1,\vef_2])}
=\{\oli{\acon(\vef_1)},\oli{\acon(\vef_2)}\}'
-\oli{\acon([\vef_1,\vef_2])}\nonumber\\
&&\hspace{0.2cm}=\{\oli{\acon(\vef_1)},\oli{\acon(\vef_2)}\}'
-\{\oli{\acon(\vef_2)},\oli{\acon(\vef_1)}\}'
-\overline{\acon([\vef_1,\vef_2])}
-\{\oli{\acon(\vef_1)},\oli{\acon(\vef_2)}\}'\nonumber\\
&&\hspace{0.2cm}=\hor(\mathcal Y_1)\cdot\bar\acon(\hor(\mathcal Y_2))-\hor(\mathcal Y_2)\cdot\bar\acon(\hor(\mathcal Y_1))
-\bar\acon([\hor(\mathcal Y_1),\hor(\mathcal Y_2)])-\nonumber\\
&&\hspace{3.5cm}-\{\bar\acon(\mathcal Y_1),\bar\acon(\mathcal Y_2)\}'=(d\bar  \acon\circ \hor-1/2\{\bar\acon,\bar\acon\}')(\mathcal Y_1,\mathcal Y_2)\nonumber
\end{eqnarray}
which implies the structure equation. For the Bianchi identity, we compute
\bea
&&d\bar\Kru^\acon\circ \hor(\mY_0,\mY_1,\mY_2)
=d(d\bar\acon
-\eha\{\bar\acon,\bar\acon\}')\circ \hor(\mY_0,\mY_1,\mY_2)
\nonumber\\
&&=-\eha d(\{\bar\acon,\bar\acon\}')(\hor(\mY_0),\hor(\mY_1),\hor(\mY_2))
\nonumber\\
&&=-\sum_{cycl} \{\oli{\acon(\vef_0)},\{\oli{\acon(\vef_1)},\oli{\acon(\vef_2)}\}'\}'
+\sum_{cycl}\{\oli{\acon([\vef_0,\vef_1])},\oli{\acon(\vef_2)}\}'
\nonumber\\
&&=\sum_{cycl} \{\oli{\acon(\vef_0)},\{\oli{\acon(\vef_1)},\oli{\acon(\vef_2)}\}'\}'
-\sum_{cycl}\{\oli{\acon(\vef_0)},\oli{\acon([\vef_1,\vef_2])}\}'
\nonumber\\
&&=\sum_{cycl} \{\oli{\acon(\vef_0)},\oli{\{\acon(\vef_1),\acon(\vef_2)\}
+\acon([\vef_1,\vef_2])}\}'=\eha\{\bar\acon,\bar \Kru^\acon\}'(\mY_0,\mY_1,\mY_2),\nonumber
\end{eqnarray}
where the summation runs over all cyclic permutations of the indices $0,1,2$, and we used the Jacobi identity for the fourth equality. This proves the corollary. \qed

\begin{defin}
\begin{rm}
For any Poisson manifold $(Z,\po)$ and any Lagrangian
submanifold $X\subset S$ of a symplectic leaf $S$,
the {\it Sternberg-Weinstein approximation} at $X$ is given by the Poisson manifold $(\ZX,\po')$, where $\ZX=\dxb=TZ|_X/TX$, and $\po'$ is the Poisson tensor defined by the Poisson algebra structure $\{\cdot,\cdot\}'$ on $C^\infty(\ZX)$ in lemma \ref{podot}, given in local Darboux coordinates by (\ref{SWw}).
\end{rm}
\end{defin}

\begin{cor}\label{linspannenauf}
The subbundle $\SX=TS_{|X}/TX\subset \ZX$ is a symplectic leaf, which contains the zero section, identified with $X$, as a Lagrangian submanifold. The bundle morphism
\bes
(-\ow_X^\flat)^*:\SX=(\XO)^*\longrightarrow \T X,
\end{equation*}
where $\ow^\flat_X=\ow^\flat_S|_{TX}:TX\to \XO$, 
is a symplectomorphism of $\SX$ with $\T X$ and its canonical symplectic structure. The subspace $\fra C_{\SX}=\{f\in C^\infty(\ZX)|f|_{\SX}=0\}$ is an ideal of $C^\infty(\ZX)$, and the subspace $\baSE(\fx)$ of the functions on $\dxb$ defined by sections of $\fx$ is an ideal of $\baSE(\jx)$. Furthermore, the canonical projection
\beq
\chela:\ZX=\dxb\longrightarrow \dfb\cong\gs(\Pa_X),
\label{cheladefin}
\end{equation}
is a Poisson morphism for the Poisson structure on $\dfb$ defined in lemma \ref{thelinearapprox} by the linear Poisson structures on each fiber. That is, if $f,g \in C^\infty(\dfb), n\in (\dfb)_x$,
\beq
\{\chela^*f,\chela^*g\}(\chela^{-1}(n))=\{f|_{(\dfb)_x},g|_{(\dfb)_x}\}_x(n),\label{linapr}
\end{equation}
where $\{\cdot,\cdot\}_x$ denotes the Poisson bracket on $(\dfb)_x\cong \gs$.
Finally, the Sternberg-Weinstein approximation of $\ZX$ at $X\subset \SX$ is canonically isomorphic to $\ZX$. 
\end{cor}

\noindent
{\sc Proof:}
The first two statements are clear from the construction or easily checked in the local Darboux coordinates. In the same way, we can check the last assertion, taking into account that for every vector bundle $M\to N$, there are obvious canonical vector bundle isomorphisms $TM|_{o(N)}/T(o(N))\cong V(M)|_{o(N)}\cong M$, where $o(N)\cong N$ is the image of the zero section. It follows from proposition \ref{admissible} that $\mathfrak C_{\SX}$ is an ideal, and from lemma \ref{bracketrestrictedtoX} that $\baSE(\fx)\subset \baSE(\jx)$ is an also an ideal.

As to the remaining assertions, it suffices to check them on the characteristic subspace $\tau^*C^\infty(X)\oplus \baSE(\fx)$, where the last term denotes the vertically linear functions in $\chela^*C^\infty(\dfb)\subset C^\infty(\dxb)$ defined by the sections of $\fx$. Since (\ref{derdefin}) and $\fx=\ker (\po_X^\sharp)$ imply $\{\tau^*C^\infty(X),\baSE(\fx)\}=0$, and since the bracket on sections of $\fx$ is given by the fiberwise Lie algebra bracket, (\ref{linapr}) follows from the construction of the bracket. \qed 

\begin{remark}
\begin{rm}
We can see here that $\ZX$ is locally equivalent to $\T X\times\gs$, and in particular, transversally linear at $\SX=\T X$. Since $Z$ is locally equivalent to $V\times N$, and there is always a local symplectomorphism of $V$ and $\T X$, we see that $Z$ is locally Poisson equivalent to $\ZX$ iff it is linearizable at $V$. It is easy to see (corollary \ref{SW=SW}) that the Sternberg-Weinstein phase space is naturally isomorphic to its Sternberg-Weinstein approximation, which motivated our definition.  \bems
\end{rm}
\end{remark}

\subsection{The Wong system}

A Hamiltonian on $Z$ can induce a Hamiltonian system on the Sternberg-Weinstein approximation of $Z$ at $X$ which is precisely of the Wong type, that is, given by a metric and an $\jx$-connection form on $X$. 

\begin{lem}\label{zerllem}
Let $M\to N$ be a vector bundle, and denote $o:N\to
o(N)\subset M$ the zero
section. There is a canonical bundle isomorphism $TM|_{o(N)}\cong TN\oplus_N M.$
\end{lem}

\noindent
{\sc Proof:} For every $x\in N$, 
$T_{o(x)}M=T_{o(x)}(o(N))\oplus V_{o(x)}(M)\cong T_xN\oplus M_x$
is a natural decomposition induced by the canonical identifications $o(N)\cong N$ and $V(M)\cong M\oplus_N M$ for any vector bundle $M$.  \qed

\begin{theo}\label{theo1}
Let $X\subset S\stackrel{i_S}{\hookrightarrow}Z$ be a Lagrangian submanifold of the symplectic leaf $(S,\ow_S)$ of the Poisson manifold $(Z,\po)$. Every Hamiltonian $H\in C^\infty(Z)$ with $dH|_X=0$
uniquely defines a bundle morphism
$t^VH: \hspace{0.1cm}TS|_X/TX\to \jx$.
If the composition
$t^VH_0= (Ti_S)^*\circ t^VH:\hspace{0.1cm}TS|_X/TX\cong(\XO)^* \to \XO$
is invertible, then the bundle morphisms
\bea
\emet_H^{\flat} &=&(-\ow^\flat_X)^*\circ (t^VH_0)^{-1}\circ (-\ow^\flat_X):\hspace{0.1cm}TX
\longrightarrow \T X\nonumber\\
\econ_H^{\flat} &=& t^VH\circ (t^VH_0)^{-1}\circ(-\ow^\flat_X): \hspace{0.6cm}
TX\longrightarrow \TXo\nonumber
\end{eqnarray}
where $\ow^\flat_X=\ow^\flat_S|_{TX}:TX\to\XO$, define a metric $\emet_H$ and an $\jx$-connection form
$\econ_H$ on  $X$, as well as the principal connection
form $\Ahl_H=\Ahl^{\econ}$ on $\Pa_X$.
\end{theo}

\noindent
{\sc Proof:} The tangent map of $dH|_S:S\to \jb$ yields bundle morphism
\bes
\Tan (dH)|_{TS|_X}: TS|_X\longrightarrow (dH|_X)^* T(\TU)=TS|_X\oplus_X \jb|_X,
\end{equation*}
where the decomposition is given by the above lemma. The first component is the identity by the definition of jets. The second yields the bundle morphisms
\bea
pr_2 \circ \Tan (dH)|_{TS|_X}:&& TS|_X\longrightarrow \jb|_X\label{Buend1}\\
(Ti_S)^*\circ pr_2 \circ \Tan (dH)|_{TS|_X}:&& TS|_X\longrightarrow \T S|_X\label{Buend2}.
\end{eqnarray}
Since $dH_X=0$, these maps vanish on $TX\subset TS|_X$. On the other hand, since $d(i^*_S dH)=d(d(H|_S))=0$, the bundle morphism (\ref{Buend2}) defines symmetric bilinear forms on the fibers of $TS|_X$ whose kernels contain the vectors in $TX$. Thus, it induces symmetric bilinear forms on the fibers of the quotient bundle $TS|_X/TX$ and an associated bundle morphism $TS|_X/TX\to (TS|_X/TX)^*\cong \XO$. Consequently, the morphism (\ref{Buend2}) takes values in $\XO$. From these facts, we see that the morphism (\ref{Buend1},\ref{Buend2}) induce the announced bundle morphisms 
\bea
t^VH:& \hspace{0.1cm}TS|_X/TX&\longrightarrow \hspace{0.2cm}\TXo\nonumber\\
t^VH_0 = (Ti_S)^*\circ t^VH:&\hspace{0.1cm}TS|_X/TX\cong(\XO)^* &\longrightarrow \hspace{0.2cm}\XO.\nonumber
\end{eqnarray}
Now, if $t^VH_0$ is invertible, the bilinear form
$\emet_H$ will be well-defined, non-degenerate and symmetric.
Thus, $\emet_H$ is a metric on $X$.
On the other hand, 
\bes
\ank_X \circ \econ_H^\flat=Ti_S\circ (-\ow^\flat_S)^{-1}\circ (Ti_S)^*\circ t^VH\circ (t^VH_0)^{-1}\circ (-\ow^\flat_X)=Id_{TX}
\end{equation*}
by construction. Thus, $\econ_H$ is an $\jx$-connection form on $X$, inducing a principal connection form on $\Pa_X$. \qed
\vspace{0.4cm}

\noindent
The definitions of $t^VH$, $t^VH_0$, $\emet_H$ and $\econ_H$ are
summarized in the following diagram:
\bea
\jb|_X \quad\longrightarrow \hspace{0.1cm}\TXo \longrightarrow &\XO& \stackrel{-\ow^\flat_X}{\longleftarrow} \hspace{0.2cm} TX\nonumber\\
\quad {\scriptstyle pr_2\circ \Tan(dH)|_{TS|_X}} {\nwarrow} \quad\qquad{\scriptstyle t^VH} {\nwarrow} & {\scriptstyle t^VH_0} {\Big \uparrow}{\Big \downarrow} {\scriptstyle t^VH_0^{-1}}& \quad\qquad {\Big \downarrow} {\scriptstyle \emet_H}\nonumber\\
TS|_X \longrightarrow &TS|_X/TX&\stackrel{-(\ow^\flat_X)^*}{\longrightarrow} \T X\nonumber
\end{eqnarray}
\bes
\text{where}\hspace{1.3cm}\econ_H^{\flat} = t^VH\circ (t^VH_0)^{-1}\circ (-\ow^\flat_X): \hspace{0.1cm}
TX\longrightarrow \TXo.
\end{equation*}

\begin{cor} \label{coro1}
Any section $h\in\Sekt(\jb)$ with $h|_X=0$ and such that $d(i^*_S h)|_X=0$ uniquely defines a bundle morphism $t^Vh: TS|_S\to \jx$. If the composition $t^Vh_0=(Ti_S)^*\circ t^Vh$ is invertible, then $t^Vh$ uniquely defines a metric,
an $\jx$-connection form and an associated principal connection form
on $\Pa_X$.
\end{cor}

\begin{cor}
Given in addition a linear connection in the bundle
$\jb$, which provides a splitting
$T\jb=TS\oplus_{S} V\jb$
over $S$, a section $h\in\Sekt(\jb)$ with $d(i_S^* h)=0$ naturally defines a metric and an $\jb$-connection
form on the leaf $S$.
\end{cor}

Given a Hamiltonian $H$ on $Z$, theorem \ref{theo1} should allow us to define a system analogous to the Wong system on $\ZX$. In deed, we have the dual bundle morphism
\bes
(\econ_H^\flat)^*: \ZX=\dxb\longrightarrow \T X.
\end{equation*}
On the other hand, the metric $\emet_H$ defines a Hamiltonian $\HX$ on $\T X$. Thus, we can define a Hamiltonian system on $\ZX$ by 
\beq
\HSW=\HX\circ (\econ_H^\flat)^*, \qquad \text{where}\quad \HX(p)=p(\emet_H^{\flat-1}(p)) \quad \forall p\in \T X.
\label{WongHam}
\end{equation}
Notice that the Hamiltonian $\HX$ is always defined even if $t^VH_0$ is not invertible. However, the definition of $\HSW$ requires the invertibility.

The equations of motion of this system are written as follows. The bundle morphism over $X$ defined by
\bea
&&\pomo_H=((\econ_H^\flat)^*,\chela):\hspace{0.2cm} \ZX\longrightarrow \bfx=\T X\oplus_X \dfb \cong\gs(\bPa)\nonumber\\
&&\qquad\qquad\hspace{1.1cm}\text{where}\qquad \bPa=\T X\times_X \Pa_X\nonumber
\end{eqnarray}
is easily seen to be a diffeomorphism. It can be used to induce an $H$-depending Poisson structure $\po'_H=(\pomo_H)_*\po'$ on $\bfx$. Then, $\HX$ can be pulled back to a function $\HXH$ on $\bfx$, defining a Hamiltonian system on it, and by construction, the system $(\bfx, \po'_H, \HXH)$ is equivalent to the system above. In addition, $\bfx$ is naturally fibred over $\T X$ and $X$, which allows to write down the equations of motion in a natural physical coordinates if we can find an expression for $\po_H$. 

Let $(pr_1,pr_2):\bfx\to \T X\times\dfb$ denote the natural projections. They induce the affine bundle morphism $(pr_{1*}\oplus_{TX}pr_{2*}):T(\bfx)\stackrel{\sim}{\to} T(\T X)\times_{TX} T(\dfb)$. This allows to define a graded algebra map $\adj^\pi:\Wek(\T X)\to \Wek(\bfx)$ by 
\bea
\mathcal Z=\adj^\pi(\mathcal Y)&\Longleftrightarrow & pr_{1*}(\mathcal Z)=\mathcal Y, \quad pr_{2*}(\mathcal Z)=\adj(\pi_{X*}(\mathcal Y)) \nonumber\\
\adj(\vef)&=&(\chela_*\circ \ada)(\econ_H)(\vef)=-\chela_*X_{\oli{\econ_H(\vef)}}\quad\quad \forall \vef\in \Wek^1(X),\nonumber
\end{eqnarray}
where the Hamiltonian vector field is defined by $\po'$ and projectable by $\chela$ since $\bar\Sekt(\fx)\subset\bar\Sekt(\jx)$ is an ideal. Since corollary \ref{horistproj} implies that $(\epro_{\dfb})_*\circ \adj=Id_{TX}$, where $\epro_{\dfb}:\dfb\to X$ is the bundle projection, the map is well defined. Furthermore, the vertical restriction $V(\bfx)\cong V(\T X)\oplus_XV(\dfb)$ induces injections 
\bea
\Sekt(\wedge^n\T X\otimes_X \fx)\ni\varphi&\longmapsto&\overrightarrow{\varphi} \in \Sekt(\wedge^n V(\bfx))\hookrightarrow \Wek^n(\bfx)\nonumber\\
\Sekt(\wedge^n V(\dfb))\ni w&\longmapsto& \overleftarrow{w}\in\Sekt(\wedge^n V(\bfx))\hookrightarrow\Wek^n(\bfx),
\nonumber
\end{eqnarray}
where in the first line, we used also the natural injections $\wedge^n\Sekt(\T X)\to\wedge^n V(\T X)$ and $\Sekt(\fx)\to \bar{\bar \Sekt}(\dfb) \stackrel{pr_2^*}{\longrightarrow}C^\infty(\bfx)$ as constant vertical multi-vector fields and as vertically linear functions, respectively. 

\begin{prop}
The Poisson structure $\po_H$ on $\bfx$ is given by
\beq
\po_H=\adj^\pi\circ w_{\T X}+\overrightarrow{\Kru}^{\econ_H}+\overleftarrow{w}_{\dfb},
\label{poHdefin}
\end{equation}
where $w_{\T X}$ is the Poisson tensor on $\T X$, $\Kru^{\econ_H}$ is the curvature of $\econ_H$, and $w_{\dfb}$ is the Poisson tensor on $\dfb$.
\end{prop}

\noindent
{\sc Proof:} Let us determine the Poisson structure $\po'_H$ on $\bfx$ on the characteristic set of functions on $\bfx$ given by the functions on $\T X$ induced by functions and vector fields on $X$, and functions induced by sections of $\fx$ on $\dfb$. The dual bundle isomorphism $\pomo_H^*$ decomposes into a direct sum as
\bea
\pomo_H^*=(\econ_H^\flat,l^*):\hspace{0.1cm}(\bfx)^*=
TX\oplus_X\fx&\stackrel{\sim}{\longrightarrow}
&\econ_H^\flat(TX)\oplus_X \fx\cong\jx.\nonumber
\end{eqnarray}
It follows that for $f_i\in C^\infty(X)=\Wek^0(X)$, $\vef_i\in \Wek^1(X)$, and $\mathcal V_i\in \Sekt(\fx)$, $i=1,2$, we have, denoting the corresponding functions on $\bfx$ by a double bar,
\bes
\pomo_H^*\bar{\bar f}_i=\epro^*f_i\qquad\quad \pomo_H^*\bar{\bar \vef}_i=\oli{\econ_H(\vef_i)}\qquad\quad\pomo_H^*\bar{\bar{\mathcal V}}_i=\bar{\mathcal V}_i.
\end{equation*}
Denoting the Poisson bracket defined by $\po'_H$ by $\{\cdot,\cdot\}'_H$, this implies that
\bea
\{\bar{\bar f}_1,\bar{\bar f}_2\}'_H
&=&(\pomo_H^{-1})^*\{\epro^*f_1,\epro^*f_2\}'=0=\{\bar{\bar f}_1,\bar{\bar f}_2\}_{\T X}\nonumber\\
\{\bar{\bar \vef}_1,\bar{\bar f}_2\}'_H
&=&(\pomo_H^{-1})^*\{\oli{\econ_H(\vef_1)},\epro^*f_2\}'
=(\pomo_H^{-1})^*\epro^*((\ank_X\circ \econ_H(\vef_1))\cdot f_2)\nonumber\\
&=&(\pomo_H^{-1})^*\epro^*(\vef_1\cdot f_2)=\oli{\oli{\vef_1\cdot f_2}}=\{\bar{\bar \vef}_1,\bar{\bar f}_2\}_{\T X}\nonumber\\
\{\bar{\bar f}_1,\bar{\bar {\mathcal V}}_2\}'_H
&=&(\pomo_H^{-1})^*\{\epro^*f_1,\bar{\mathcal V}_2\}'=-(\pomo_H^{-1})^*\epro^*((\ank_X\circ \mathcal V_2)\cdot f_1)=0\nonumber\\
\{\bar{\bar \vef}_1,\bar{\bar {\mathcal V}}_2\}'_H
&=&(\pomo_H^{-1})^*\{\oli{\econ_H(\vef_1)},\bar{\mathcal V}_2\}'= (\pomo_H^{-1})^*\oli{\{\econ_H(\vef_1),\mathcal V_2\}}=\oli{\oli{\{\econ_H(\vef_1),\mathcal V_2\}}}\nonumber\\
\{\bar{\bar \vef}_1,\bar{\bar \vef}_2\}'_H
&=&(\pomo_H^{-1})^*\{\oli{\econ_H(\vef_1)},\oli{\econ_H(\vef_2)}\}'
\nonumber\\
&=&(\pomo_H^{-1})^*(\oli{\Kru^{\econ_H}(\vef_1,\vef_2)}
+\oli{\econ_H([\vef_1,\vef_2])}) \nonumber\\
&=&\oli{\oli{\Kru^{\econ_H}(\vef_1,\vef_2)}}
+\oli{\oli{[\vef_1,\vef_2]}}=\oli{\oli{\Kru^{\econ_H}(\vef_1,\vef_2)}}
+\{\bar{\bar \vef}_1,\bar{\bar \vef}_2\}_{\T X}\nonumber\\
\{\bar{\bar{\mathcal V}}_1,\bar{\bar{\mathcal V}}_2\}'_H
&=&(\pomo_H^{-1})^*\{\bar{\mathcal V}_1,\bar{\mathcal V}_2\}' =\oli{\oli{\{\mathcal V_1,\mathcal V_2\}}}=\{\bar{\bar{\mathcal V}}_1,\bar{\bar{\mathcal V}}_2\}_{\dfb}\nonumber
\end{eqnarray}
where we also wrote $\{\cdot,\cdot\}_{\T X}$ and $\{\cdot,\cdot\}_{\dfb}$ for the Poisson bracket on $\T X$ and $\dfb$, respectively, omitting obvious restrictions. By testing on the characteristic set of functions, it follows that the Poisson tensor $\po_H$ is given by (\ref{poHdefin}). For example,
 \bea
\po_H(d\bar{\bar\vef}_1,d\bar{\bar\vef}_2)&=&
(pr_{1*}\circ \adj^\pi\circ w_{\T X})(d\bar{\bar\vef}_1|_{\T X} ,d\bar{\bar\vef}_2|_{\T X})+\overrightarrow{\Kru}^{\econ_H}(d\bar{\bar\vef}_1 ,d\bar{\bar\vef}_2)\nonumber\\
&=&\{\bar{\bar \vef}_1,\bar{\bar \vef}_2\}_{\T X}+\oli{\oli{\Kru^{\econ_H}(\vef_1,\vef_2)}}=\{\bar{\bar \vef}_1,\bar{\bar \vef}_2\}'_H,\nonumber 
\end{eqnarray}
where we used that by construction, $\overrightarrow{\varphi} (d\bar{\bar\vef})=d\bar{\bar\vef}(\overrightarrow{\varphi})=
\oli{\oli{\varphi(\vef)}}$ for all $\vef\in \Wek(X)$ and $\varphi\in \Sekt(\T X\otimes_X \fx)$. \qed

\begin{defin}
\begin{rm}
The equivalent systems $(\ZX,\po',\HSW)$ and $(\bfx, \po'_H, \HXH)$ will be called the {\it Wong system} and the {\it gauged Wong system} associated to the Hamiltonian system $(Z,\po,H)$ at $X$, respectively.
\end{rm}
\end{defin}

Wong's equations (\cite{Wong}) can now be easily written down. We refer to the   literature for their discussion. In particular, \cite{RatiuPerl} provides a detailed calculation and discussion of the Poisson structure (\ref{poHdefin}) for the original Sternberg-Weinstein phase space and the (left) gauged Wong system. Note that the original Wong system is associated to itself.

\subsection{The Einstein-Mayer system} \label{SektTayCoe}

If $t^VH_0$ in theorem \ref{theo1} in degenerate, it is still possible to define an approximated system on the Sternberg-Weinstein approximation of the underlying Poisson manifold.  

\begin{theo}\label{theo2}
Let $X\subset S\stackrel{i_S}{\hookrightarrow} Z$ be a Lagrangian submanifold of the symplectic leaf $(S,\ow_S)$ of the Poisson manifold $(Z,\po)$. Every Hamiltonian $H\in C^\infty(Z)$ such that $dH|_X=0$ uniquely and naturally defines a section
$j^VH: X\to S^2(\jx)$
of symmetric bilinear forms on the fibers of $\dxb$. If the associated bundle morphism $j^VH^\sharp:\dxb\to \jx$ is invertible, then the inverse bundle morphism
defines an associated nondegenerate field $\kal$ of scalar products on the fibers of $\jx$. These scalar products naturally determine, and are determined by,
a triple $(\emet_H,\econ_H,\scalg_H)$ of fields on $X$, where $\emet_H$ is a metric, $\econ_H$ is an $\jx$-connection form, determining a principal connection form $\Ahl_H$ on $\Pa_X$, and $\scalg_H$ is a field of scalar products on the associated vector bundle $\fx=\g(\Pa_X)$.
\end{theo}

\noindent
{\sc Proof:} The differential of $H\in C^\infty(Z)$ is a section $dH:Z\to \T Z$, whose first jet and tangent map yield the maps
\bea
\jet^1(dH)|_X: &&\quad\hspace{0.2cm} X\longrightarrow \T Z|_X\otimes_X (dH|_X)^* T(\T Z)=
\nonumber\\
&&\qquad\qquad \qquad\qquad=\T Z|_X\otimes_X (TZ|_X\oplus_X \TxU),\nonumber\\
\Tan(dH)|_{TZ|_X}:&& TZ|_X\longrightarrow (dH|_X)^* T(\T Z)=TZ|_X\oplus_X \TxU\nonumber
\end{eqnarray}
respectively, where lemma \ref{zerllem} provides the decompositions since $dH|_X=0$.
By definition, the first components are given by the identity. The second yield maps
\bes
pr_2\circ \jet^1(dH)|_X:X\to \otimes^2(\T Z|_X) \qquad pr_2\circ \Tan(dH)|_{TZ|_X}:TZ|_X\to\TxU,
\end{equation*}
where the second map is the bundle morphism corresponding to the tensor field given by the first map.

Since $dH|_X=0$, $pr_2\circ \Tan(dH)|_{TZ|_X}$ vanishes on $TX\subset TZ|_X$. On the other hand, since $d(dH)=0$, $pr_2\circ \jet^1(dH)|_X$ defines symmetric bilinear forms on the fibers of $TZ|_X$ whose kernels contain the vectors of $TX$. Thus, it induces symmetric bilinear forms on the fibers of the quotient bundle $TZ|_X/TX=\dxb$ and an associated bundle morphism $\dxb\to \jx$. Thus, $pr_2\circ \Tan(dH)|_{TZ|_X}$ takes values in $\jx$, and we obtain an induced field of forms and a corresponding bundle morphism
\beq
j^VH: X \to S^2(\jx)\qquad\qquad j^VH^\sharp: \dxb\longrightarrow \TXo
\label{Buend3}
\end{equation}
as claimed. If the morphism is invertible, we obtain the bundle morphism
\bes
\kal^\flat=j^VH^{\sharp-1}:\TXo\longrightarrow \dxb
\end{equation*}
corresponding to a nondegenerate field $\kal$ of scalar products on the fibers of $\TXo$.

Restriction of the bundle morphism in (\ref{Buend3}) to the subspaces $TS|_X/TX\stackrel{Ti_S}{\hookrightarrow} \dxb$ yields bundle morphisms
\bea
j^VH^\sharp\circ Ti_S&=&t^VH: \quad\qquad TS|_X/TX\quad\qquad \longrightarrow \TXo\nonumber\\
j^VH^\sharp_0=(Ti_S)^*\circ j^VH^\sharp\circ Ti_S&=&t^VH_0: TS|_X/TX\cong(\XO)^*\longrightarrow \XO\nonumber
\end{eqnarray}
where $t^VH$ and $t^VH_0$ are defined as in the proof of theorem \ref{theo1}.  Since $j^VH^\sharp$ is invertible and corresponds to a symmetric form, $j^VH^\sharp_0$ must be invertible, too. In fact, it corresponds to a field
$j^VH_0:X\to S^2\XO$
of nondegenerate symmetric bilinear forms on the fibers of $TS|_X/TX$. By theorem \ref{theo1}, this also means that $H$ uniquely defines a metric $\emet_H$ on $X$ and an $\jx$-connection form $\econ_H$, given by
\bea
\emet_H^{\flat} &=&(-\ow^\flat_X)^*\circ j^VH^{\sharp-1}_0\circ (-\ow^\flat_X):\hspace{0.6cm}TX
\longrightarrow \T X\nonumber\\
\econ_H^{\flat} &=& j^VH^\flat\circ Ti_S\circ j^VH^{\sharp-1}_0\circ(-\ow^\flat_X): \hspace{0.1cm}
TX\longrightarrow \TXo.\nonumber
\end{eqnarray}
In addition, $\econ_H$ yields a principal connection form $\Ahl_H$ in $\Pa_X$.

On the other hand, $\kal^\flat$ induces an invertible bundle morphism
\bes
\scalg_H^{\flat}:\fx \longrightarrow \dfb
\end{equation*}
by restriction and projection. But this is just the bundle morphism associated to a field of scalar products $\scalg_H$ in the fibers of $\fx\cong \g(\Pa_X)$. That is, $\scalg_H$ is a section of the associated bundle $(\Pa_X\times S^2(\gs))/Aut(\g)$ for the canonically induced action on scalar products. The symmetry of $\scalg_H$ follows from the symmetry of $\kal$.

Conversely, a symmetric bilinear form is determined by its restrictions to complementary subspaces. In particular, since $\kal$ is nondegenerate, it is determined on the fibers of $\TXo$ by its restriction to the fibers of the subbundle $\fx$ and their orthogonal complements with respect to $\kal$. On $\fx$, $\kal$ coincides with by definition with $\scalg_H$. For the subbundle consisting of the complements, the map
$(Ti_S)^*:(\fx)^{\bot_{\kal}}\mapsto \XO$
is obviously a bundle isomorphism. Thus, on the complementary subspace, $\kal$ is determined by $j^VH^\sharp_0$ and thus, by $\emet_H$. Finally, 
\bes
(\fx)^{\bot_{\kal}}=\kal^{\flat-1}((\fx)^0)=j^VH^\sharp(TS|_X/TX),
\end{equation*}
and thus, the splitting is precisely determined by $j^VH^\sharp\circ Ti_{S}$ and thus, by the $\jx$-connection form $\econ_H$. Thus, the triples $(\emet_H, \econ_H,\scalg_H)$ and the field $\kal$ determine each other.
\qed
\vspace{0.4cm}

The definitions of $j^VH^\sharp$, $j^VH^\sharp_0$, $\kal^\flat$, $\scalg_H^\flat$ and $\emet_H^\flat$ can be summarized in the following commutative diagram, built up of natural inclusions and projections:
\beq
\begin{CD}
\fx @>>> \TXo @>(Ti_S)^*>> \XO @<-\ow_X^\flat<< TX \\
@VV\scalg_H^\flat V
@A \kal^\flat{\Big \downarrow} A j^VH^\sharp A
@A j^VH^\sharp_0 A{\Big \downarrow}j^VH^{\sharp-1}_0 A
@VV\emet_H^\flat V\\
\dfb @<<< \dxb @<Ti_S<< T S_{|X}/TX @>(-\ow_X^\flat)^*>> \T X
\end{CD}\label{diag20}
\end{equation}
\bes
\text{and}\hspace{1.3cm}\econ_H^\flat=j^VH^\sharp\circ Ti_S\circ j^VH^{\sharp-1}_0\circ (-\ow_X^\flat).
\end{equation*}

\begin{cor}
Let $X\subset S\stackrel{i_S}{\hookrightarrow}Z $ be a Lagrangian submanifold of the symplectic leaf $S$ of the Poisson manifold $Z$. Any 1-form $h\in\Frm^1(Z)$ with $h|_X=0$ and such that $dh|_X=0$ uniquely defines a field
$j^Vh: X\rightarrow S^2(\jx)$ of symmetric bilinear forms on the fibers of $\dxb$. If the associated bundle morphism $j^Vh^\sharp:\dxb\to \jx$ is invertible, then the inverse bundle morphism
defines an associated nondegenerate field $\kal$ of scalar products on the fibers of $\jx$ and a triple $(\emet_h,\econ_h,\scalg_h)$ of fields on $X$, where $\emet_h$ is a metric, $\econ_h$ is an $\jx$-connection form, determining a principal connection in $\Pa_X$, and $\scalg_h$ is a field of scalar products on the associated vector bundle $\fx=\g(\Pa_X)$.
\end{cor}

The section $j^VH:X\to S^2(\jx)$ defines a Hamiltonian function $\HEM$ on the Sternberg-Weinstein approximation $(Z'=\dxb,\po')$ to $(Z,\po)$ at $X$. It follows from the definitions that 
\beq 
\HEM=\oli{j^VH}=\HSW+\chela^*\HL,
\label{SWHa}
\end{equation}
where $\HSW$ is the Hamiltonian of the Wong system defined in (\ref{WongHam}), $\chela$ is the natural projection (\ref{cheladefin}), and $\HL$ is the vertically quadratic function on $\dfb$ defined by the bundle morphism $\scalg_H^{\flat-1}$. Thus, the Hamiltonian system of $\HEM$ is obtained from the Wong system by adding the pull-back of a term quadratic on the fibers of $\dfb$.

\begin{defin}
\begin{rm}
The equivalent systems $(Z',\po',\HEM)$ and $(\bfx,\po'_H,\HXH+pr_2^*\HL)$, where $pr_2:\bfx\to \dfb$ is the canonical second factor projection, will be called the
{\it Einstein-Mayer system} and the {\it gauged Einstein-Mayer system} associated to the Hamiltonian system $(Z,\po,H)$ at $X$, respectively. 
\end{rm}
\end{defin}

\begin{remark}\label{1932}
\begin{rm}
In 1932, Einstein and Mayer considered a unified theory of gravitation and electricity which was based on an alternative tangent bundle to the four-dimensional space-time manifold  (\cite{EinsteinMayer1, EinsteinMayer2}). Besides motivating our denomination, this work could be regarded as a precursor of Lie algebroid ideas in physics before their apparition in mathematics.
The Wong system can be regarded a special case of the  Einstein-Mayer system (cf remark \ref{ambiguity}), or as an approximation to it since it exploits only part of the first jet of $dH$. Note that it is the invertibility of $j^VH^\sharp$ that yields the splitting (\ref{SWHa}).
Furthermore, the Einstein-Mayer system is naturally related to Kaluza-Klein theory by means of a special symplectic realization, which we will construct in section \ref{Kaluzasection}. This also motivated Einstein's and Mayer's work. \bems
\end{rm}
\end{remark}

\section{Reduced approximation and scalar fields}

Theorem \ref{theo2} states that if the values of the field $j^VH$ defined by a Hamiltonian $H$ are nondegenerate, there is a well-defined field $\scalg_H$ of scalar products on the fibers on $\fx$, the {\it scalar fields}. In order to identify such scalar fields with Higgs fields, we have to modify our formalism so that these fields contain an irreducible representation of the right structure group. This can be achieved by a constraint.

Let $S\subset \co\stackrel{i_\co}{\hookrightarrow} Z$,
where $i_\co$ is the embedding of a {\it locally closed coisotropic} constraint submanifold $\co$. By proposition \ref{lokgeschlossen}, the sub-characteristic distribution ${\sf N}(\co)=\po^\sharp((T\co)^0)\cap T\co=\po^\sharp((T\co)^0)$ has constant dimension, and by proposition \ref{soprjectierts}, it is integrable to a sub-characteristic foliation $\Iq=\Iq(\co)$. If this is transversal to the symplectic leaves of $Z$, and if the quotient space possesses a manifold structure such that the projection is a submersion, then proposition \ref{reductionexists} assures that there is a Poisson bracket induced  by the projection
\beq
\qq:\co\longrightarrow \Zhig=\co/\Iq
\label{quotbysubchar}
\end{equation}
on $C^\infty(\Zhig)\cong C^\infty(\co)^{\Iq}=\{f\in C^\infty(\co)|f=\text{const on the leaves of}\hspace{0.1cm} \Iq\}$. If $\Zhig$ fails to be a manifold, as in many physical situations, the induced Poisson bracket on the latter subspace is still defined and yields the {\it reduced Poisson algebra} of $\PA=C^\infty(Z)$
\beq
\KA=C^\infty(\co)^{\Iq}=i_\co^*\No\cong\No/\Co, \qquad\qquad \No=\Noa_{\PA}(\Co),
\label{constraintidenti}
\end{equation}
as in proposition \ref{admissible} with $\Co=\Co_\co=\{f\in \PA|f_{|Q}=0\}$ as constraint algebra.

\subsection{Reduced Sternberg-Weinstein approximation }

The map $(Ti_{\co})^*:\T Z|_\co\to \T \co$ induces the exact sequence of bundle morphisms
\bea
%&&T\co|_V \longrightarrow TU|_T\nonumber\\
&0\longrightarrow (T\co)^0_{|X} \longrightarrow \T Z|_X\stackrel{(Ti_{\co})^*}{\longrightarrow} \T \co|_X\longrightarrow 0,&\label{qexseq}
\end{eqnarray}
and $(Ti_{\co})^*$ can be seen as a quotient map since
$\T \co|_X\cong \TxU/(T\co)^0_{|X}$.
Let
\bea
\cx=(T\co)^0_{|X}\quad& \jq=(TX)^{0}\subset \T \co_{|X}\quad\hspace{0.2cm}&  \quad\fq=(TS)_{|X}^0\subset \T \co_{|X}\nonumber\\
&\djq=T\co_{|X}/TX=(\jq)^*&  \quad\dfq=T\co_{|X}/TS_{|X}=(\fq)^*\nonumber
\end{eqnarray}
We call $\cx$ the {\it annihilator bundle}, the remaining bundles the {\it (dual) Lie algebroid quotient} and {\it (dual) Lie algebra quotient bundle at $X$}, respectively. From (\ref{qexseq}), we obtain the exact sequences and canonical bundle isomorphisms 
\bea
%&&\ti{(TV)}^0 \longrightarrow \TVO\nonumber\\
&0\longrightarrow \cx \longrightarrow \jx \longrightarrow \jq\longrightarrow 0&\quad\qquad\jq\cong \jx/\cx  \label{projQ2}\\
&0\longrightarrow \cx \longrightarrow \fx \longrightarrow \fq\longrightarrow 0&\quad\qquad \fq\cong \fx/\cx.\nonumber
\end{eqnarray}

\begin{lem}\label{linspannen2}
The submanifold $\djq\subset \dxb$ is given as the common zero level set of the functions in $\baSE(\cx)\subset C^\infty(\dxb)$ defined by differentials $d\Co|_X\in \Sekt(\cx)$ as in corollary \ref{HamVFlin}. Consequently, the fibers of $(T\djq)^0\subset \T \dxb$ are spanned at each point of $\djq$ by the differentials of these functions. In particular, the submanifold $\djq\subset \dxb$ is coisotropic. 
\end{lem}

\noindent
{\sc Proof:} The first two statements are obvious since the subspaces $(\cx)_x\subset (\fx)_x$ are spanned by the differentials $d\Co_x$ at each point $x\in X$. Corollary \ref{linspannenauf} further implies that the Sternberg-Weinstein bracket of two functions in $\baSE(\cx)$ is given by the pointwise Lie algebra bracket in the fibers of $\cx$.  Since $\co$ is coisotropic and thus, $\Co$ forms a subalgebra of $\PA$, we have $\{df_x,dg_x\}_x=(d\{f,g\})_x\in (\cx)_x$ for all $f,g\in \Co$ and $x\in X$. This finally shows that $\djq$ is coisotropic.  \qed 

\begin{defin}
\begin{rm}
The {\it Sternberg-Weinstein constraint algebra} and {\it admissible function algebra} are the subalgebras of $\PA'=C^\infty(\dxb)$
\bes
\Co'=\{f\in \PA'|f|_{\djq}=0\} \qquad \text{and}\qquad
\No'=\Noa_{\PA'}(\Co').
\end{equation*}
The pair $(\co',\KA')$, where $\co'\subset \ZX$ and the reduced Poisson algebra $\KA'$ are given by
\beq
i_{\co'}:\co'=\djq\to \dxb=\ZX \quad \quad \quad \KA'=i_{\co'}^*\No'\cong\No'/\Co',\label{redSWapp}
\end{equation}
is called the {\it reduced Sternberg-Weinstein approximation} of $Z$ at $X$ by $\co$.
\end{rm}
\end{defin}

Let us denote by $\ti\chela:\djq\to\dfq$ the dual to the inclusion $\fq\to \jq$. Obviously, 
\beq
\ti\chela^*C^\infty(\dfq)
=i_{\co'}^*(\chela^*C^\infty(\dfb)). 
\label{chelarel}
\end{equation}
The fact that the Sternberg-Weinstein bracket of functions on $\dxb$ which are pull-backs of functions in $\dfb$ only depends on their restrictions to the fibers now allows us to characterize, by means of two additional assumptions, the elements of $\ti\chela^*C^\infty(\dfq)\cap\KA'$ which are vertically polynomial. 

First, let us recall from the proof of lemma \ref{linspannen2} that $(\cx)_x$ is a Lie subalgebra of $(\fx)_x$ for all $x\in X$, and that the fibers  of $\fx$ are isomorphic to a fixed Lie algebra $\g$. We will make the following assumption:
\beq
\text{\it The fibers of $\cx$ are all isomorphic to some subalgebra $\ha\subset \g$}.
\label{assumpt}
\end{equation}
We will further make the stronger assumption that for every 
fiber there exists an isomorphism which is the restriction of an isomorphism $a\in (\Pa_X)_x$. If $\Na$ denotes the stabilizer subgroup of $\ha$ in $Aut(\g)$, this means that:  
\beq
\text{\it The Lie-frame bundle $\Pa_X$ can be reduced to $\Nau=\Na$.}
\label{assumpt2}
\end{equation}

\begin{defin}  
\begin{rm}
If (\ref{assumpt2}) is valid, the reduced subbundle of $\Pa_X$ given by 
\bes
\Pred\to X\qquad\quad (\Pred)_x=\{a\in \Pa_x| a:\ha \to (TQ)^0_x\} \qquad \forall x\in X,
\end{equation*}
with structure group $\Nau$, is called the {\it constraint Lie frame bundle}.
%Let $\Ze=\Zen_{\Nau}(\g/\ha)$ denote the (normal) common stabilizer subgroup of all elements of $\g/\ha$ under the induced action of $\Nau$. \
Let $S^i(\fq)$ and $S^i(\dfq)$ denote the $i$-th symmetric powers. We identify them, by setting 
\bes
S^i(\fq)=(\Pred\times S^i(\g/\ha))/\Nau \qquad  \quad S^i(\dfq)=(\Pred\times S^i(\ha^0))/\Nau
\end{equation*}
where $\ha^0\subset \gs$ is the annihilator subspace, with associated vector bundles to $\Pred$ for the naturally induced action of $\Nau$ on $\g/\ha$ and $\hs$, respectively. 
\end{rm}
\end{defin}

\begin{defin}\label{kerndef}
\begin{rm} 
Let $\Lc$ be the analytic subgroup defined by $\ha\subset \g$ in some Lie group with Lie algebra $\g$, and let $S^i(\g/\ha)^\Lc$ 
and $S^i(\ha^0)^\Lc$ 
denote the subspaces of invariant elements under the (well-defined) induced $Ad_*(\Lc)$- and $Ad^*(\Lc)$-action. If $\admo_i: \Nau\to Gl(S^i(\g/\ha)^\Lc)$ are the naturally induced group homomorphisms, we define the {\it i'th quotient principal bundle} and {\it quotient structure group} by
\bes  
\ti\Pa_i=\Pred/\Stab_i\quad\text{and}\quad \Nq_i=\Nau/\Stab_i, \qquad\text{where}\quad\Stab_i=\ker \admo_i.
\end{equation*}
We can define subbundles of $S^i(\fq)$ and $S^i(\dfq)$ as the mutually dual bundles
\beq
\Fred_i=(\ti\Pa_i \times S^i(\g/\ha)^\Lc)/{\Nq_i}  
\quad\qquad \Fred_i^*=(\ti \Pa_i \times S^i(\hs)^\Lc)/{\Nau}.
\label{deffred2}
\end{equation}
For $i=1$, the quotient principal bundle, structure group, and vector subbundle
\bes
\Pq=\ti\Pa_1=\Pred/\Stab_1\qquad  \Nq=\Nq_1=\Nau/\Stab_1\qquad \Fred_1=(\Pq \times (\g/\ha)^\Lc)/{\Nq} 
\end{equation*}
are called {\it reduced Lie frame bundle}, {\it reduced structure group}, and {\it reduced Lie algebra bundle}, respectively. 
\end{rm}
\end{defin}

\begin{lem}\label{Adcinv}
The vertically polynomial functions in $\ti\chela^*C^\infty(\dfq)\cap \KA'$ are those defined on $\djq$ by the sections of $\oplus_{i\geq 0}\Fred_i$. In particular, 
\beq
\baSE(\fq)\cap\KA'=\baSE(\Fred_1),\quad \text{and}\quad \Fred_1=(\Pq\times \ka)/\Nq \quad\text{with} \quad \ka=\fra{N}_{\g}(\ha)/\ha,
\label{freddef}
\end{equation}
where $\fra{N}_{\g}(\ha)$ denotes the idealizer of $\ha$ in $\g$, and $\baSE(\fq)\subset \ti\chela^* C^\infty(\dfq)$ as before.
\end{lem}

\noindent
{\sc Proof:} Let $\bar s\in \ti \chela^*C^\infty(\dfq)$ be a vertically polynomial function defined by $s\in \Sekt(S^\bullet(\fq))$, and, using (\ref{chelarel}), let $\bar s=i_{\co'}^*\hat {\bar s}$, where $\hat {\bar s}\in \chela^*C^\infty(\dfb)$ is the vertically polynomial function defined by $\hat s\in \Sekt(S^\bullet(\fx))$. As in corollary \ref{linspannenauf}, let $f\in \Co$ induce the section $df|_X$ of $\fx$ and the corresponding function $\bar{df}|_X$.  As usual, we can interpret  $\hat s$, $df|_X$ and $s$ as equivariant maps
\bea
&&\qquad\underline{\hat s}:\quad\Pred \to S^\bullet(\g)\subset C^\infty(\gs)\qquad\quad \underline{df}|_X:\Pred \to \ha\subset C^\infty(\gs)\nonumber\\
&&\underline{s}=i_{\hs}^*\circ \underline{\hat s}:\Pred \to S^\bullet(\g/\ha)\subset C^\infty(\hs), \label{sfunc}
\end{eqnarray}
where $i_{\hs}:\hs\to \gs$ denotes the inclusion. In deed, the restriction to $\hs$ of a polynomial on $\gs$ defined by an element of $S^\bullet(\g)$ is precisely given by the polynomial defined by the natural image of that element in $S^\bullet(\g/\ha)$. 

Let now $m\in (\dxb)_x$ and $n=\chela(m)=[p,d]\in \gs(\Pred)$, $p\in (\Pred)_x, d\in\gs$. Then,
\bea
\{\hat {\bar s},\bar{df}|_X\}'(m)=X_{\bar{df}|_X}(m)\cdot \hat {\bar s}
&=&\{\hat {\bar s}|_{(\fx)_x},\bar{df}|_X|_{(\fx)_x}\}_x(\chela(m))\nonumber\\
&=&\{\underline{\hat s}(p), \underline{df}|_X(p)\}_{\gs}(d)\nonumber\\
&=&X_{\underline{df}|_X(p)}\cdot(\underline{\hat s}(p))(d)\nonumber\\
&=&d/dt|_{t=0} (Ad^*(-t\hspace{0.7mm} \underline{df}|_X(p)))^*\underline{\hat s}(p)(d)\nonumber
%\\
%&=&d/dt|_{t=0} Ad_*(-t\hspace{0.7mm} df|_X(p))(\hat s(p))(d),\nonumber
\end{eqnarray}
where $\{\cdot,\cdot\}_x$ and $\{\cdot,\cdot\}_{\gs}$ denote the Poisson brackets on $(\fx)_x$ and $\gs$, respectively, $Ad^*$ the coadjoint action, and we used proposition \ref{sympleavescoadact}. Of course, this is independent of the choice of $p$ and $d$ because of the equivariance. 

Now, since the differentials $df_x$ span $(\cx)_x\cong \ha$ at each point $x\in X$, and since by lemma \ref{linspannen2}, the differentials of the functions $\bar{df}|_X$ span $(T\co')^0$ at each point of $\co'$, this implies that
\bea
\bar s\in \KA'&\Longleftrightarrow&\hat {\bar s}\in \No'\hspace{0.1cm}\Longleftrightarrow\hspace{0.1cm}\{\hat {\bar s},\Co'\}'\subset \Co'\nonumber\\
&\Longleftrightarrow&\{\hat {\bar s},\bar{df}|_X\}'(m)=0\quad \hspace{3cm}\forall f\in \Co,m\in \co'=\djq\nonumber\\
&\Longleftrightarrow& d/dt|_{t=0} (Ad^*(-t\hspace{0.7mm} \underline{df}|_X(p)))^*\underline{\hat s}(p)(d)=0\quad\qquad \forall  p\in \Pred, d\in\hs \nonumber\\
&\Longleftrightarrow& ((Ad^*(-t\hspace{0.7mm} \underline{df}|_X(p)))^*\underline{\hat s}(p)-\underline{\hat s}(p))(d)=0\quad \qquad\forall t\in \R\nonumber\\
&\Longleftrightarrow& ((Ad^*(\Lc))^*\underline{\hat s}(p)-\underline{\hat s}(p))(d)=0\quad \hspace{1.7cm}\forall p\in \Pred, d\in\hs\nonumber\\
&\Longleftrightarrow& (Ad^*(\Lc)|_{\hs})^*(\underline{\hat s}(p)|_{\hs})=Ad_*(\Lc)(\underline{\hat s}(p)|_{\hs})=\underline{\hat s}(p)|_{\hs}\quad \hspace{0.2cm}\forall p\in \Pred\nonumber\\
&\Longleftrightarrow& Ad_*(\Lc)\circ \underline{s}=\underline{s} \hspace{0.1cm}\Longleftrightarrow \hspace{0.1cm}s\in \baSE(\oplus_{i\geq 0} \Fred_i)\nonumber
\end{eqnarray}
where $Ad_*(\Lc)$ denotes the induced action on $S^\bullet(\g/\ha)\cong S^\bullet(\g)|_{\hs}\subset C^\infty(\hs)$. This proves the first assertion. For $i=1$, we have 
\beq
[n]\in \ka \Longleftrightarrow ad(\ha)[n]=0\Longleftrightarrow Ad_*(\Lc)([n])=[n]\Longleftrightarrow [n]\in (\g/\ha)^\Lc, \label{inva=norm}
\end{equation}
which completes the proof of the lemma. We see also that $\Fred_1$ is the maximal subbundle of $\fq$ whose fibers posses an induced Lie algebra structure. This justifies our denomination for $\Fred_1$. \qed

\subsection{The reduced Einstein-Mayer and Wong systems}

Let $H\in \PA$ such that $dH|_X=0$ be an admissible Hamiltonian.  We want to show that there are reduced fields on $X$ determined by $\Hq=i_{\co}^*H$, which define an admissible Einstein-Mayer Hamiltonian on $\co'$ if $\Hq\in \KA$.

\begin{theo}\label{redHinv}
Let $X\subset S\stackrel{i_S}{\hookrightarrow} Z$ be a Lagrangian submanifold of the symplectic leaf $(S,\ow_S)$ of the Poisson manifold $(Z, \po)$, and $\co$ a constraint manifold as above. Let $\Hq\in C^\infty(\co)$ be such that $d\Hq|_X=0$. Then, there is a natural field 
$j^V\Hq: X\to S^2(\jq)$
of symmetric bilinear forms on the fibers of $\djq$ defined by $\Hq$. If $\Hq=i_\co^*H$ for $H\in \PA$,
then $\Hq'_2=i_{\co'}^* \HEM$, where $i_{\co'}:\co'\to \ZX$ is the inclusion in (\ref{redSWapp}), $\HEM$ is given by (\ref{SWHa}), and $\Hq'_2=\oli{j^V\Hq}\in C^\infty(\co')$. Furthermore,
$\Hq\in \KA$ implies $\Hq'_2\in \KA'$.
\end{theo}

\noindent
{\sc Proof:} If $d\Hq|_X=0$, we can obtain the field $j^V\Hq$ from the first jet of $d\Hq:\co\to \T\co$ in the same way as $j^VH$ was obtained from $H$. Without restricting the generality, we can always assume that $\Hq=i_\co^* H$ for some $H\in \PA$ with $dH|_X=0$. Then, the bundle morphism $j^VH^\sharp$ yields the bundle morphism $j^V\Hq^\sharp$ by restriction of the corresponding symmetric bilinear form according to
\beq
{j^V}\Hq^\sharp=(Ti_{\co})^*|_X\circ {j^V}H^\sharp \circ Ti_{\co}|_X:\hspace{0.1cm}
\djq\longrightarrow \jq.
\label{diag22}
\end{equation}
This is equivalent to $\Hq'_2=i_{\co'}^* \HEM$. We need only to show that $\Hq\in \KA$ implies $\Hq'_2\in\KA'$ or, equivalently, $H\in \No$ implies $\HEM \in \No'$.

Since $\co$ is coisotropic, the Hamiltonian vector field $X_f$ of a function $f\in \Co$ is tangent to $\co$. Thus, its flow defines a local family of diffeomorphisms
$\exp(tX_f):\co\to \co, t\in \R$. By abuse of notation, we omit the restriction of $X_f$ to $\co$.  There are  the canonically induced cotangent bundle morphism $\T\exp(tX_f)$ and first order jet prolongation $\jet^1(\T\exp(tX_f)):\jet^1(\T \co)\to\jet^1(\T \co)$ (cf e.g. \cite{GMS}, p. 45f) over $\T\exp(tX_f)$ and $\exp(tX_f)$.
%\bes
%\begin{CD}
%\jet^1(\T \co)@>\jet^1(\T\exp(tX_f))>>\jet^1(\T \co)
%@VVV @VVV\\
%\T \co  @>\T\exp(tX_f)>> \T \co\\
%@VVV @VVV\\
%\co @>\exp(tX_f)>> \co.
%\end{CD}
%\end{equation*}
Now, since $df|_S\in (T\co)^0|_S\subset (TS)^0$, $X_f$ vanishes on $S$. Thus $\exp(tX_f)$ induces the identity map on $S$ and, in particular, on $X$. Thus, we can restrict all of our bundle morphisms to $X$. Furthermore, since $\T\exp(tX_f)$ is linear in the fibers of $\T \co$, it maps the image of the zero section identically onto itself. Thus, we obtain a bundle morphism
\beq
\begin{CD}
\jet^1(\T \co)|_{o_X}@>\jet^1(\T\exp(tX_f))|_{o_X}>>\jet^1(\T \co)|_{o_X}
\end{CD}
\label{diag31}
\end{equation}
over $Id_{o_X}$ and $Id_X$, respectively, where $o_X\cong X\subset \T \co|_X$ denotes the image of the restriction of the zero section of  $\T\co$ to $X$.

On the other hand, we have the commutative diagram
\beq
\begin{CD}
\T \co|_X\otimes_{o_X}T(\T \co)|_{o_X}@>\sim>> \T \co|_X\otimes_X (T\co|_X\oplus_X \T \co|_X)\\
@AAA @AAA\\
\jet^1(\T \co)|_{o_X}@>\sim>> \T \co|_X\otimes_X \T \co|_X
\end{CD}\label{diag30}
\end{equation}
where the upper identification is provided by the natural splitting of $T(\T \co)$ over the zero section of $\T \co$ as in lemma \ref{zerllem}, the left vertical map is the natural inclusion, the right vertical map is obtained by requiring that the first component be the identity map of $\T \co|_X$ in $\T \co|_X\otimes_X\T\co$,  and the lower identification is then defined by commutativity.  It is easy to see that under this last identification, (\ref{diag31}) corresponds to the bundle morphism
\beq
\otimes^2\T\exp(tX_f)|_X: \quad\T \co|_X\otimes_X \T \co|_X \to \T \co|_X\otimes_X \T \co|_X
\label{2tensor}
\end{equation}
over $Id_X$, which is precisely the induced action of $\exp(tX_f)$ on covariant 2-tensors on $\co$, restricted to $X$. Note that since $X_f|_\co$ is tangent to $\co$, and $X$ is preserved, there are also induced bundle morphism on $\jq$ and $\djq$ and there respective tensor powers.

Now, by the defining property of the first order jet prolongation, we have for the section $d\Hq:\co\to \T \co$
\bea
\jet^1(\T\exp(tX_f))\circ \jet^1(d\Hq) \circ \exp^{-1}(tX_f)&=&\jet^1(\T\exp(tX_f)\circ d\Hq \circ \exp^{-1}(tX_f))\nonumber\\
&=&\jet^1(\exp(-tX_f)^*d\Hq)\nonumber\\
&=&\jet^1(d(\exp(-tX_f)^*\Hq))\label{jetH}.
\end{eqnarray}
Under the identifications of diagram (\ref{diag30}) ($d\Hq|_X=dH|_X=0$), (\ref{2tensor}) and the induced bundle morphisms on $\jq$, this yields (since $X_f|_X=0$)
\beq
\otimes^2\T\exp(tX_f)|_X\circ j^V\Hq =j^V(\exp(-tX_f)^*\Hq) \hspace{0.3cm}\qquad \forall f\in \Co,\label{covabed}
\end{equation}
or equivalently,
\bea
\Lie_{X_f}(j^V\Hq)&=&j^V(-X_f\cdot \Hq) \label{invariance}\\
i_{\co'}^*(\oli{\Lie_{X_f}(j^VH)})&=&i_{\co'}^*(\oli{j^V(-X_f\cdot H)})\nonumber\\
i_{\co'}^*\{H'_2,\bar{df}|_X\}'&=&\oli{j^V(i_\co^*\{f,H\})}\label{infcovbed}
\hspace{0.3cm}\qquad \qquad\forall f\in \Co.
\end{eqnarray}
Here, we used (\ref{diag22}), as well as corollary \ref{HamVFlin}. Notice that the Lie derivative is well-defined with the help of an arbitrary extension around $X$.

Finally, we know from lemma \ref{linspannen2} that the differentials of the functions of the form $df|_X$ span $(T\co')^0\subset\T Z'|_{\co'}$ at each point of $\co'$. Hence, it follows from (\ref{infcovbed}) that $H\in \No$ implies $\HEM \in \No'$ and thus, $\Hq'_2\in\KA'$.  \qed
\vspace{0.4cm}

In the same way as we obtained a metric and a $\jx$-connection form on $X$ from $j^VH$, we can obtain corresponding reduced fields from $j^V\Hq$. Indeed, composing the maps of diagram \ref{diag20} with $Ti_{\co}|_X$ and $(Ti_{\co})^*|_X$ (or the induced maps), taking into account (\ref{projQ2}), we obtain the diagram
\beq
\begin{CD}
\fq @>>> \jq @>>> \XO @<-\ow_X^\flat<<  TX \\\
@VV\scalg_{\Hq}^{\flat} V @A\kaq^{\flat} {\Big \downarrow}Aj^V\Hq^\sharp A @A j^V\Hq^\sharp_0 A{\Big \downarrow}j^V\Hq^{\sharp-1}_0 A @VV\emet_{\Hq}^\flat V\\
\dfq @<<< \djq @<Ti_S<< T S_{|X}/TX @>(-\ow_X^\flat)^*>> \T X
\end{CD}
\label{diag21}
\end{equation}
\beq
\text{with}\quad \econ_{\Hq}^{\flat}\hspace{0.1cm}=\hspace{0.1cm}j^V\Hq^\sharp\circ Ti_S \circ j^V\Hq^{\sharp-1}_0\circ (-\ow_X^\flat)\hspace{0.1cm}=\hspace{0.1cm}(Ti_{\co})^*\circ \econ_H^\flat, \label{diag21anex}
\end{equation}
where the reduced fields are denoted by tildes as $\kaq$, $\econ_{\Hq}$ etc (with  $j^V\Hq_0=j^VH^\sharp_0$ and $\emet_{\Hq}=\emet_H$).
The form $\econ_{\Hq}$ on $X$ will be called the {\it $\jq$-connection form} defined by $\Hq$. More precisely, we have the following result.

\begin{theo}\label{theo17}
If the morphism $j^V\Hq^\sharp$ defined by $\Hq=i_\co^*H$ as in (\ref{diag22}) is invertible, it
uniquely and naturally defines a bundle morphism
$\kaq^{\flat}$, as well as bundle morphisms
$\scalg_{\Hq}^{\flat}$, $\emet_{\Hq}^\flat$ and $\econ_{\Hq}^{\flat}$ as in
(\ref{diag21}) and (\ref{diag21anex}) above, which correspond to bundle metrics $\kaq$ on $\jq$ and $\scalg_{\Hq}$ on $\fq$, a metric  $\emet_{\Hq}$ and an $\jq$-connection form $\econ_{\Hq}$ on $X$, respectively. $\kaq$ and the triple $(\emet_{\Hq}$,$\econ_{\Hq}$,$\scalg_{\Hq})$ determine each other.

Furthermore, suppose that $\Hq\in \KA$, and that the assumptions (\ref{assumpt},\ref{assumpt2}) are valid. Then, for every $\vef\in \Wek(X)$, $\oli{i_{\vef}\econ_{\Hq}}\in \baSE(\jq)\cap \KA'$, and $\econ_{\Hq}$ defines a principal connection form $\Ahl_{\Hq}$ on $\Pq$ taking values in $ad(\ka)$.
%=\fra{N}_{ad(\g)}(\ha)/ad(\ha)
Finally, the field  $\scalg_{\Hq}$ corresponds to a section of
$\Fred_2^*$ defined in (\ref{deffred2}).
\end{theo}

\noindent
{\sc Proof:} The first assertion of the theorem is a consequence of theorem \ref{theo2} and the definitions of diagram (\ref{diag21}) above. We only need to prove the second assertion.

Let $\Hq\in \KA$, that is, $\Hq=i_\co^*H$ with  $H\in \No$. (\ref{covabed}) implies that for all $f\in \Co$
\bea
&&\T\exp(tX_f)|_X\circ j^V\Hq^\sharp \circ T\exp(-tX_f)|_X\circ Ti_S\circ j^VH^{\sharp-1}_0\circ (-\ow_X^\flat)=\nonumber\\
&&\hspace{6cm}=
j^V\Hq^\sharp \circ Ti_S\circ j^VH^{\sharp-1}_0\circ (-\ow_X^\flat)\nonumber\\
&&\hspace{2.87cm}\T\exp(tX_f)|_X\circ\econ_{\Hq}^\flat = \econ_{\Hq}^\flat\nonumber
\end{eqnarray}
since $\exp(-tX_f)$ induces the identity on $S$. Using corollary \ref{HamVFlin}, this implies
\bea
&&\oli{\Lie_{X_f}(\inn_{\vef}\econ_{\Hq})}=i_{\co'}^*(\oli{\Lie_{X_f}(\inn_\vef\econ_H)})
=i_{\co'}^*\{\oli{\inn_\vef\econ_H},\bar{df}|_X\}'=0 \nonumber\\
%&=&=i_{\co'}^*[\inn_\vef\econ_H,df|_X]_{\jx}=0
&&\Longleftrightarrow \hspace{0.1cm}\oli{\inn_\vef\econ_H}\in \baSE(\jx)\cap \No' \nonumber\\
&&\Longleftrightarrow \hspace{0.1cm}\oli{\inn_\vef\econ_{\Hq}}=i_{\co'}^*(\oli{\inn_\vef\econ_H})\in \baSE(\jq)\cap \KA'\qquad\forall f\in \Co,\vef\in \Wek^1(X).\label{dreh}
\end{eqnarray}
Since $\baSE(\fq)\cap\KA'=\baSE(\Fred_1)$ by lemma \ref{Adcinv}, it follows from this that $\econ_{\Hq}$ defines an adjoined action on $\baSE(\Fred_1)$ by the reduced bracket on $\KA'$ and thus, a covariant derivative on $\Sekt(\Fred_1)$. Since by definition, the structure group $\Nq$ acts effectively on the standard fiber of $\Fred_1$, this determines a principal connection form on $\Pq$. We claim that this is a form 
\bes
\Ahl_{\Hq}:\Pq\longrightarrow ad(\ka)
%={\fra N}_{ad(\g)}(\ha)/ad(\ha) 
\hspace{0.1cm}\subset\hspace{0.1cm} {\nq}=Lie({\Nq})=\fra{n}/\stab_1,
\end{equation*}
where $\fra{n}=Lie(\Nau)$ and $\stab_1=Lie(\Stab_1)$. Since the map $\admo_1:\Nau\to Gl(\ka)$ of definition \ref{kerndef} is in fact to $Aut(\ka)$, and since the covariant derivative is defined by an adjoined action, $\Ahl_{\Hq}$ must indeed take values in the inner derivations of $\ka$. Elementary verification confirms that
$(ad(\g)\cap \fra{n})/(ad(\g)\cap \stab_1)=ad(\fra{N}_{\g}\ha)/ad(\ha)=ad(\ka)$. Alternatively, we deduce form (\ref{dreh}) and lemma \ref{Adcinv} that the form $\Ahl_H=\Ahl^{\econ_H}$ on $\Pa_X$ takes values in  $ad(\g)\cap \fra{n}=ad(\fra{N}_{\g}\ha)$, and consequently, restricts to $\Pred$. The equivariance of principal connection form implies that the composition of $\Ahl_H|_{\Pred}$ with the induced Lie algebra map $Lie(\admo_1)|_{ad(\fra{N}_{\g}\ha)}: ad(\fra{N}_{\g}\ha)\to ad(\ka)$ is constant on the fibers of the projection $pr:\Pred\to\Pq$, and $ Lie(\admo_1)\circ \Ahl_H|_{\Pred}=pr^*\Ahl_{\Hq}.$

From (\ref{invariance}), we deduce that for $H\in \No$, we have
\beq
\Lie_{X_f}(j^V\Hq)=0=\Lie_{X_f}\kaq\qquad \qquad \forall f\in \Co.
\label{etainv}
\end{equation}
by duality. Recall that the Lie derivative is well-defined with the help of an arbitrary extension. Now, $\kaq\in \Sekt(S^2\djq)$ can be 
seen as a function on $\jq$ and thus, restricted to a function on $\fq$. Since the flow of $X_f$ preserves $\fq$, it has an induced action on $C^\infty(\fq)$, and in particular, on the restriction of the function defined by $\kaq$, which is defined by $\scalg_{\Hq}\in \Sekt(S^2\dfq)$. 
Hence, from (\ref{etainv}) follow 
\bes
\Lie_{X_f}\scalg_{\Hq}=0 \hspace{0.1cm}\forall f\in \Co
\hspace{0.1cm}\Leftrightarrow\hspace{0.1cm}\Hq_{20}'\in \baSE(S^2\fq)\cap \KA'=\baSE(\Fred_2)\hspace{0.1cm}\Leftrightarrow\hspace{0.1cm} \scalg_{\Hq}\in \Sekt(\Fred_2^*),
\end{equation*}
where $\Hq_{20}'$ is the function defined by $\scalg_{\Hq}^{\flat-1}$ on $\dfq$, and we used corollary \ref{HamVFlin}, as well as lemmas \ref{linspannen2} and \ref{Adcinv}. This completes the proof of the theorem. \qed

\begin{cor}
There is a well-defined curvature $\Kru_{\Hq}=\Kru^{\econ_{\Hq}}$ of $\econ_{\Hq}$, given by
\bes
\Kru_{\Hq}(\vef_1,\vef_2)=\{\econ_{\Hq}(\vef_1),\econ_{\Hq}(\vef_2)\}-\econ_{\Hq}([\vef_1,\vef_2]) \qquad \forall \vef_1,\vef_2\in \Wek^1(X),
\end{equation*}
taking values in $\Fred_1$. Furthermore, the curvature form of $\Ahl_{\Hq}$ takes values in $ad(\ka)$.
\end{cor}

\begin{cor}
Let $\Hq'_1=\Hq'_0\circ (\econ_{\Hq}^\flat)^*$, where $\econ_{\Hq}^\flat:TX\to \jq$, and $\Hq'_0$ denotes the Hamiltonian defined by $\emet_{\Hq}^{\flat-1}$ on $\T X$. Then, $\Hq'_1\in \KA'$.
\end{cor}

\noindent
The restriction of $\pomo_H$ to $\co'$ is obviously given by the bundle isomorphism  
\bes
\pomo_{\Hq}=\pomo_H|_{\djq}=((\econ_{\Hq}^\flat)^*,\ti \chela):\djq\longrightarrow\T X\oplus_X \dfq=:\fq^\pi\subset \bfx.
\end{equation*}
Thus, $\fq^\pi$ is a coisotropic submanifold of $(\bfx,\po_H)$. We set $\KA'_{\Hq}=(\pomo_{\Hq}^{-1})^*(\KA')$.

\begin{defin}
\begin{rm}
The triples $(\co',\KA',\Hq'_1)$ and $(\fq^\pi, \KA'_{\Hq},pr_1^*\Hq'_0)$ are called the {\it reduced} and {\it reduced gauged Wong systems} associated to $(Z,\po,H;\co)$ at $X$. The  triples $(\co',\KA',\Hq'_2)$ and $(\fq^\pi, \KA'_{\Hq},pr_1^*\Hq'_0+pr_2^*\Hq'_{20})$ are called the {\it reduced} and {\it reduced gauged Einstein-Mayer systems} associated to $(Z,\po,H;\co)$ at $X$.  
\end{rm}
\end{defin}

\subsection{Structure of the scalar fields}

Here we characterize the elements $S^2(\hs)^C$ which appeared as the values of the scalar fields $\scalg_{\Hq}$ defined by $\ti H$, i.e., $Ad_*(\Lc)$-invariant scalar products on $\g/\ha$, in the 
following example of Manton \cite{Manton}:
\begin{gather*}
\g=\mathfrak{so}(3)\oplus \mathfrak{su}(3)\qquad\qquad \hspace{0.5cm}\ha\hspace{0.1cm}=\Delta(\mathfrak{so}(2)\oplus \mathfrak{u}(1))\cong \mathfrak{u}(1)\hspace{0.3cm}\\
G=SO(3)\times SU(3)\qquad\qquad \Lc=\Delta(SO(2)\times U(1))\cong U(1),
\end{gather*}
where $\Delta$ we denoted the diagonal with respect to the standard identifications $SO(2)\cong U(1)$ and $\mathfrak{so}(2)\cong \mathfrak{u}(1)$. A possible realization is, for $a\in \lbrack0,2\pi\lbrack$,
\bes
\Lc=\left\{h^a:=\left(
\begin{pmatrix}
\cos(a) &-\sin(a)& 0\\
\sin(a) & \cos(a) & 0\\
0 & 0 & 1
\end{pmatrix},
\begin{pmatrix}
e^{ia} & 0 & 0\\
0 & e^{ia} & 0\\
0 & 0 & e^{-2ia}
\end{pmatrix}
\right)\right\}
\subset SO(3)\times SU(3),
\end{equation*}
with $\Lc\cong U(1)$ via $h^a\mapsto e^{ia}$, and with the isomorphism $\mathfrak{so}(3)\cong\mathfrak{su}(2)$ given by 
\beq
\fra{so}(3)\ni
\begin{pmatrix}
 0 & -a &  c\\
 a &  0 & -b\\
-c &  b &  0
\end{pmatrix}
\longmapsto
\frac{1}{2}\begin{pmatrix}
ia & b+ic \\
-b+ic & -ia
\end{pmatrix}
\in\fra{su}(2),
\label{standforsuso}
\end{equation}
\bea
\g&=&\left\{\left(
%\frac{1}{2}
\begin{pmatrix}
ia & \beta \\
-\bar\beta & -ia
\end{pmatrix},
%\frac{1}{2}
\begin{pmatrix}
 i(g+f)& \gamma &  \delta\\
 -\bar\gamma &  i(g-f) & \varepsilon\\
-\bar\delta &  -\bar\varepsilon &  -2ig
\end{pmatrix}
\right)\right\} \hspace{0.2cm}=\hspace{0.2cm}\fra{su}(2)\oplus \fra{su}(3)\nonumber\\
\ha&=&\hspace{0.2cm}\left\{\left(
%\frac{1}{2}
\begin{pmatrix}
ia & 0 \\
0 & -ia
\end{pmatrix},
%\frac{1}{2}
\begin{pmatrix}
 ia & 0 &  0 \\
 0 &  ia & 0\\
 0 &  0 &  -2ia
\end{pmatrix}
\right)\right\}\hspace{0.2cm}\hspace{1cm}= \hspace{0.2cm}\Delta(\fra{u}(1)\oplus\fra{u}(1))\nonumber
\end{eqnarray}
with $a,f,g\in \R$ and $\beta=b+ic,\gamma,\delta,\varepsilon\in \C$. That is, we have eleven real or, alternatively, three real and for complex linear coordinate functions on $\g$. These form a basis of $\gs$, and obviously
\bea
\ha^0= \langle \beta,f,g-a,\gamma,\delta,\varepsilon\rangle&\subset&
\gs=\langle a,\beta,f,g,\gamma,\delta,\varepsilon\rangle.
\nonumber
\eea
The coadjoint action of $h^a\in \Lc$ on $\gs$ reads explicitly
\bes
Ad^*(h^a):\left\{
\begin{matrix}
a\longmapsto a& g\longmapsto g&\beta \longmapsto e^{-ia}\beta&\delta\longmapsto e^{-3ia}\delta\\
f\longmapsto f&\gamma\longmapsto \gamma&&\varepsilon\longmapsto e^{-3ia}\varepsilon
\end{matrix}
\right. .
\end{equation*}
Generally speaking, every irreducible representation of $U(1)$ is isomorphic to one of the (complex) representations
\bes
\rho_n:\Lc\cong U(1)\longrightarrow U(1)\subset GL(1,\C)\qquad h^a\longmapsto e^{nia}\qquad n\in \Z
\end{equation*}
for precisely one $n\neq 0$, while $\rho_0$ decomposes into two copies of the trivial one-dimensional real representation.  Thus, our decomposition reads
\bes
\ha^0=(\R)^4\oplus \C \oplus (\C)^2=\langle f,\gamma_1,\gamma_2,g-a\rangle \oplus \langle \beta\rangle \oplus \langle \delta,\epsilon\rangle
\end{equation*}
where the three terms contain the subrepresentations  with $n=0$, $n=-1$ and $n=-3$, respectively, and
the exponents count respective multiplicities. 

Recall that there is a canonical isomorphism (cf e.g. \cite{Hein}, III 1.10)
\beq
\otimes^2(\hs)^\Lc\cong Hom_\Lc(\g/\ha,\hs)
\label{schur1}
\end{equation}
of the space of invariant bilinear forms on $\g/\ha$ with the space of $\Lc$-equivariant homomorphisms of $\g/\ha$ and its dual representation. Also, by Schur's lemma, a homomorphism of irreducible representations is either an isomorphism, or zero. 

Now, the trivial real one-dimensional representation is clearly isomorphic to its dual representation, and for every (underlying real) representation $\rho_n $, $n\neq 0$ of $\Lc$, the map $z\mapsto \delta(z)$ with $\delta(z)(w)=\Re(\bar z w)$ for $z,w\in \C$ provides an isomorphism with the dual representation, corresponding to the invariant real scalar product  $(z,w)\mapsto \Re(\bar zw)$ induced by the natural hermitian form on $\C$. Thus, every irreducible representation is isomorphic to its dual and thus, not isomorphic to the dual of any non-isomorphic representation. In addition, the space of such isomorphisms is one-dimensional in all cases. In deed, this is trivial for the trivial representation. For $\rho_n$ with $n\neq 0$, given another $\Lc$-invariant real scalar product on $\C$, it is necessarily the real part of a hermitian form on $\C$, and the composition of the two corresponding isomorphisms would yield an equivariant isomorphism of $\C$ as a complex representation, which by Schur's lemma must be a multiple of the identity. Hence, the space of hermitian forms is complex one-dimensional, as is the space of symmetric forms.   

As a result, the $Ad_*(\Lc)$-invariant scalar products on $\g/\ha$ are precisely the elements $S^2(\hs)$ corresponding to the quadratic function on $\g/\ha$ given in matrix notation by
\bes
q=(f\hp\gamma_1\hp\gamma_2\hp (g-a))
A
(f\hp \gamma_1\hp\gamma_2\hp (g-a))^T + l|\beta|^2+
(\bar\delta\hp\bar \varepsilon)
\mathcal B
(\delta\hp\varepsilon)^T,
\end{equation*}
where $T$ denotes transposition, $A\in Sym(4,\R), l\in \R$, and $\mathcal B\in Herm(2,\C)$, that is, $A$ and $\mathcal B$ are symmetric and hermitian matrices, respectively.  Thus,
\beq
S^2(\ha^0)^\Lc\cong Sym(4,\R)\oplus \R \oplus Herm(2,\C)\cong \R^{10}\oplus \R \oplus \R^4.
\label{hnulldec}
\end{equation}

\begin{remark}
\begin{rm}
We notice that in \cite{Talmadge}, Higgs fields are modelled by Hermitian forms which appear in the interaction term between spinor fields. \bems
\end{rm}
\end{remark}

Next, we want to decompose the space (\ref{hnulldec}) into irreducible subspaces under the action of the reduced structure group. Note that since $G$ is semi-simple, $Aut(\g)=Ad(\g)$, and since $G$ and $\Lc$ are both connected, $N_G(\Lc)/\Lc=N_G(\ha)/\Lc$.
It follows that $\Nq=Ad_*(N_G(\Lc)/\Lc)$, and thus, it suffices to consider the coadjoint action of $\ti G=N_G(\Lc)/\Lc$ or $\ka=\mathfrak{N}_{\g}(\ha)/\ha$ on $\hs$. It can be shown (\cite{Coquereaux}) that (locally)
\bea
\ti G&\stackrel{loc}{\cong}&N_{SO(3)}(SO(2))/SO(2)\times Z_{SU(3)}(U(1))\cong\Z_2\times U(2)\nonumber\\
&\stackrel{loc}{\cong}& \left\{
\begin{pmatrix}
u & 0\\
0 & \det^{-1}(u)
\end{pmatrix},
 u\in U(2)\right\}\cong U(2)\nonumber\\
\ka
%=\mathfrak{N}_{\g}(\ha)/\ha
&\cong& \mathfrak{N}_{\mathfrak{so}(3)}(\mathfrak{so}(2))/\mathfrak{so}(2)\times \mathfrak{Z}_{\mathfrak{su}(3)}(\mathfrak{u}(1))
%\nonumber\\
\cong\left\{
%\left(
%\frac{1}{2}
%\begin{pmatrix}
%0 & 0 \\
%0 & 0
%\end{pmatrix},
%\frac{1}{2}
\begin{pmatrix}
 U &  0\\
0 & -tr(U)
\end{pmatrix}
%\right)
, U\in \fra{u}(2)\right\}\cong\mathfrak{u}(2)
%=\left\{
%\left(
%\begin{pmatrix}
%0 & 0 \\
%0 & 0
%\end{pmatrix},
%\begin{pmatrix}
% i(g+f)& \gamma &  0\\
% -\bar\gamma &  i(g-f) & 0\\
%0 & 0 &  -2ig
%\end{pmatrix}
%\right)
%\right\}
\nonumber
\end{eqnarray}
which act only on the $\fra{su}(3)$ part. In particular, $a$ and $\beta$, and thus $l|\beta|^2$ and $l$
are unaffected by $\ti G$. Furthermore, recall that $U(2)$, which is isomorphic to a semi-direct product $SU(2)\rtimes U(1)$, is doubly covered by the map
\beq
SU(2)\times U(1)\stackrel{2:1}{\rightarrow}U(2)\quad
\left(
\begin{pmatrix}
\mu&\nu\\
-\bar \nu&\bar \mu
\end{pmatrix},
\begin{pmatrix}
e^{ig} & 0\\
0&e^{ig}
\end{pmatrix}
\right)
\mapsto
\begin{pmatrix}
e^{ig}\mu&e^{ig}\nu\\
-e^{ig}\bar \nu&e^{ig}\bar \nu
\end{pmatrix}\label{coversuuu}
\end{equation}
where $|\mu|^2+|\nu|^2=1$ as usual, with the corresponding Lie algebra isomorphisms given by $\fra{su}(2)\oplus \fra{u}(1)\stackrel{\sim}{\rightarrow}\fra{u}(2)$.
%\bes
%\fra{su}(2)\oplus \fra{u}(1)\stackrel{\sim}{\rightarrow}\fra{u}(2) \quad
%\left(
%\begin{pmatrix}
%if&\gamma\\
%-\bar\gamma &-if
%\end{pmatrix},
%\begin{pmatrix}
%ig & 0\\
%0&ig
%\end{pmatrix}
%\right)
%\mapsto
%\begin{pmatrix}
%i(g+f)&\gamma\\
%-\bar \gamma&i(g-f)
%\end{pmatrix}.
%\end{equation*}
Recall that $SU(2)\times U(1)$ would be the structure group of standard electroweak gauge theory, with $\fra{su}(2)\oplus \fra{u}(1)$ as structure algebra.

%Now, the decomposition $\fra{u}(2)=\fra{su}(2)\oplus \fra{u}(1)$ is already the decomposition into irreducible subspaces of the adjoint representation of $U(2)$. 
%In deed, for $u=sz\in U(2)$ with $s\in SU(2)$ and $z\in U(1)$ as in (\ref{coversuuu}), $U=Y+ig\in \fra{su}(2)\oplus \fra{u}(1)=\fra{u}(2)$, we have
%\bes
%Ad_*(u)(U)=Ad_*(sz)(Y+ig)=Ad_*(s)(Y)+ig.
%\end{equation*}
%In order to calculate the coadjoint $\ti G$-action on $\ha^0$ %and the terms $Sym(4,\R)$ and $Herm(2,\C)$, 

Let now $n\in\ti G$, and $X\in \fra{su}(3)$, and $u=sz\in U(2)$ such that
%, $U=Y+ig\in \fra{u}(2)=\fra{su}(2)\oplus \fra{u}(1)$ as above, and $v\in \C^2$ 
such that
\bes
n=
\begin{pmatrix}
u&0\\
0&\det^{-1}(u)
\end{pmatrix}=
\begin{pmatrix}
sz&0\\
0&z^{-2}
\end{pmatrix} \quad\qquad s\in SU(2), z\in U(1).
% \qquad \qquad
%X=
%\begin{pmatrix}
%Y+ig&v\\
%-\bar v^T&-2ig
%\end{pmatrix}.
\end{equation*}
Then, the action of $n$ on $S^2(\hs)^C$, parameterized by the triple $(A,l, \mathcal{B})$, reads
% Since diagram (\ref{diag45}) implies that the matrix of $Ad_*(s)$ is precisely $Ad_*(s)$ seen as an element of $SO(3)$, $n$ acts as follows:
\bea
A=\begin{pmatrix}
A'& w\\
w^T& d \end{pmatrix}
&\longmapsto&
\begin{pmatrix}
Ad_*(s)A'Ad_*^{-1}(s)& Ad_*(s)w\\
w^T Ad_*^{-1}(s)& d
\end{pmatrix}
\nonumber\\
\mathcal{B}&\longmapsto& s \mathcal{B}s^{-1}\qquad \qquad l\hspace{0.1cm}\longmapsto\hspace{0.1cm} l\nonumber,
\end{eqnarray}
where $A'\in Sym(3,\R)$, $w\in \R^3$ and $d\in\R$. Because of the canonical identification
$Herm(2,\C)_0\ni X\mapsto iX\in  Antiherm(2,\C)_0=\fra{su}(2)$,
where the subscript 0 denotes the traceless subspace, the action on $\mathcal B$ decomposes in fact into a trivial one-dimensional and the adjoint representation of $SU(2)$. That is, the space of invariant symmetric two forms on $\ha^0$ as in (\ref{hnulldec}) decomposes under $\ti G$ as
\bea
S^2(\ha^0)^\Lc&=&(Sym(3,\R)_0\oplus \R\oplus \R^3\oplus \R)\oplus \R \oplus (Herm(2,\C)_0\oplus \R)\nonumber\\
&=&(\R^5)\oplus (\R)^4\oplus (\R^3)^2\nonumber,
\end{eqnarray}
where the trivial representations $(\R)$ correspond to $tr(A')$, $d$, $l$ and $tr(\mathcal B)$, and $(\R^3)$ is the coadjoint representation of $SU(2)$.

\section{Local Kaluza-Klein realization}

Finally, we would like to study the link, provided by the idea of symplectic realization, of our constructions with classical gravitational, Yang-Mills and Kaluza-Klein theory (\cite{Kaluza,Klein,Kerner}), which also inspired Einstein's and Mayers' work (\cite{EinsteinMayer1,EinsteinMayer2}). 
In addition to our earlier notation, let $\rho:(\Uh,\ow)\to (U,\po)$ be a symplectic realization of dimension $d_{min}$ at $z_0\in X\cap U$ of an open subset $U\subset Z$, whose level sets define the foliation $\mathcal \fas$ in $\Uh$ with connected leaves. Theorem \ref{essentialunique} implies its {\it local uniqueness} up to symplectomorphism. As before, $V=S\cap U$, but suppose $X\subset V$ for simplicity. Let the surjective submersion $\rho|_{\Vh}:\Vh=\rho^{-1}(V)\to V$ be defined as in (\ref{fibration3}), and suppose that {\it $\Vh$ is a fiber bundle with standard fiber $\fas$}. 
%that is, there are an open covering $\{V_i,i\in I\}$ of $V$, and  diffeomorphisms $\Vh_i\cong V_i\times \fas$, where $\Vh_i=\rho^{-1}(V_i)$. 
Note that $\Uh$ does not need to be split. Then, we define the restricted bundle $Y\subset \Vh$ over $X$ by
\beq
Y=\rho^{-1}(X)\stackrel{\yp}{\longrightarrow} X\qquad \qquad \yp=\rho|_Y.
\label{fiberbundle}
\end{equation}
We call $i_Y:Y\to \Uh$ the inclusion.

\subsection{The Kaluza-Klein realization}
\label{Kaluzasection}

The total space $Y$ is a Lagrangian submanifold of $\Uh$, since it is coisotropic as the preimage of a coisotropic submanifold, and from lemma \ref{Faserbuendel}, it follows that $2\dim Y=2\dim X+2\dim \fas =\dim V +\dim \Nh=\dim \Uh$. Thus, the Sternberg-Weinstein approximation to $\Uh$ at $Y$ is given by
\bes
\Uh'=T\Uh_{|Y}/TY\stackrel{(-\ow^\flat_Y)^*}{\longrightarrow} \T Y,\qquad\ow_Y^\flat=\ow^\flat|_{TY}:TY \to (TY)^0, 
\end{equation*}
where $(-\ow^\flat_Y)^*$ is a symplectomorphism for the canonical symplectic structure on $\T Y$ as in corollary \ref{linspannenauf}. 

\begin{prop}\label{MWdualpairprop}
The projection $\rho':\Uh'\to \UX$ induced by $T\rho|_Y$ yields a symplectic realization of $\UX$. If the quotient $\Upsilon=\Uh/\mathcal \fas^\bot$ by the orthogonal foliation to $\mathcal \fas$ has a manifold structure such that we have dual pair $\Upsilon\stackrel{\la}{\longleftarrow}\Uh\stackrel{\rho}
{\longrightarrow}  U$, there is a projection $\la'$ induced by $T\la|_Y$ and a dual pair
\beq
\Upsilon'\stackrel{\la'}{\longleftarrow}\Uh'\stackrel{\rho'}{\longrightarrow} \UX,
\label{MWdualpair}
\end{equation}
where $\UX$ is the Sternberg-Weinstein approximation of $U$ at $X\cap U$, and $\Upsilon'$ is the Sternberg-Weinstein approximation of $\Upsilon$ at $l_0$. In addition, we have $\Upsilon'\cong\gsl$.
\end{prop}

\noindent
{\sc Proof:}
If $X\not\subset U$, then $\UX$ is of course the restriction of $\ZX=\dxb$ to $X\cap U$. So we can suppose $X\subset U$ as above. Since $\rho$ is a surjective submersion, the same is true for $\rho'$. For local Darboux coordinates $(x^\mu,p_\mu,r_a)$ on $U$, inducing local Darboux coordinates $(x^\mu,[\dot p_\mu],[\dot r_a])$ on $\UX$ as in proposition \ref{podot}, it is always possible to choose Darboux coordinates $(\hat x^\mu=\rho^*x^\mu,\hat p_\mu=\rho^* p_\mu,\hat y^a,\hat p_a)$ on $\Uh$, inducing Darboux coordinates $(\hat x^\mu,[\dot{\hat p}_\mu]=(\rho')^*[\dot p_\mu],\hat y^a,[\dot{\hat p}_a])$ on $\ZX\cong \T Y$ which are holonomic with respect to coordinates $(\hat x^\mu|_Y, \hat y^a|_Y)$ on $Y$. In these coordinates, direct calculation shows that $\rho'$ induces the Sternberg-Weinstein Poisson structure.   

Furthermore, suppose that $\Upsilon=\Uh/\mathcal \fas^\bot$ is a manifold such that $\la$ becomes a submersion. As before it follows that (\ref{MWdualpair}) forms a dual pair. Lemma \ref{Upsilonistnull} implies that the symplectic leaf in $\Upsilon$ corresponding to $V$ and thus, any Lagrangian submanifold inside it is reduced to a point $l_0$, and $\Upsilon'=T_{l_0}\Upsilon\cong\gsl$ (cf (\ref{wegenantiiso})).  \qed

\begin{defin}
\begin{rm}
The Sternberg-Weinstein approximation $\Uh'\cong \T Y$ will be called the {\it Kaluza-Klein realization} of $Z$ at $z_0$. We call $\wep:\Uh'\to Y$ the natural projection.
\end{rm}
\end{defin}

\begin{cor}\label{SW=SW}
The Sternberg-Weinstein approximation of the Sternberg-Wein-stein phase space $\T P/G$ associated to a principal fiber bundle $P\to B$ with structure group $G$ at $B\subset \T B$ is naturally isomorphic to $\T P/G$.
\end{cor}

\noindent
{\sc Proof:} Obviously, $\T P$ is a global symplectic realization of $\T P/G$ of dimension $d_{min}$. Furthermore, the canonical isomorphism $(\T P)'=T(\T P)|_P/TP\cong \T P$ is symplectic and compatible with the projection maps $\rho:\T P\to \T P/G$ and $\rho':T(\T P)|_P/TP\to T(\T P/G)|_B/TB$ since $\rho$ is defined by a fiberwise linear action of $G$. Thus, it induces a natural Poisson equivalence of the quotients. \qed

\begin{lem}\label{actionlem}
Suppose that the surjective Poisson submersion $\la':\Uh'\to \Upsilon'=\gsl$ in proposition \ref{MWdualpairprop} exists. Then, the fibers of $\rho'|_Y$ are given by the orbits of a local right action of the connected and simply connected Lie group $\Lg_1$ with Lie algebra $\gl$ on $Y$, and the fibers of $\rho'$ are given by the orbits of the canonical Hamiltonian lift to $\Uh'\cong\T Y$ of this local action, which in addition is locally free. Furthermore, there is a natural bundle morphism $\Te:Y\to \Pa_X$ over $Id_X$ which is locally equivariant with respect to the canonical homomorphism $Ad:\Lg_1\to Aut(\g)$.
\end{lem}

\noindent
{\sc Proof:} It follows from the proofs of propositions \ref{momentintegriert} and  \ref{linreal} that $\la'$ induces a local right action of the Lie group $\Lg_1$ on $\Uh '$, and that the fibers of $\rho'$ are given by the orbits of this local action. The  fundamental vector fields of $D\in \gl\subset C^\infty(\gsl)$ is the Hamiltonian vector field of  $-(\la')^*D$. Since $\la'$ is a surjective, it follows that these Hamiltonian vector fields span $T\Uh'$ at each point and thus, the action is locally free. For $y\in Y$, $\la'|_y:T_y\Uh/T_yY\cong\T_yY\to \gsl$ is a linear map, an thus, $-(\la')^*D|_{\T yY}$ is a linear function, that is, an element of $T_yY$. Hence, $-(\la')^*D$ is a vertically linear function on $\Uh'\cong\T Y$ given by a vector field $D_Y\in \Wek(Y)$. 

On the other hand, it is well-known that the Hamiltonian vector field of a function defined on $\T Y$ with its canonical symplectic structure by a vector field on $Y$ is precisely the unique Hamiltonian lift of this vector field. In particular, $D_Y=X_{-(\la')^*D}|_Y$. Thus, the local action of $\Lg_1$ on $\T Y$ is the unique Hamiltonian lift of the restricted local action on $Y$. Consequently, we can define the map 
\bes
\Te:Y\to \Pa_X\qquad \Te(y)(D)=(pr_2\circ (\ow^{\flat-1}\circ\canco)^{-1}\circ D_Y)(y)\qquad \forall D\in\g
\end{equation*}
where $\canco$ is the bundle morphism defined in (\ref{kmorphismus}). Note that $(\ow^{\flat-1}\circ\canco)(Y\times_X(TV)^0_{|X})=VY$ so that we have in deed $\Te(y):\g\to \fb_{\yp(y)}$, and since $\ow^\flat$ and $D\to D_Y$ and are Lie algebra anti-automorphisms (since $\g$ is anti-isomorphic to $\gl$), and $\rho$ is Poisson a morphism, this is a Lie algebra isomorphism. 
On the other hand, $\Te$ is equivariant since 
%$((Ad_*(g)(D))_Y(y)=R_{g*}^{-1}\circ D_Y(yg)$, 
\bea
\Te(yg)(D)&=&(pr_2\circ (\canco)^{-1}\circ \ow^\flat\circ D_Y)(yg)\nonumber\\
&=&(pr_2\circ (\canco)^{-1}\circ \ow^\flat\circ TR_g\circ (Ad_*(g)(D))_Y)(y)\nonumber\\
&=&(pr_2\circ (\canco)^{-1}\circ (\rho'_{yg,g})^*\circ \ow^\flat\circ (Ad_*(g)(D))_Y)(y)\nonumber\\
(\Te(y)\circ Ad_*(g))(D)&=&(pr_2\circ (\canco)^{-1}\circ \ow^\flat\circ((Ad_*(g)(D))_Y)(y)\nonumber
\end{eqnarray}
where $R_g:Y\to Y $ denotes the action of $g\in \Lg_1$ on $Y$ whenever it is defined, and 
\bes
\rho'_{yg,g}=\rho'|_{T_{y}\Uh/T_{y}Y}^{-1}\circ \rho'|_{T_{yg}\Uh/T_{yg}Y}:\hspace{0.1cm} T_{yg}\Uh/T_{yg}Y\to T_{y}\Uh/T_{y}Y
\end{equation*}
since we saw above $(TR_g)^*\circ(-\ow^\flat_Y)^*=(-\ow^\flat_Y)^*\circ \rho'_{yg,g}$. We then used  $\canco|_{Y\times_X(TX)^0}=(\wep,\rho')^*$ for last two equalities (cf lemma \ref{rhodaul}). \qed

\begin{prop}\label{KKdynproj}
Suppose that $\la:\Uh\to \Upsilon$ exists. Let $H\in C^\infty(U)$ be a Hamiltonian such that $dH|_X=0$, and $\KK\in C^\infty(\Uh)$ such that $d\KK|_Y=0$ and $\rho_*X_\KK=X_H$. Then, the Einstein-Mayer systems $(\Uh',\ow',\KEM)$ and $(\UX,\po',\HEM)$ are $\rho'_*$-related. Under the usual non-degeneracy assumption,  $\KEM=\bar{\ymet}^{-1}$, where $\ymet$ is a metric on $Y$, and we wrote $\ymet^{-1}$ for the section of $S^2(TY)$ defining $\ymet^{\flat-1}$. Furthermore, we have a decomposition
\beq
\bar{\ymet}^{-1}=(\rho')^*\bar{\kal}^{-1}+(\la')^*\bar{\iota}^{-1}, \label{covKaluzadec}
\end{equation}
where $\iota$ is a scalar product on $\gl$, and the term $(\rho')^*\bar{\kal}^{-1}$ corresponds to a $\Lg_1$-invariant metric for the (local) $\Lg_1$-action of lemma \ref{actionlem}.
\end{prop}

\noindent
{\sc Proof:} If $\la$ exists, 
Lemma \ref{Aufspaltungslemma} yields a decomposition $\KK=\rho^*H +\la^* \OHa$ for some $\OHa\in C^\infty(\Upsilon)$. Thus $d\KK=\rho^*dH +\la^* d\OHa$, and similarly as in the proof of theorem \ref{redHinv}, this implies $\KEM=(\rho')^*\HEM +(\la')^* \OHa'_2$, where $\OHa'_2$ is the Einstein-Mayer approximation of $\OHa$ at $l_0$. This yields (\ref{covKaluzadec}) with $\bar{\iota}^{-1}=\OHa'_2$, and $\rho'_* X_{\KEM}=X_{\HEM}$. The term $(\rho')^*\bar{\kal}^{-1}$ is a $\Lg_1$-invariant function which is quadratic on the fibers of $\Uh'\cong \T Y$, and thus, corresponds to a  $\Lg_1$-invariant metric on $Y$. \qed

\begin{remark}\label{ambiguity}
\begin{rm}
Recall that given a function $\KK$ as above, the decomposition of lemma \ref{Aufspaltungslemma} is not unique. In the Einstein-Mayer approximation, the fields $\econ_H$ and $\emet$ are obviously well-defined, while the fields $\scalg_H$ and $\iota$ are defined up to a Casimir function.
If we can choose $H$ and $\OHa$ such that the field $\scalg_H$ vanishes, then the projection of the Kaluza-Klein metric dynamics coincides with the Sternberg-Weinstein system, and we will call $\ymet$  (or $\kal$) of {\it Sternberg-Weinstein type} (e.g. in proposition \ref{KKmetricprop}). If we can make the choice such that $\iota$ vanishes, then we will say that $\ymet$ is of {\it invariant type} (e.g. if $\KK=\rho^*H$). If both cases happen simultaneously, we will then say that $\ymet$ (or $\kal$) is of {\it ad-type}. This happens iff $\scalg_H$ and $\iota$ both correspond to an $ad(\g)$-invariant scalar product. Other cases will be called of {\it mixed type}.\bems
\end{rm}
\end{remark}

\paragraph{Nonlinear metric Hamiltonians.}

Darboux's theorem implies that the local symplectic realization $\Uh$ is locally isomorphic to its Sternberg-Weinstein approximation $\Uh'\cong\T Y$, and it can be shown that the symplectomorphisms can be forced to induce the identity on $Y$ identified with the zero section in $\T Y$. Thus, we have an inclusion $i_{\Uh}:\Uh\hookrightarrow \T Y$ identifying $\Uh$ with an open neighborhood of the zero section in $\T Y$. 
Consequently, it becomes simultaneously a symplectic realization of $U$ and its Sternberg-Weinstein approximation $\UX$.
If $\KK\in C^\infty(\Uh)$ is such that $\rho_*X_{\KK}=X_H$ for a $H\in C^\infty(U)$, the inclusion $i_{\Uh}$  allows us to assume that $\KK$ is well-defined by a metric on $Y$, that is, that $\KK=\KEM$. Then, proposition \ref{KKdynproj} implies that we also have $\rho'_*X_{\KK}=X_{\HEM}$. In summary:
\bea
&(\UX,\po',X_{\HX})\stackrel{(\rho',\rho'_*)}{\longleftarrow}
(\Uh,\ow,X_\KK)\stackrel{(\rho,\rho_*)}{\longrightarrow}
(U,\po,X_H)&\nonumber
\end{eqnarray}

Hence, we can consider the Poisson structure $\po$ as a nonlinear deformation of the Poisson structure $\po'$, obtained by a nonlinear deformation in $\Uh$ of the foliation $\mathcal \fas'$ defined by $\rho'$ to the foliation $\mathcal \fas$ defined by $\rho$. Equally, the metric Hamiltonian system defined by $H$ appears as a nonlinear deformation of the Sternberg-Weinstein system defined by $\HEM$. The fact that both systems are projections of a system defined by the same metric $\ymet$ restricts the possible deformations to the natural subclass. In particular, it can be shown that the equations of motion obtained in this way are natural non-linear analogues to the Wong equations, involving a non-linear Yang-Mills potential and field strength.  Here, we will just state an interesting result which is an easily proved in local coordinates.

\begin{defin}
\begin{rm}
A metric $\ymet$ for which the induced Hamiltonian vector field
projects to a Hamiltonian vector field on $U$
will be called a {\it projectable  metric}. A Hamiltonian $H$ obtained from a projectable metric
will be called a {\it metric Hamiltonian}.
\end{rm}
\end{defin}

\begin{theo}\label{theo3}
If $Z$ is linearizable at the leaf $S\supset V$, then, locally and in linear coordinates, any metric Hamiltonian $H$ defines equations of motion which coincide with the equations of motion of the corresponding Einstein-Mayer approximated system. That is, the dynamics is entirely given by the symmetric bilinear form $\kal$ defined in theorem \ref{theo2}. 
\end{theo}

\begin{remark}
\begin{rm}
Since the Lie algebras common in physics are semi-simple of compact type, or semidirect products with $\R$, theorems \ref{linearization1} and \ref{theo3} imply that {\it the Wong equations are the most general Hamiltonian equations of motion for particles obtained as the projection of geodesic motion on some higher-dimensional space}. \bems
\end{rm}
\end{remark}

\subsection{Local $\g$-valued principal connection forms}

Since our $\jx$-connection forms are locally $\g$-valued (cf definition \ref{trivialalpha}), we expect them to be related to $\g$-valued principal connection forms. This is in deed the case.

\begin{lem} \label{invform=connect}
%Let $\rho:\Uh\to U$ be a minimal symplectic realization of $U\subset Z$ such that $V=U\cap S\neq\emptyset$ as in (\ref{minimalreal}), defining the foliation $\mathcal \fas$ of $\Uh$. 
Let $\theta:V\to \T V\otimes_V \jb_{|V}$ be the restriction to $V$ of an $\jb$-connection
form. There is
an associated connection on
the bundle $\rho|_{\Vh}:\Vh\to V$ given by the bundle morphism
\bea
&&\hat \theta=-\ow^{\flat-1}\circ \canco \circ (Id_{\Vh},\theta^\flat):
\quad \Vh\times_V TV \to T\Vh,
\label{ausformconnection}\\
\text{where}&&\canco:\Uh\times_U \T U\to (T\mathcal \fas)^0\subset \T \Uh\label{kmorphismus}
\end{eqnarray}
is the canonical bijective morphism of fibred manifolds over $U$ induced by the surjective submersion $\rho:\Uh\to U$.
\end{lem}

\noindent
{\sc Proof:}
Because of lemma \ref{Faserbuendel}, ${\mathcal{\fas}}\cap \Vh$,
i.e. the fibers of $\rho|_{\Vh}:\Vh\to V$,
form the characteristic foliation of the coisotropic submanifold
$\Vh\subset W$. This implies that
$\ow^{\flat-1}(T{\mathcal \fas})^0|_{\Vh})=T\Vh$. Thus, the bundle
morphism $\hat \theta$ is well defined.
In order to show that it defines a connection in $\rho:\Vh\to V$,
%i.e. a splitting of the corresponding exact sequence
%(\ref{zusammenhangssequTang}),
it remains to show that
\beq
(\tau_{\Vh},T\rho)\circ \hat \theta=Id_{\Vh\times_V TV}.
\label{nochzuzeigen}
\end{equation}
But for any $v\in V$, $u\in \rho^{-1}(v)$, (\ref{nochPoiscond}) implies that
\bes
T_u\rho\circ (-\ow^{\flat-1})_u\circ (\canco)_u \circ \theta^\flat_v=(\ank)_{v}\circ
(\theta^\flat)_v
= Id_{T_vV}\nonumber
\end{equation*}
since $\theta$ is an $\jb$-connection form. Since both sides of
(\ref{nochzuzeigen}) are bundle morphisms over
$Id_{\Vh}$, this proves the lemma.
\hfill $\square$

\begin{lem}\label{rhodaul}
Let $\acon:X\to \T X\otimes_X \jx$ be an $\jx$-connection
form. There is
an associated  connection on
the bundle $\rho:Y\to X$ defined in (\ref{fiberbundle}) given by the bundle morphism
\beq
\hat \acon=-\ow^{\flat-1}\circ \canco\circ (Id_{Y},\acon^\flat):
\quad Y\times_X TX \to TY,
\label{restrcitedconnection}
\end{equation}
where $\canco$ is the bundle morphism (\ref{kmorphismus}), whose restriction to $Y\times_X (TX)^0$ is precisely the dual to $\hat \rho=(\wep,\rho')$.
\end{lem}

\noindent
{\sc Proof:}
Lemma \ref{invariant-restricted} allows us to look at
$\acon$ as the restriction of an $\jb$-connection form $\theta$,
and we have $\hat \acon=\hat \theta \circ (i_Y,Ti_{X})$.
Since $(\tau_{\Vh},T\rho)\circ \hat \theta=
Id_{\Vh\times_V TV}$, it follows that
$(\tau_{Y},T\yp)\circ \hat \acon=Id_{Y\times_X TX}$,
and in particular, this
shows that $\hat \acon$ takes its values in $TY$. Thus, $\hat \acon$ defines a connection on $Y$. By definition, the restriction $\canco:Y\times_X (TX)^0\to(TY)^0$ is dual to $\hat \rho=(\wep,\rho'):T\Uh|_Y/TY\to Y\times_X TZ|_X/TX$. \qed

\begin{lem}\label{princlem}
Suppose that $\la'$ exists and is complete. Then, the bundle $Y$ is a fiber bundle with homogeneous standard fiber $\Lg_{y_0}\backslash \Lg_1$, where $\Lg_{y_0}$ is the discrete stabilizer subgroup of a point $y_0\in\fas$, and the $\Lg_1$-action corresponds to the right action of $\Lg_1$ on $\Lg_{y_0}\backslash \Lg_1$. In particular, if $G_{y_0}$ is normal, then $Y$ is a principal bundle with structure group $\Lg=\Lg_{y_0}\backslash \Lg_1$. Furthermore, we have natural isomorphisms $\fx\cong \g(Y)$ and $\bfx\cong \T X\times_X \gs(Y)$, with respect to the (co)adjoint action of $\Lg$. 
\end{lem}

\noindent
{\sc Proof:} From our definition of $Y$, it follows that there are local trivializations $Y_i\cong X_i\times \fas$, where $\{X_i,i\in I\}$ is an open covering of $X$, $Y_i=\rho^{-1}(X_i)$, and $\fas=\rho^{-1}(z_0)$. On the other hand, the local $\Lg_1$-action constructed in lemma \ref{actionlem} extends to a global action if $\la'$ is complete. Since we assumed that $\fas$ was connected, this yields a locally free transitive action on $\fas$, and thus, via the choice of $y_0\in\fas$, a diffeomorphism $\fas\cong \Lg_{y_0}\backslash\Lg_1$ as claimed. Note this map will be $\Lg_1$-equivariant since it is induced by the restriction of a Poisson morphism. If $\Lg_{y_0}$ is normal, then $Y$ becomes a principal fiber bundle with structure group $\Lg$, and the bundle morphism $\Te$ of lemma \ref{actionlem} induces the isomorphisms of associated bundles as stated in the lemma defined by $\g(Y)\ni[y,D]\mapsto [\Te(y),D]\in \fx$ etc.  \qed

\begin{prop}\label{princprop}
Under the hypotheses of lemma \ref{princlem}, suppose that $G_{y_0}$ is a normal subgroup. If $\acon$ is an $\jx$-connection form, then the connection on $Y$ induced by the associated bundle morphism $\hat \acon$ defined in (\ref{restrcitedconnection}) is a principal connection corresponding to the $\gl$-valued form $\hat A$ on $Y$ given by the bundle morphism
\bea
&&\hat A^\flat = pr_2\circ(-\ow^{\flat-1}\circ \lanco)^{-1}\circ (Id_{TY}-\hat \acon \circ (\tau_Y,T\yp)):\hspace{0.1cm}TY\to \T_{l_0}\Upsilon=\gl\nonumber\\
&&\text{where}\qquad\lanco:\Uh\times_\Upsilon \T\Upsilon \to (T\mathcal \fas^\bot)^0\subset \T \Uh\nonumber
%\label{lancodefin}
\end{eqnarray}
is the canonical bijective morphism of fibred manifolds over $\Upsilon$ induced by the surjective submersion $\la:\Uh\to \Upsilon$. Furthermore, $ad \circ \hat A^\flat=(\Te^*\Ahl^\acon)^\flat$.
\end{prop}

\noindent
{\sc Proof:} First, we note that $\hat A$ can be defined as a principal connection form on a covering space if $Y$ is not a principal bundle. Here we regard only the special case.

In the proof of lemma \ref{actionlem}, we saw that the fundamental vector field induced by $D\in \gl\subset C^\infty(\Upsilon')$ at $y\in Y$ was given by
\bea
&&D_Y(y)=X_{-{\la'}^*D}(y)=-\ow^{\flat-1}\circ \lanco(y,D)\label{princcon}\\
&&\text{where} \quad -\ow^{\flat-1}\circ \lanco:\hspace{0.1cm}Y\times \T_{l_0}\Upsilon\longrightarrow(T\Vh)^0_{|Y}\longrightarrow VY\nonumber
%\label{Adefincon}
\end{eqnarray}
is a bundle isomorphism since $\Vh=\la^{-1}(l_0)$ (lemma \ref{Upsilonistnull}) is coisotropic with  a characteristic foliation given by the fibers of $\yp=\rho|_Y$ (lemma \ref{Faserbuendel}). But (\ref{princcon}) implies that $\hat A(I)_Y=(Id_{TY}-\hat \acon \circ (\tau_Y,T\yp))(I)$ for all $I\in TY$. Thus, the vertical projection corresponding to $\hat \acon$ is given by $I\to\hat A(I)_Y$. Thus, $\hat A$ will be a principal connection form corresponding to $\hat \acon$ iff $\hat \acon$ is equivariant, that is, if $TR_g\circ \hat \acon=\hat \acon \circ (R_g,Id_{TX})$ for all $g\in \Lg$. But this follows form a similar calculation as at the end of the proof of lemma \ref{actionlem}. 

Finally, the fibers of the $Ad_*$-equivariant bundle morphism $\Te:Y\to\Pa_X$, are precisely the orbits of the center $Z=\ker Ad_*\subset \Lg$.
Thus, the $\Lg$-equivariance of $\hat A$ implies that $ad\circ \hat A^\flat$ is constant on these fibers. Furthermore, it is easy to see that $ad \circ \hat A$ and $\Ahl^\acon$ induce the same covariant derivative on $\fx\cong \g(Y)$, and thus, $ad \circ \hat A^\flat=(\Te^*\Ahl^\acon)^\flat$ as claimed. \qed

\begin{cor}
The $\gl$-valued curvature 2-form $\hat F=d\hat A +\eha [\hat A,\hat A]$ of the principal connection $\hat A$ corresponding to $\acon$ is given by the bundle morphism
\bes
\hat F^\flat =-pr_2\circ (\lanco)^{-1}\circ \canco\circ (\tau^2_Y,(\Kru^\al)^\flat\circ \wedge^2T\yp):\hspace{0.1cm} \wedge^2 TY \to \T_{l_0}\Upsilon=\gl,
\end{equation*}
where $\tau^2_Y:\wedge^2 TY\to Y$ is the natural map, and $(\Kru^\acon)^\flat:\wedge^2TX\to \fx$ is the bundle morphism defined by curvature of $\acon$. Furthermore, $ad (\hat F)=\Te^*\Fk^\acon$, where $\Fk^\acon$ denotes the curvature 2-form of $\Ahl^\acon$.
\end{cor}

\noindent
{\sc Proof:} The map $(Id_{TY},\acon^\flat)\circ (\tau_Y,T\yp)=(\tau_Y,\acon^\flat\circ T\yp)$ is the horizontal projection corresponding to $\hat A$. For $\vef_1,\vef_2\in \Wek^1(Y)$, we have, writing $T\yp(\vef_1)$ for $T\yp\circ \vef_1$ etc.,
\bea
&&(Id_{TY}-\hat \acon \circ (\tau_Y,T\yp))([\hat \acon\circ (\tau_Y,T\yp)(\vef_1), \hat \acon\circ (\tau_Y,T\yp)(\vef_2)])=\nonumber\\
&&\qquad=[\hat \acon\circ (\tau_Y,T\yp)(\vef_1), \hat \acon\circ (\tau_Y,T\yp)(\vef_2)]-\nonumber\\
&&\qquad\hspace{3cm}-\hat \acon \circ (\tau_Y,T\yp)([\hat \acon\circ (\tau_Y,T\yp)(\vef_1), \hat \acon\circ (\tau_Y,T\yp)(\vef_2)])\nonumber\\
&&\qquad=[\hat \acon\circ (\tau_Y,T\yp)(\vef_1), \hat \acon\circ (\tau_Y,T\yp)(\vef_2)]-\hat \acon ([(\tau_Y,T\yp)(\vef_1),(\tau_Y,T\yp)(\vef_2)])\nonumber\\
&&\qquad=[\hat \acon\circ (\tau_Y,T\yp)(\vef_1), \hat \acon\circ (\tau_Y,T\yp)(\vef_2)]-\hat \acon\circ (\tau_Y,T\yp) ([\vef_1,\vef_2])\nonumber\\
&&\qquad=[-\ow^{\flat-1}\circ \canco\circ (\tau_Y,\acon^\flat\circ T\yp)(\vef_1),-\ow^{\flat-1}\circ \canco\circ (\tau_Y,\acon^\flat\circ T\yp)(\vef_2)]+\nonumber\\
&&\qquad\hspace{4cm}+\ow^{\flat-1}\circ \canco\circ (\tau_Y,\acon^\flat\circ T\yp)([\vef_1,\vef_2])\nonumber\\
&&=-\ow^{\flat-1}\circ \canco\circ (\tau^2_Y(\vef_1\wedge \vef_2),\{\acon^\flat\circ T\yp(\vef_1),\acon^\flat\circ T\yp(\vef_2)\}-\acon^\flat \circ T\yp([\vef_1,\vef_2]))\nonumber\\
&&=-\ow^{\flat-1}\circ \canco\circ (\tau^2_Y(\vef_1\wedge \vef_2),\{\acon(T\yp(\vef_1)),\acon(T\yp(\vef_2))\}-\acon([T\yp(\vef_1),T\yp(\vef_2)]))\nonumber\\
&&=-\ow^{\flat-1}\circ \canco\circ (\tau^2_Y(\vef_1\wedge \vef_2),(\Kru^\acon)^\flat(T\yp(\vef_1)\wedge T\yp(\vef_2)))\nonumber
\end{eqnarray}
since $-\ow^{\flat-1}\circ \canco$ is a Lie algebra isomorphism. On the other hand,
\bea
\hat F^\flat(\vef_1,\vef_2)&=& d\hat A(\hat \acon\circ (\tau_Y,T\yp)(\vef_1), \hat \acon\circ (\tau_Y,T\yp)(\vef_2))\nonumber\\
&=&-\hat A([\hat \acon\circ (\tau_Y,T\yp)(\vef_1), \hat \acon\circ (\tau_Y,T\yp)(\vef_2)])\nonumber\\
&=&-pr_2\circ(\lanco)^{-1}\circ \canco\circ (\tau^2_Y(\vef_1\wedge \vef_2),(\Kru^\acon)^\flat(T\yp(\vef_1)\wedge T\yp(\vef_2)))\nonumber
\end{eqnarray}
This last assertion follows easily. \qed

%The natural homomorphism $Ad:G\to Aut(\g)$ induces a homomorphism of principal bundles $\hat{Ad}:Y\to (Y\times_G Aut(\g))$, where the $g\in G$ act on $Aut(\g)$ via left multiplication by $Ad(g)$ (cf e.g. \cite{Percacci}, prop. 1.5.1). Note that this indeed makes $Y\times_G Aut(\g)$ into a principal $Aut(\g)$-bundle. Thus, it is isomorphic to $\Pa_X$ because all our bundles are trivial over        $X$. Furthermore, there is an induced isomorphism of associated bundles $\Check{Ad}:\gl(Y)\to \g(\Pa_X)$.

\begin{remark}
\begin{rm}
Recalling that $\jx$-connection forms are related to {\it global} $ad(\g)$-valued principal connection forms, we could say that localization allows the passage from the adjoined structure group to the group itself. A similar transition appears in the duality of certain quantized gauge theories. \cite{tHooft}
\bems
\end{rm}
\end{remark}

\subsection{Reduced Kaluza-Klein realization}

Let $V\subset \co\stackrel{i_\co}{\hookrightarrow} U$ be a locally closed coisotropic constraint manifold. Then
\bes
i_{\ch}:\ch=\rho^{-1}(\co) \hookrightarrow \Uh
\end{equation*}
is obviously a coisotropic submanifold in $\Uh$. Let $\Ihq$ denote the characteristic foliation of $\ch$, and suppose that there is a reduction of $\Uh$ via $\ch$ given by 
\beq
\cm:\ch\longrightarrow \ch/\Ihq=:\cb, \qquad \text{such that}\quad (\cb,\ti \ow)
\label{reducWQ}
\end{equation}
is a symplectic manifold with the reduced symplectic form $\ti \ow$. If we define {\it constraints} and {\it admissible functions} by 
\bes
\Ch=\{f\in C^\infty(\Uh)|f_{|\ch}=0\}
\qquad \qquad\Noh=\Noa_{C^\infty(\Uh)}(\Ch),
\end{equation*}
we obtain the canonical isomorphisms $i_{\ch}^*\Noh\cong\Noh/\Ch\cong C^\infty(\cb)$.

Since $\ch$ is a union of leaves of the foliation $\mathcal \fas$ defined by $\rho$, $\Ihq$ is a subfoliation of the orthogonal foliation $\mathcal \fas^\bot$, which we suppose to  define a submersion $\la:\Uh\to \Upsilon$, and $\Fgr^\bot\subset \Noh$. Consequently, there is a unique map $\tila:\cb\to\Upsilon$ such that $\la\circ i_{\ch}=\tila\circ\cm$, every $\la^*f\in\Fgr^\bot$ reduces to $\tila^*f\in C^\infty(\cb)$, and $\tila^*$ is an isomorphism of Poisson algebras. Furthermore, the orbits of the flow of the Hamiltonian vector fields of $\ti\Fgr^\bot=\tila^*C^\infty(\Upsilon)$ are precisely given by $\mathcal \cf:=\cm(\mathcal \fas\cap{\ch})$. Since $\tila$ is not surjective, this will be a nonregular foliation of $\cb$ in general, and the image of 
\bes
\cp:\cb\longrightarrow \hig:=\cb/\mathcal \cf
\end{equation*}
will not be a manifold. However, we can define
$C^\infty(\hig)$ and an isomorphic subalgebra of $C^\infty(\cb)$ by
\bes
C^\infty(\hig)=(\cp^*)^{-1}C^\infty(\cb)\qquad \quad
\ti \Fgr=\fra\Zen_{C^\infty(\cb)}(\ti\Fgr^\bot)=\cp^*C^\infty(\hig)
%\{f\in C^\infty(\cb)|f \hspace{0.2cm}\mbox{const. on the leaves of $\mathcal\cf$}\}
.
\end{equation*}
On the other hand, $\ch=\rho^{-1}(\co)$ also implies that
$\Co\cong \Ch\cap \Fgr$ and
$\No\cong \Noa_{C^\infty(\Uh)}(\Ch\cap \Fgr)\cap \Fgr=\Noh \cap \Fgr$, and thus, $\hig$ can be identified with the image of the map $\qq$ of (\ref{quotbysubchar}). With (\ref{constraintidenti}) (for $\hig$ instead of $\Zhig$), we get the commutative diagrams:
\beq
\begin{CD}
\Upsilon\hspace{0.1cm}\stackrel{\la}{\longleftarrow}\hspace{0.1cm}\ch@>\rho>>\co\\
\tila\nwarrow\hspace{0.2cm}@VV\cm V @VV\qq V\\
\qquad\quad\cb@>\cp >> \hig
\end{CD}\qquad\qquad\quad
\begin{CD}
i_{\ch}^*(\Noh\cap \Fgr)@<\sim<< \ti\PA\\
@AA\wr A @AA\wr A\\ 
\ti \Fgr @<\sim<< C^\infty(\hig)
\end{CD}
\label{high}
\end{equation}

\begin{lem}\label{KKreduc}
Let $\KK\in C^\infty(\Uh)$ be a Hamiltonian such that $\rho_* X_\KK=X_H$ for some Hamiltonian $H\in C^\infty(U)$. Then, $\KK$ is admissible for $\ch$ iff $H$ is admissible for $\co$. In this case, the reduced Hamiltonian $\Ga\in C^\infty(\cb)$ is given as $\Ga=\ti\rho^*\Hq_{\hig}+\tila^*\OHa$, where $\Hq=\qq^*\Hq_{\hig}$, and $\ti\rho^*\Hq_{\hig}$ is precisely the reduction of $\rho^*H$ by $\ch$.
\end{lem}

\noindent
{\sc Proof:} Lemma \ref{Aufspaltungslemma} gives us the decomposition $\KK=\rho^*H+\la^*\OHa$, where $\OHa\in C^\infty(\Upsilon)$. $\No\cong \Noh \cap \Fgr$ implies that $\rho^*H$ is admissible for $\ch$ iff $H$ is admissible for $\co$. On the other hand, $\Fgr^\bot\subset \Noh$
 shows that every $\la^*\OHa\in \Fgr^\bot$ is admissible for $\ch$. This implies the remaining assertions. \qed
\vspace{0.4cm}

Hence, we may think of 
%$\cb$ and 
the dynamics induced by $\Ga$ on $\ti\Fgr$ as a ''symplectic realization'' of 
%the Poisson algebra on $C^\infty(\hig)$ and 
the reduced dynamics induced by $H$ on $\KA$.

\begin{prop}\label{redapproxKKreal}
Let $\ch=\rho^{-1}(\co)\subset \Uh$ be a coisotropic constraint manifold as above such that there is a reduction $(\cb,\ti \ow)$ of $\Uh$ by $\ch$ as in (\ref{reducWQ}). Then  
\bes
i_{\ch'}:\ch'=T\ch|_Y/TY\longrightarrow \Uh'
\end{equation*}
is a coisotropic submanifold. There is a reduced manifold of $\Uh'$ by $\ch'$ given by the Sternberg-Weinstein approximation of $\cb$ at $\qq(Y)$, that is, by
\bes
\cb'=T\cb_{|\ti Y}/T\ti Y\cong\T \ti Y, \qquad \text{where}\qquad
\ti Y:=\cm(Y)\subset \cb
\end{equation*}
is a Lagrangian submanifold, and the identification with $\T \ti Y$ is given by $(\ti \ow^\flat|_{T\ti Y})^*$. The projection defined by the characteristic foliation $\Ihq'$ of $\ch'$ is given by the map $\cm':\ch'\to \cb'$ naturally induced by $T\cm|_{Y}$. Furthermore, the restriction of local $\Lg_1$-action of lemma \ref{actionlem} to $\ch'$ projects to a local $\Lg_1$-action on $\cb'$ which is the Hamiltonian lift of the restricted projected action on $\ti Y$. There is a natural bijection $\ti\rho':\hig'\stackrel{\sim}{\to}\co'/\Iq'$ of the orbit space $\hig'$ with the set of leaves of the sub-characteristic distribution $\Iq'$ of $\co'$. 
\end{prop}

\noindent
{\sc Proof:} Since $\ch'=(\rho')^{-1}(\co')$, $\ch'$ is coisotropic. Furthermore, in local Darboux coordinates providing an identification $\Uh\cong\Uh'\cong\T Y$, it is easy to see that $\ker \ow'|_{\ch'}=T\Ihq|_Y/TY=\ker \cm'$, and obviously, $\cm'(\ch')=T\cb|_{\ti Y}/T\ti Y=\cb'$, where $\ti Y=\cm(Y)$ is a Lagrangian submanifold as the reduced image of a Lagrangian submanifold of $\Uh$. If $\la':\Uh'\to \gsl$ exists, we see that there is an induced map $\tila':\cb'\to \gsl$ such that $\la'\circ i_{\ch'}=\tila'\circ \cm'$. Thus, every $(\la')^*D, D\in \gl$ yields a reduced Hamiltonian $(\tila')^*D$ on $\cb'$, 
and as in the proof of lemma \ref{actionlem}, this yields a local $\Lg_1$-action which is given by the unique Hamiltonian lift of the action on $\ti Y$, and at the same time the projection of the restricted $\Lg_1$-action on $\ch'$. 
However, this action will not be locally free if $\tila'$ is not surjective. The last assertion follows from the commutativity of (\ref{high}); in fact, it means that there is a Sternberg-Weinstein approximated analogue of these diagrams. \qed

\begin{cor}\label{hombuendel}
The orbits of the restricted projected $\Lg_1$-action on $\ti Y$ define a regular foliation of $\ti Y$. Under the hypotheses of lemma \ref{princlem} and proposition \ref{princprop}, $\ti Y$ is a fiber bundle over $X$ with homogeneous fibers diffeomorphic to $\Lc^-\backslash\Lg$, where $\Lc^-$ is the analytic subgroup of $\Lg$ defined by $\ha\subset \g$. The projected $\Lg$-action in given by the natural right action of $\Lg$ on $\Lc^-\backslash\Lg$.  
\end{cor}

\noindent
{\sc Proof:} Lemma \ref{linspannen2} implies that $T\Iq$ and  $T\Ihq'$ are spanned at each point by the Hamiltonian vector fields of the functions $\ti s$ and $(\rho')^*\ti s$ for all $s\in \Sekt(\cx)$, respectively. Since $V'=\SX\cap \UX\subset \co'$ and thus $(T\co')^0_{|V'}\subset (TV')^0$, the Hamiltonian vector field of $\ti s$ vanishes on $X$. Thus, that of $(\rho')^*\ti s$ must be tangent the fibers of $\rho'$ on $Y$. On the other hand, and as in the proof of lemma \ref{actionlem}, we see that the restriction to $Y_x$ depends only on $s(x)\in (\cx)_x$ for all $x\in X$ and corresponds to an infinitesimal Lie algebra action of $(\cx)_x$ on $Y_x$. Fixing a point $y\in (\cx)_x$, we get (local) isomorphisms $\Lg\cong Y_x$ and $\Te(y):\ha\cong(\cx)_x$. Under these identifications, the fiber of $\Ihq'$ through $y$ is locally given by the orbit of the local $\Lc^-$-action obtained by integrating the infinitesimal $\ha$-action on $\Lg$, that is, the natural left action, since $\ha\subset \g$ corresponds to {\it right} invariant vector fields on $\Lg$. Thus, the $\Lg_1$-orbits in $\ti Y=Y/(\Ihq'\cap Y)$ are locally isomorphic to $\Lc^-\backslash\Lg$, and thus, the foliation is regular. Under the additional hypotheses, $\ti Y$ is a fiber bundle over $X$ with homogeneous fibers isomorphic to $\Lc^-\backslash\Lg$. \qed

\begin{prop}
Let $H\in C^\infty(U)$ such that $dH|_X=0$ be admissible for $\co$, and $\KK\in C^\infty(\Uh)$ such that $d\KK|_Y=0$ and $\rho_* X_\KK=X_H$. Suppose that there is a reduction $(\cb,\ti \ow)$ as in (\ref{reducWQ}). Then, the reduction of the Hamiltonian defining the Einstein-Mayer system to $\KK$ on $\Uh'$ is precisely the Hamiltonian $\ti\KEM$ defining the Einstein-Mayer system to the reduced Hamiltonian $\Ga$ of $\KK$ on $\cb'$. Under the usual non-degeneracy assumption, $\ti\KEM=\bar{\ti \ymet}^{-1}$, where $\ti\ymet$ is a metric on $\ti Y$, and we wrote $\ti \ymet^{-1}$ for the section of $S^2(T\ti Y)$ defining $\ti \ymet^{\flat-1}$. Furthermore, we have a decomposition
\beq
\bar{\ti \ymet}^{-1}=(\ti \rho')^*\bar{\ti \kal}^{-1}+(\tila')^*\bar{\iota}^{-1}, \label{covreddec}
\end{equation}
where $\iota$ is a scalar product on $\gl$. The function $ \bar{\ti \kal}^{-1}\in \No'$ is seen as a function on $\hig'$ here. The term $(\ti \rho')^*\bar{\ti \kal}^{-1}$ corresponds to a $\Lg_1$-invariant metric for the (local) $\Lg_1$-action of lemma \ref{actionlem}. 
\end{prop}

\noindent
{\sc Proof:} 
Proposition \ref{KKdynproj} yields a decomposition  $\KEM=(\rho')^*\HEM +(\la')^* \OHa'_2$ or, under the non-degeneracy assumption, 
$\bar{\ymet}^{-1}=(\rho')^*\bar{ \kal}^{-1}+(\la')^*\bar{\iota}^{-1}$.
Applying lemma \ref{KKreduc} in the Sternberg-Weinstein approximated situation, we obtain the decomposition (\ref{covreddec}) if we can show that the reduction of $\KEM$ is given by $\ti \KEM$. But this follows easily from $\cm^*d\Ga=i_{\ch}^*d\KK$. The last assertion follows from proposition \ref{redapproxKKreal}. \qed
\vspace{0.4cm}

Together with the results on the reduced Sternberg-Weinstein approximation, this reproduces the main results of Kaluza-Klein-theory on bundles with homogenous fibers, in particular, the {\it Reduction theorem} \cite{Coquereaux1,Coquereaux}.

\subsection*{\itshape \bfseries  Acknowledgments}

The author would like to express his gratitude to his teacher Daniel Bennequin, for continuous support and advise during the last years. He is also greatly indebted to  Paul Gauduchon, Richard Kerner, Tudor Ratiu, John Rawnsley, and Alan Weinstein for many helpful discussions and comments to the present work.

\end{document}